\pdfoutput=1 

\documentclass[11pt,letterpaper]{scrartcl}

\makeatletter
\DeclareOldFontCommand{\rm}{\normalfont\rmfamily}{\mathrm}
\DeclareOldFontCommand{\sf}{\normalfont\sffamily}{\mathsf}
\DeclareOldFontCommand{\tt}{\normalfont\ttfamily}{\mathtt}
\DeclareOldFontCommand{\bf}{\normalfont\bfseries}{\mathbf}
\DeclareOldFontCommand{\it}{\normalfont\itshape}{\mathit}
\DeclareOldFontCommand{\sl}{\normalfont\slshape}{\@nomath\sl}
\DeclareOldFontCommand{\sc}{\normalfont\scshape}{\@nomath\sc}
\makeatother

\usepackage[top=2.4cm, bottom=2.4cm, left=3.2cm, right=3cm,footskip=0.8cm]{geometry}

\usepackage[T1]{fontenc}
\usepackage[utf8]{inputenc}

\usepackage[hidelinks]{hyperref}
\usepackage[parfill]{parskip}
\usepackage[separate-uncertainty = true,multi-part-units=single]{siunitx}
\usepackage{slashed}
\usepackage{microtype}
\usepackage{csquotes}
\usepackage{graphicx}
\usepackage{amsmath}
\usepackage{amssymb}
\usepackage{braket}
\usepackage{booktabs}
\usepackage{caption}
\usepackage{subcaption}
\usepackage{todonotes}
\usepackage{grffile}

\usepackage[numbers,sort&compress]{natbib}

\usepackage{multirow}

\usepackage{fancyhdr}
\usepackage[yyyymmdd,hhmmss]{datetime}
\fancypagestyle{plain}
{
}
\usepackage{cleveref}

\usepackage[auth-sc,affil-it]{authblk}

\usepackage{scalefnt}
\newcommand{\abbrev}{\scalefont{.9}}

\newcommand{\NTHREENNLO}{\text{\abbrev N$^3$LL+NNLO}}
\newcommand{\NNLL}{\text{\abbrev N$^2$LL}}
\newcommand{\NNNLL}{\text{\abbrev N$^3$LL}}
\newcommand{\NNLO}{\text{\abbrev NNLO}}
\newcommand{\NNNLO}{\text{\abbrev N$^3$LO}}
\newcommand{\NLO}{\text{\abbrev NLO}}
\newcommand{\CSS}{\text{\abbrev CSS}}
\newcommand{\LO}{\text{\abbrev LO}}
\newcommand{\EFT}{\text{\abbrev EFT}}

\newcommand{\SM}{\text{\abbrev SM}}
\newcommand{\BSM}{\text{\abbrev BSM}}
\newcommand{\IR}{\text{\abbrev IR}}

\newcommand{\RG}{\text{\abbrev RG}}

\newcommand{\QCD}{\text{\abbrev QCD}}

\newcommand{\PDF}{\text{\abbrev PDF}}

\newcommand{\LHC}{\text{\abbrev LHC}}

\newcommand{\CMS}{\text{\abbrev CMS}}
\newcommand{\ATLAS}{\text{\abbrev ATLAS}}

\newcommand{\SCET}{\text{\abbrev SCET}}
\newcommand{\RGE}{\text{\abbrev RGE}}

\newcommand{\LHAPDF}{\text{\abbrev LHAPDF}}
\newcommand{\MCFM}{\text{\abbrev MCFM}}
\newcommand{\CuTe}{\text{\abbrev CuTe}}
\newcommand{\CuTeMCFM}{\text{\abbrev CuTe-MCFM}}

\newcommand{\NNPDFTO}{\texttt{NNPDF31\_nnlo\_as\_0118}}
\newcommand{\NNPDFTZ}{\texttt{NNPDF30\_nnlo\_as\_0118}}
\newcommand{\CTFOUR}{\texttt{CT14nnlo}}
\newcommand{\CTEIGHT}{\texttt{CT18NNLO}}
\newcommand{\MMHTFOUR}{\texttt{MMHT2014nnlo68cl}}
\newcommand{\MSTW}{\texttt{MSTW2008nnlo68cl}}

\newcommand{\MSBAR}{\text{\abbrev $\overline{\text{MS}}$}}

\newcommand{\abs}[1]{\lvert#1\rvert}
\newcommand{\xmin}{\ensuremath{x^\text{max}}}
\newcommand{\etiso}{\ensuremath{E_T^{\gamma,\text{max}}}}

\newcommand{\phistar}{\ensuremath{\phi^*}}
\newcommand{\delphi}{\ensuremath{\Delta\phi}}

\newcounter{notecount}

\makeatletter
\renewcommand\maketitle{
	\begin{center}
		{\huge\bfseries\@title\par\vspace{0.3em}}
		{\scshape\@author, \@date}
	\end{center}
}
\makeatother

\fancypagestyle{firstpage}{%
	\lhead{}
	\rhead{FERMILAB-PUB-20-272-T, IIT-CAPP-20-02}
}

\graphicspath{{figs/}}

\begin{document}

\thispagestyle{firstpage}
\title{\LARGE {Fiducial $q_T$ resummation of color-singlet processes at \abbrev{N$^3$LL+NNLO}}}

\author[1]{Thomas Becher}
\author[2,3]{Tobias Neumann}

\affil[1]{Albert Einstein Center for Fundamental Physics, Institut f\"ur Theoretische Physik\\ 
Universit\"at Bern, Sidlerstrasse 5, CH-3012 Bern, Switzerland}
\affil[2]{Fermilab, PO Box 500, Batavia, Illinois 60510, USA}
\affil[3]{Department of Physics, Illinois Institute of Technology, Chicago, Illinois 60616, USA}

\date{}
\maketitle

\vspace{1cm}

\begin{abstract}
        We present a framework for $q_T$ resummation at \NTHREENNLO{} accuracy for arbitrary 
        color-singlet processes based on a factorization theorem in \SCET{}. Our 
        implementation \CuTeMCFM{} is fully 
        differential in the Born kinematics and matches to 
        large-$q_T$ fixed-order predictions at relative order $\alpha_s^2$. It provides an efficient way to estimate uncertainties from fixed-order truncation, resummation, and parton distribution functions. In addition to $W^\pm$, $Z$ and $H$ production, also the diboson processes $\gamma\gamma,Z\gamma,ZH$
        and $W^\pm H$ are available, including decays. We discuss and exemplify 
        the framework with several direct comparisons to experimental 
        measurements as well as inclusive benchmark results. In particular, we present novel results for $\gamma\gamma$ and $Z\gamma$ at \NTHREENNLO{} and discuss in detail the power corrections induced by photon isolation requirements.
\end{abstract}

\tableofcontents
\vspace{-5mm}

\clearpage

\section{Introduction}
\label{sec:introduction}

While hadron colliders were traditionally considered discovery machines, one cannot 
deny the success of the
Large Hadron Collider (\LHC{}) experiments in Standard Model (\SM{}) precision physics.
Already today, differential measurements at sub-percent level precision are available, a prime example being the transverse-momentum ($q_T$) spectrum of the $Z$ boson.
A related precision measurement at the \LHC{} is the extraction of the $W$-boson mass by the 
\ATLAS{} collaboration
\cite{Aaboud:2017svj}, heavily relying on a precise understanding of the charged lepton transverse-momentum distribution through a template fit.

Transverse-momentum distributions and the experimentally easier to measure, but closely associated,
$\phistar$ \cite{Banfi:2010cf} distributions in electroweak boson production are key observables for \SM{} precision tests. For example, the precise measurements and predictions 
of $Z$-boson transverse-momentum spectra allow for significant constraints on \PDF{}s 
\cite{Boughezal:2017nla} and might help to resolve tensions in existing PDF fits. The large data sets of the LHC also allow 
for increasingly precise diboson production measurements 
\cite{Khachatryan:2015sga,Aad:2016ett,Aaboud:2017oem,Aaboud:2017rwm,Sirunyan:2017zjc, 
Sirunyan:2019bez,Aaboud:2019lxo,Sirunyan:2019gkh,Aaboud:2019gxl}, which are key to test the gauge 
structure of the \SM{}, as was pointed out a long time ago \cite{Hagiwara:1986vm,Hagiwara:1989mx}. 
Recent theoretical studies of such processes include 
refs.~\cite{Frye:2015rba,Franceschini:2017xkh,Grojean:2018dqj,Baglio:2020oqu}. To increase 
sensitivity to Beyond the Standard Model (\BSM{}) effects it is important to veto QCD radiation. The 
transverse momentum can be used as a kinematic variable to
veto jets, see e.g. ref.~\cite{Franceschini:2017xkh}, and is therefore relevant also in the search for \BSM{} physics. It is therefore essential
for the physics program at the \LHC{} that theoretical predictions and associated 
uncertainties for these processes are under good control.

Predictions in fixed-order perturbation theory at hadron colliders start 
with a collinear factorization theorem 
involving
parton distribution functions and a hard scattering cross section at a scale $Q$, corresponding to the invariant mass of the final-state electroweak bosons.
However, when considering the kinematical distributions of transverse momenta at 
small values, fixed-order
corrections are enhanced by large Sudakov logarithms of scale ratios $Q^2/q_T^2$. To obtain meaningful results, the
fixed-order predictions need to be improved with an all-order resummation 
of such logarithms.
In this paper we address this issue by combining the fixed-order color-singlet \NNLO{} processes in \MCFM{} 
\cite{Boughezal:2016wmq,Campbell:2016jau,Campbell:2017aul,Campbell:2019dru} with
the \SCET{}-based $q_T$ resummation at \NNNLL{} introduced in 
refs.~\cite{Becher:2010tm,Becher:2011xn,Becher:2012yn,Becher:2019bnm}. The resulting
code \CuTeMCFM{} will be made publicly available shortly at 
\url{https://mcfm.fnal.gov}. 

\paragraph{Transverse-momentum resummation in \SCET{}}
The enhanced logarithms for small transverse momenta are universal and originate from soft and collinear radiation.
An all-order exponentiation theorem for the $q_T$ distribution was first obtained in 
ref.~\cite{Collins:1984kg} and is now known as the Collins-Soper-Sterman ({\abbrev CSS}) formula.

Two sources of enhanced terms exist. First, logarithms arising due to different scales associated with the hard process
and the soft/collinear radiation, and, secondly, logarithms generated by the rapidity difference of small-$q_T$ 
emissions
from partons flying along the beams to the left and right.
In \SCET{} \cite{Bauer:2000yr,Bauer:2001yt,Beneke:2002ph}\footnote{See \cite{Becher:2014oda,Becher:2018gno,Cohen:2019wxr} for reviews.} the first kind of logarithms are resummed by solving
the renormalization group equations (\RGE{}s) of the derived factorization theorem in the limit of 
small $q_T$. This was first considered in
refs.~\cite{Gao:2005iu,Idilbi:2005er,Mantry:2009qz} without accounting for the rapidity logarithms. Later, both
sources of logarithms have been taken into account for $q\bar{q}$-initiated processes  
\cite{Becher:2010tm,Becher:2011xn} and for $gg$-initiated processes \cite{Becher:2012yn,Chiu:2012ir}, and the 
equivalence 
to the \CSS{} formula was established. Instead of a direct exponentiation \cite{Becher:2010tm}, the rapidity
logarithms can also be resummed by solving rapidity \RGE{}s \cite{Chiu:2012ir,Chiu:2011qc}.

\paragraph{Resummation codes}

A number of computer codes for transverse-momentum resummation of color-singlet processes have been developed, some of which have been made publicly available.
They differ by the achieved logarithmic precision, the possibility of fiducial cuts on the final state colorless
particles, and in subleading terms through the use of different resummation and matching 
formalisms, in particular whether the computations are performed in momentum or impact 
parameter space.

Fiducial resummation in Drell-Yan production is available through {\abbrev DYRes} 
\cite{Bozzi:2010xn,Catani:2015vma}, its new implementation {\abbrev DYTurbo} 
\cite{Camarda:2019zyx}  and {\abbrev ReSolve} \cite{Coradeschi:2017zzw} at 
\NNLL{}$^\prime$, and at \NNNLL{}
without the possibility for fiducial cuts in \CuTe{}
\cite{Becher:2011xn,Becher:2012yn}.\footnote{The papers \cite{Becher:2011xn,Becher:2012yn} achieved 
\NNLL{}, but the accuracy was extended in version 2 of the  \CuTe{} code, see {\tt 
https://cute.hepforge.org}.} In addition, there are codes such as {\abbrev  arTeMiDe} 
\cite{Bertone:2019nxa} and {\abbrev NangaParbat} \cite{Bacchetta:2019sam}, with a special focus on 
non-perturbative transverse-momentum dependent ({\abbrev TMD}) physics. In a recent paper \cite{Ebert:2020dfc} \NNNLL{} fiducial results 
have been presented based on the private code {\abbrev SCETlib}. Fiducial resummation for Higgs  production is available through {\abbrev HRes}
\cite{deFlorian:2012mx,Grazzini:2013mca} at \NNLL{}$^\prime$. Codes for resummation in $W^\pm,Z,H,\gamma\gamma$ and $ZZ$ from various authors and at different accuracies are available under the name {\abbrev Resbos/Resbos2} \cite{Wang:2012xs,Balazs:1997xd,Ladinsky:1993zn}. 

The above results are based on analytic computations of the ingredients of the factorization 
theorem for the process at small transverse momentum. An alternative numerical resummation 
technique was developed in refs.~\cite{Banfi:2001bz,Banfi:2004yd,Banfi:2014sua}. In this formalism, 
the higher emissions are computed with Monte-Carlo methods. This numerical approach was generalised 
to transverse-momentum resummation in
ref.~\cite{Monni:2016ktx} and extended to \NNNLL{} in ref.~\cite{Bizon:2017rah}. The resulting 
resummation framework ({\abbrev RadISH}) has been interfaced to fixed-order codes for different 
color-singlet processes, which provide matching to order $\alpha_s^2$ at large $q_T$ 
\cite{Kallweit:2020gva,Wiesemann:2020gbm}, and even to order 
$\alpha_s^3$ for Higgs \cite{Bizon:2017rah} and $W$ and $Z$ production 
\cite{Bizon:2018foh,Bizon:2019zgf}.

It is of course also common, especially in the experimental collaborations, to rely on parton 
showers to dress fixed-order predictions with logarithmically enhanced terms \cite{Buckley:2011ms}. 
While these showers typically give a good description of experimentally measured spectra, they do 
not systematically include higher logarithmic terms and
need to be benchmarked against analytical resummation results such as the ones in this study. 

The modern 
approach of matching and merging often achieves impressive results in  predicting shapes of distributions, but the low logarithmic accuracy can be problematic.
Cross sections differential in transverse momentum typically peak around small values, so the bulk of the cross section comes from the region that needs an all-order resummation.
Therefore, it is clear that the fixed-order and logarithmic precision in this bulk region should be as high as possible. General purpose parton shower codes typically only reach fixed 
\NLO{} accuracy and leading logarithmic accuracy in the region of small $q_T$. For normalized 
distributions this limitation can
amplify and even invalidate the formal \NLO{} perturbative accuracy achieved in the fixed-order
tail regions \cite{Alioli:2012fc}. It is therefore important to use --  or at least compare with --
predictions that have known parametric accuracy and allow for systematic estimates of uncertainties.

\paragraph{Scheme choices}

The product form of the $q_T$-factorization formula arises in transverse position 
space, also known 
as impact parameter space. Following \CSS{}, it is therefore common to perform the resummation in 
impact parameter space and then compute the Fourier integral to obtain the transverse-momentum 
spectrum. A disadvantage of this procedure is that one ends up with running couplings that are 
functions of the impact parameter $b$, which is integrated from zero to infinity in the Fourier 
integral. This makes it 
necessary to choose a prescription to avoid Landau pole singularities. In the effective theory 
approach \cite{Becher:2010tm,Becher:2011xn} based on \RG{} 
evolution, which we adopt in our work, one instead first carries out the Fourier integral and then 
sets the boundary conditions of the evolution directly in $q_T$ space. The rapidity logarithms, on 
the other hand, which do not involve a running of the coupling, are exponentiated in position space 
in the formalism of refs.~\cite{Becher:2010tm,Becher:2011xn}. A method to resum all logarithms in 
$q_T$ space has been developed in ref.~\cite{Ebert:2016gcn}, but is challenging to implement. In 
any case, performing the resummation in  different spaces simply amounts to choosing different 
boundary conditions, which induce different subleading terms and different power corrections. 

A second source of subleading differences, on top of the choice of resummation space, is the 
matching to the fixed-order predictions \cite{Catani:1992ua}. A robust estimation of 
perturbative uncertainties therefore benefits from fully matched
results in different matching schemes and resummation formalisms. To some extend these
effects can be estimated within one framework, of course. For example, in our study we use a transition function to match our resummed results to fixed-order predictions. Varying this function provides a flexible way to estimate matching uncertainties.
One could furthermore deliberately choose to include different subleading terms in the resummation. Overall our \NTHREENNLO{} resummation framework allows for the estimation of \QCD{} uncertainties through variation of the renormalization, resummation and factorization scales (\enquote{scale uncertainties}), 
\PDF{}+$\alpha_s$ uncertainties, and matching uncertainties by varying the transition function. The combination of these should capture the bulk of uncertainties associated with a perturbative \QCD{} prediction.

\paragraph{Overview of the paper}

In this work we present a \SCET{}-derived transverse-momentum resummation framework and publicly available implementation \CuTeMCFM{} to
calculate fully matched predictions with fiducial cuts at \NTHREENNLO{} ($\alpha_s^2$ relative to 
the Born). The name \CuTeMCFM{} was chosen to emphasize that the implementation is based on 
refs.~\cite{Becher:2010tm,Becher:2011xn} as the earlier public code \CuTe{}. However, while we 
performed various cross checks against this earlier code, \CuTeMCFM{} is a new and completely 
independent implementation of the underlying equations. The code follows the same philosophy as 
ref.~\cite{Becher:2019bnm}, in that it uses an existing fixed-order code to compute the 
process-dependent parts of the resummation formula. Interfacing to \MCFM{} provides an efficient 
way of studying different processes and allows us to  take into account the decays of the 
electroweak bosons as well as cuts on the decay products.

While implemented in \MCFM{}, the code written for this study is not closely tied to \MCFM{}, so that it could easily 
be reused or integrated in other situations, for example as a stand-alone extension of the interface to event 
files used in ref.~\cite{Becher:2019bnm}, that is currently limited to \NNLL{} and quark-antiquark 
initiated 
processes.  The only essential input ingredients are the Born matrix element, the hard function at relative 
order $\alpha_s$ or $\alpha_s^2$, and numerically stable fixed-order predictions at $q_T>0$ for 
matching.

Relative to the Born-level boson production process, our framework achieves  $\alpha_s^2$ accuracy both at small and 
large $q_T$ through a consistent power counting of $\alpha_s$ and large logarithms. 
We demonstrate our implementation with fully matched kinematical distributions in $q_T$, 
$\phistar$, 
and with distributions in the azimuthal angle difference $\delphi$ between bosons.
We estimate scale uncertainties, \PDF{} uncertainties and matching uncertainties, and also address the impact of fiducial cuts on the size of subleading power corrections in the $q_T$ factorization. We include a detailed discussed of power corrections in processes involving photons, in which they are enhanced through the required photon isolation cuts.

In \cref{sec:calculation} we describe our framework and setup in detail. We discuss the
factorization theorem, the resummation of large logarithms through \RG{} evolution and
exponentiation of rapidity logarithms, the estimation of scale uncertainties and \PDF{} 
uncertainties,
matching to fixed-order predictions and differences to the code 
\CuTe{}. We discuss in detail subleading power corrections from fiducial cuts and photon isolation.
In addition, we provide details about the technical implementation, for example the ability to pre-generate beam-function grids.

In \cref{sec:results} we compare with the \CuTe{} code for $Z$ and $H$ production. 
We then show results for various processes with fiducial cuts and in comparison with experimental
measurements. For $Z$ production we compare with measurements at 
\SI{13}{\TeV}
and \SI{8}{\TeV}. For $Z\gamma$ production we compare with recent experimental data at 
\SI{13}{\TeV} and
 show novel results that have previously only been considered in fixed-order perturbation theory.
For diphoton production we compare against data at \SI{7}{\TeV} and recent data at 
\SI{8}{\TeV} and
improve upon previous predictions at \NNLL{}. We show results for Higgs production, both inclusively as part of our comparison with \CuTe{}, and in the $H\to \gamma\gamma$ decay mode with fiducial cuts.
We do not compare against the measured $q_T$ distribution for Higgs production that still has large uncertainties. The comparison would require a careful analysis of multiple production channels and top-quark mass effects, among other things, which go beyond the scope of our study.
 We furthermore compare with one of the  few direct $W$~boson transverse
momentum measurements. Resummation for the remaining processes $ZH$ and $W^\pm H$ is prepared in our code and ready for use. We conclude in \cref{sec:conclusions} and present an outlook for future studies based on this work.

\section{Resummation framework and implementation}
\label{sec:calculation}

\paragraph{Factorization formula} The $q_T$ resummation underlying our framework \CuTeMCFM{} has 
been derived in 
\SCET{} in refs.~\cite{Becher:2010tm,Becher:2011xn,Becher:2012yn}, where large logarithms
of argument $q_T/Q$ are resummed through \RG{} evolution of hard function and beam functions,
and rapidity logarithms are directly exponentiated through the collinear-anomaly formalism. 

The production of multiple weak bosons in this formalism has been detailed in 
ref.~\cite{Becher:2019bnm}. As in this work, we consider the production of $N$ weak bosons with 
momenta $\{ \underline{q} \} = 
\{ q_1, q_2, \dots, q_N\}$. The total boson momentum is denoted by $q^\mu = q_1^\mu + \dots + q_N^\mu$ and the resummation formalism is valid in the region where the transverse momentum $q_T = \sqrt{-q_\perp^2} $ is much smaller than the invariant mass $Q^2 = q^2$ of the electroweak final state.

The cross section is a sum of contributions from individual partonic channels $i,j\in{q,\bar{q},g}$. Up to terms suppressed by powers of $q_T$, these channels factorize as
\begin{multline}
\mathrm{d}\sigma_{ij}(p_1,p_2,\{\underline{q} \}) = \int_0^1 \mathrm{d}\xi_1 \int_0^1 \mathrm{d}\xi_2\, 
\mathrm{d}\sigma^0_{ij}(\xi_1 p_1,\xi_2 p_2,\{\underline{q}\})\, 
\mathcal H_{ij}(\xi_1 p_1,\xi_2  p_2,\{\underline{q} \},\mu) \,\cdot \\
\frac{1}{4\pi}\int\mathrm{d}^2x_\perp \,
e^{-iq_\perp x_\perp} 
\left(\frac{x_T^2 Q^2}{b_0^2}\right)^{-F_{ij}(x_\perp,\mu)} B_i(\xi_1,x_\perp,\mu) \cdot 
B_j(\xi_2,x_\perp,\mu)\,, 
\label{eq:fact}
\end{multline}
where $p_1$ and $p_2$ are the incoming hadron momenta. The cross section 
$\mathrm{d}\sigma_{ij}$ is fully differential in the electroweak momenta $\{ \underline{q} \}$.

 The beam functions $B_i$ and $B_j$ encode the soft and collinear emissions at low transverse momentum (or more precisely large transverse separation $x_\perp$) and the indices $i$ and $j$ and the momentum fractions $\xi_1$ and $\xi_2$ refer to the partons which enter the hard process after these emissions. The hard Born-level process has the differential cross section  $\mathrm{d}\sigma^0_{ij}$ and the 
hard-function as $\mathcal{H}_{ij}$ collects the associated virtual corrections. The collinear 
anomaly leads to the $Q^2$-dependent factor within the Fourier-integral over the transverse 
position $x_\perp$. The perturbatively calculable anomaly exponent $F_{ij}$ is also referred to as 
the rapidity anomalous dimension in the framework of ref.~\cite{Chiu:2012ir}.  In case of 
gluon-gluon initiated processes ($i=j=g$), a second product
of beam functions is added as required \cite{Catani:2010pd,Becher:2012yn}. Lastly, we have defined
$b_0=2e^{-\gamma_E}$, where $\gamma_E$ is the Euler-Mascheroni constant, and $x_T^2=-x_\perp^2$.

The hard function and the Born cross section are the only process-dependent ingredients 
in formula \eqref{eq:fact}. Since the hard function corresponds to
the \MSBAR{}-renormalized loop corrections to the Born amplitude and the
implementations 
of \NNLO{} corrections
in \MCFM{} are based upon a \SCET{}-derived factorization for jettiness $\tau$ \cite{Stewart:2010tn}, the 
\MSBAR{}-renormalized hard 
functions are readily available. Furthermore, the processes associated with $\tau>0$ correspond
to those with $q_T>0$ needed for the fixed-order matching, and are already well-tested and
numerically stable in the singular limits.

The hard function involves logarithms of the ratio $\mu^2/Q^2$, which are minimized with a choice of
$\mu = \mu_h^2 \sim Q^2$, but inside the beam functions the natural choice is $\mu\sim q_T$.  To 
avoid large logarithms of $q_T^2/Q^2$ one chooses $\mu_h\sim Q$ in the hard function and then 
evolves it down to 
the resummation scale $\mu \sim q_T$ using the \RG{}.  This evolution can be solved analytically to obtain a hard
function evolution factor $U(Q^2,\mu_h,\mu)$ with cusp anomalous dimension and quark and gluon 
anomalous dimensions as essential ingredients, see ref.~\cite{Becher:2009qa} for details. At 
\NNNLL{} we make use of the recent calculation of the four-loop cusp anomalous dimension 
\cite{Moch:2018wjh,Henn:2019swt,vonManteuffel:2020vjv}.

The appearance of the power-like dependence on the hard scale $Q^2$ 
from a re-factorization of regularized beam functions has been discussed extensively in 
refs.~\cite{Becher:2010tm,Becher:2011xn}, where the associated anomaly exponent $F_{ij}$ was first 
extracted to two-loop accuracy. For resummation
at \NNNLL{} we use the three-loop result of refs.~\cite{Li:2016ctv,Vladimirov:2016dll}. 


\paragraph{Improvement at very small \boldmath $q_T$ \unboldmath}
It is natural to rewrite the anomaly as a function of the logarithm $L_\perp = \log(x_T^2 \mu^2 / b_0^2)$ and the quantity
\begin{equation}
	\eta_i = \frac{C_i\alpha_s(\mu)}{\pi} \log \frac{Q^2}{\mu^2}\,,
\end{equation}
where $C_i = C_F$ for quark-antiquark initiated processes and $C_i=C_A$ for gluon-gluon initiated 
processes. For the choice $\mu\sim q_T$, as appropriate for the beam functions, we should count 
$\eta_i \sim 1$. In ref.~\cite{Becher:2011xn} the role of the anomaly exponent inside the $x_\perp$ 
integral at very small $q_T$ was analyzed in detail. Instead of the Fourier exponential, the large 
$x_\perp$ behavior of the integrand is driven by the anomaly and by the double logarithms 
$L_\perp^2$ inside the beam function. In the limit $q_T \to 0$, the $x_\perp$ integral becomes 
Gaussian and can be analyzed with a saddle point approximation, an observation that was made very 
early by Parisi and Petronzio \cite{Parisi:1979se}. The appropriate value of $\mu$ in this limit is 
denoted by $q^*$ and given by the value for which $\eta_i$ becomes equal to one 
\cite{Becher:2011xn}. Consequently one has
\begin{equation}
	q^* = Q^2 \exp\left( -\pi / C_i / \alpha_s(q^*) \right)
\end{equation}
and we solve for $q^*$ numerically in our setup for each integration \enquote{event}. It is
the characteristic scale of the process for very small $q_T$ and in practice well in the 
perturbative regime.
The physical picture behind this formalism is that instead of soft radiation recoiling against the high-$Q^2$ 
system, the radiation for $q_T\to0$ consists of \QCD{} emissions at a scale $q^*$ recoiling
against each other. For on-shell 
$Z$ production $q^*$ is about \SI{2}{\GeV} and for Higgs production around \SI{8}{\GeV}.

To achieve uniform perturbative accuracy also for $q_T\to0$, it has been observed 
that one should 
count $L_\perp \sim 1/\sqrt{\alpha_s}$ \cite{Becher:2011xn}. This was called improved power 
counting to distinguish it from the standard counting $L_\perp \sim 1$ relevant at moderately small 
$q_T$. To implement this power counting, it is important to factor out the enhanced double-logarithmic part of the beam functions. To this end we work with the functions $\bar{B}_i$ which are 
defined through
\begin{equation}
	B_i(\xi_i,x_\perp,\mu) = e^{h_i(L_\perp,\alpha_s)} \bar{B}_i(\xi_i,x_\perp,\mu)\,,
\end{equation}
where $h_i(L_\perp,\alpha_s)$ is provided by the solution of the \RGE{}
\begin{equation}
	\frac{\mathrm{d}}{\mathrm{d}\log \mu} h_i(L_\perp,\alpha_s) = C_i \gamma_\text{cusp} L_\perp 
		- 2 \gamma^i(\alpha_s)\,,
\end{equation}
with boundary condition $h_i(L_\perp,\alpha_s) = 0$. For the cusp anomalous dimension 
$\gamma_\text{cusp}$ and the quark and gluon anomalous dimensions 
$\gamma^i$ see refs.~\cite{Becher:2009qa,Becher:2019avh}.  The functions $\bar{B}_i$ are then 
implemented numerically in our code.
 
The modified beam functions $\bar{B}_i$ can be factorized further into a convolution 
\begin{equation}\label{eq:beammod}
	\bar{B}_i(\xi,x_\perp,\mu) = \sum_j \int_\xi^1 \frac{\mathrm{d}z}{z} \bar{I}_{i\leftarrow 
	j}(z,x_\perp,\mu) f_j(\xi/z,\mu)\,,
\end{equation}
of perturbative kernels $\bar{I}_{i\leftarrow j}(z,x_\perp,\mu)$ with the standard \PDF{}s 
$f_j(\xi,\mu)$. For our resummation at \NNNLL{} we need the kernel function at two loops, which were
computed in refs.~\cite{Gehrmann:2012ze,Gehrmann:2014yya}. After the double-logarithmic part has 
been removed, the beam functions only depend polynomially on $L_\perp$. We are therefore able to 
perform the Fourier integral independently
of the rest of the beam functions over the combined anomaly factor and the relevant powers of $L_\perp$.

In our code, we expand each individual ingredient to a common accuracy, according to our improved logarithmic and $\alpha_s$ power counting. Explicitly these are the hard 
function $\mathcal H$, the exponent of the hard function evolution $U$, the combined
collinear anomaly and double-logarithmic exponent $h_i$, and the product of beam functions $B_i\cdot B_j$.
Overall we achieve an accuracy of $\alpha_s$ relative to Born level
for \NNLL{} resummation and $\alpha_s^2$ relative to Born level for \NNNLL{} resummation, respectively. 
In the improved counting $\sqrt{\alpha_s} \sim 1/L_\perp \sim \epsilon $, we include terms up to 
$\epsilon ^{3}$. To also achieve higher accuracy for very small $q_T \to 0$, one would need to 
include additional terms in the beam functions. These terms are predicted by the \RGE{} and were 
included in version two of the \CuTe{} code which achieves $\epsilon ^{5}$ accuracy. Numerically 
their effect is small.

\paragraph{Matching to fixed order}
\label{par:matching}
A simple additive matching prescription
\begin{equation}
\left.\frac{\mathrm{d}\sigma^{\text{N$^3$LL}}}{\mathrm{d}q_T}\right|_\text{naively matched to 
\NNLO{}} 
= \frac{\mathrm{d}\sigma^{\text{N$^3$LL}}}{\mathrm{d}q_T} +
\underbrace{\frac{\mathrm{d}\sigma^\NNLO{}}{\mathrm{d}q_T} - 
	\left.\frac{\mathrm{d}\sigma^\NNNLL{}}{\mathrm{d}q_T}\right|_\text{exp. to \NNLO{}}
}_\text{matching correction $\Delta\sigma$}\,
\end{equation}
combines the resummed result at small $q_T$ with the fixed-order predictions at larger $q_T$, but 
suffers from two problems. First of all, the fixed-order result is only recovered up to 
higher-order 
terms. While formally not a problem, the leftover higher-order terms can induce unphysical 
behavior. We should therefore switch off the resummation at large $q_T$, which we implement using a 
transition function $t(x)$ with $x=q_T^2/Q^2$. This function is constructed so that $t(x)=1 
+\mathcal{O}(x)$ near $x=0$ and $t(x\geq 1)= 0$. The intermediate behavior is such that it smoothly 
switches the resummation off as $x\to 1$. A similar problem arises for small $q_T$. The matching 
corrections are power suppressed, but can become numerically unstable and suffer from large 
unresummed logarithms. For this reason, we switch the matching off at very small $q_T$, below a 
cutoff scale $q_0\lesssim \SI{1}{\GeV}$. The following modified matching prescription
\begin{equation}\label{eq:matchingmod}
\left.\frac{\mathrm{d}\sigma^{\text{N$^3$LL}}}{\mathrm{d}q_T}\right|_\text{matched to \NNLO{}} 
=  t(x) \left( \frac{\mathrm{d}\sigma^{\text{N$^3$LL}}}{\mathrm{d}q_T} + 
\left.\Delta\sigma\right|_{q_T>q_0} \right)
+ (1-t(x)) \frac{\mathrm{d}\sigma^\NNLO{}}{\mathrm{d}q_T}\,
\end{equation}
addresses both issues discussed above. Since we match on the level of the differential cross section, the fully inclusive fixed-order result is only restored within the nominal
perturbative accuracy, and not exactly. For inclusive $Z$ production it was found that the difference between resumming and matching the spectrum or the cumulant, which would preserve the integrated fixed-order result, are numerically small \cite{Becher:2011xn}. A detailed comparison of the two approaches can be found in ref.~\cite{Bertolini:2017eui}.

Choosing an appropriate transition region has to be done in dependence
of the process and the kinematical cuts. This is necessary in order not to include resummation in a region where it is no longer valid. While it could be considered a drawback to have to manually choose the transition region, respectively transition function, we believe that it offers clear advantages: The transition is performed transparently and we can guarantee which parts of the fully matched resummation are included in which kinematical region. Contributions where the $q_T$ resummation clearly becomes invalid, for example due to kinematical thresholds, can be fully excluded.

Below, we discuss the matching procedure in detail for the diboson processes $\gamma\gamma$ and $Z\gamma$ where kinematical thresholds require switching off the resummation relatively early. To choose the transition region, we first evaluate the size of the matching corrections relative to the (naively) matched result for each process and set of cuts. These relative corrections should be small in the resummation region, at worst of order one. Comparing results, we then try to identify a matching window in which the resummed and fixed-order results agree well enough that the transition between them can be performed reliably.

Within our setup one can easily implement any desired transition function
or even implement other matching procedures. All our results in this study are obtained with a suitably parametrized sigmoid function. Following a choice in \CuTe{}, we first define
\[
s(x;l,r,u) = \left (1 + \exp\left(\log\left(\frac{1-u}{u}\right) \frac{x-m}{w}\right) \right 
)^{-1}\,,\quad
m = (r+l)/2\,,\quad w = (r-l)/2\,.
\]
The function $s(x)$, parametrized by $l,r,u$, is defined to be $s(l)=1-u$ and $s(r)=u$.
In terms of this sigmoid, our transition function $t(x; x^\text{min},x^\text{max},u)$, where $x=q_T^2/Q^2$, is then defined by
\begin{equation}\label{eq:transition}
t(x; x^\text{min},x^\text{max},u) = \left\{\begin{array}{lr}
1 , & \text{for } x < x^\text{min}\\
\frac{s(x; x^\text{min}, x^\text{max},u)}{s(x^\text{min}; x^\text{min}, x^\text{max},u)}, & 
\text{otherwise}
\end{array}\right\}\,.
\end{equation}
This ensures that below $x^\text{min}=(q_T^\text{min}/Q)^2$ only the naively matched result is 
used, and at
$x^\text{max}$
for small $u\ll1$ the transition function is approximately $u$. In practice it makes sense to set 
the transition
function to zero below a small threshold like $10^{-3}$ without a noticeable discontinuity.
This has the advantage that the deteriorating resummation and matching corrections do not impact 
the region of 
large $q_T$ at all.
Our default choices in the remainder of this paper are $x^\text{min}=0.001$, and
$u=0.001$.

\begin{figure}
	\centering
	\includegraphics{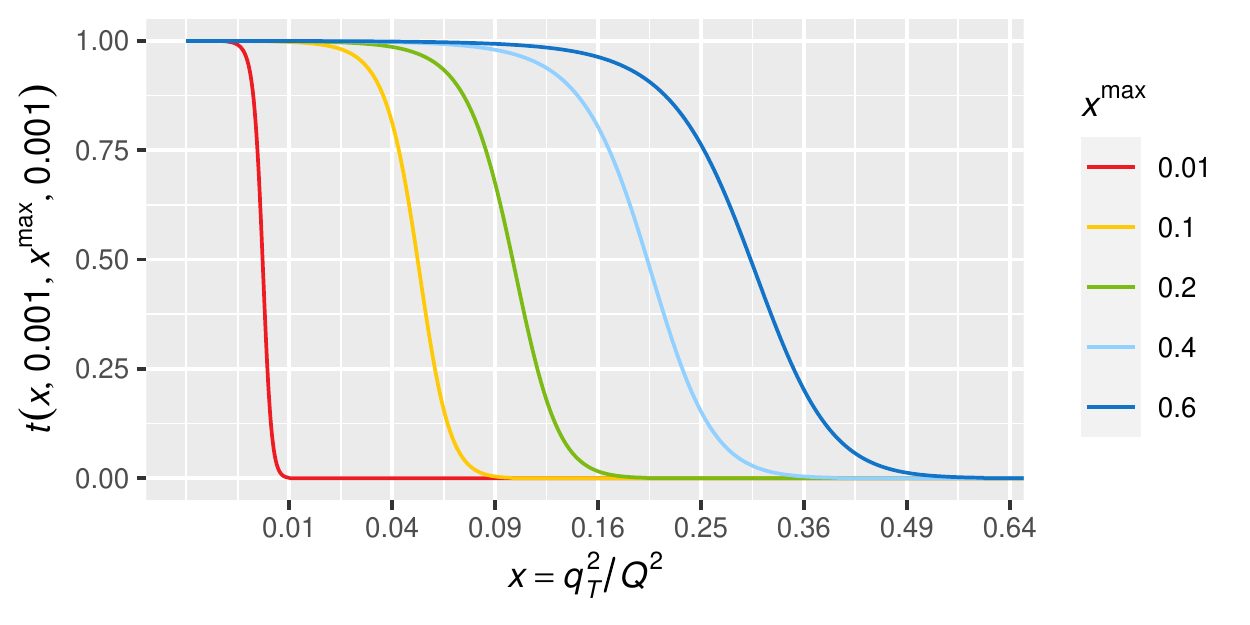}
	\caption{The transition function defined in \eqref{eq:transition} for different values of the parameter $\xmin$ which determines the position of the 
	transition. The $x$-axis is displayed on a square-root scale 
		to guide the eye on 
		the quadratic $q_T$-dependence.}
	\label{fig:transition}
\end{figure}

For the fiducial results studied here, we find that without the presence of a threshold or presence of photons,  
power-suppressed corrections are of order $q_T^2/Q^2$, and the size of the 
matching corrections is well-behaved up to relatively large values of
$q_T^2/Q^2$. Concretely, we
find that values of $\xmin=0.4$ and $\xmin=0.6$ can be used and allow us to estimate
the effect of the matching. For the processes with photons and with experimental cuts inducing additional thresholds, we have to start the transition much 
sooner. This is discussed
in detail in the sections for the $\gamma\gamma$ and $Z\gamma$ predictions.
We plot all transition functions used in our study in \cref{fig:transition}.

\paragraph{Power corrections and recoil effects}

The factorization theorem in \cref{eq:fact} is derived strictly in the limit $q_T\to0$ with power corrections that scale like $q_T^2/Q^2$ for fully inclusive 
production of a large-$Q^2$ system. Through the matching to fixed-order predictions, subleading power corrections  are automatically included to all powers in $q_T/Q$, but of course not resummed.  Since the 
factorization theorem is 
a function of $q_T^2$, it is most natural to consider the cross section 
$d\sigma/dq_T^2$. In fixed-order perturbation theory, the inclusive cross section for $q_T>0$ takes 
the form
\begin{equation}
\frac{d\sigma}{dq_T^2} =   \frac{A}{q_T^2} + B + \dots\,,
\end{equation}
where the coefficients $A$ and $B$ depend logarithmically on $q_T^2$. 
The leading logarithms at the $n$-th order in these coefficients have the form $\alpha_s^n(\mu) \ln^{2n-1}(q_T^2/\mu^2)$. The terms contained in $A$ are captured by the factorization formula, while the contributions in $B$ and all other power-suppressed terms are added through the matching correction $\Delta \sigma$.

Resummation cures the divergence of the cross section and the quantity ${d\sigma}/{dq_T^2}$ takes a 
finite value for $q_T\to 0 $. A detailed discussion of the intercept for $q_T\to 0 $ was given in 
ref.~\cite{Becher:2011xn}; in this context the $\epsilon$-expansion discussed earlier plays a 
crucial role. Much less is known about effect of resummation on the power corrections, but first 
leading-logarithmic resummed results for power-suppressed contributions indicate that Sudakov 
suppression is present also in this case \cite{Penin:2014msa,Moult:2018jjd, 
Beneke:2019mua,Bahjat-Abbas:2019fqa,Liu:2020tzd,Liu:2020wbn}. Since we do not resum the power-suppressed matching 
corrections, their computation becomes unreliable at low $q_T$ because higher-order terms are 
enhanced by large logarithms and they can start to numerically compete with the resummed, 
Sudakov-suppressed leading-power cross section for $q_T\to 0$. We should therefore switch off the 
matching at very low $q_T$, which is achieved using a hard cutoff $q_T>q_0$ in 
\cref{eq:matchingmod}. This is also necessary for numerical stability, as we will discuss in 
detail in \cref{sec:results}. 

Experimentally one usually measures $d\sigma/dq_T  = 2q_T 
\,d\sigma/dq_T^2$. The extra factor of $q_T$ then suppresses also the power corrections.
However, cuts on the leptonic final state can enhance power corrections and lead to a weaker 
suppression. For the Drell-Yan process subleading corrections of $\mathcal{O}(q_T/Q)$ to fiducial 
cross sections 
can be accounted for by working with the exact lepton tensor, as has been recently demonstrated in 
ref.~\cite{Ebert:2020dfc}. This is possible because (azimuthally symmetric) corrections to the hadronic 
tensor are suppressed with $q_T^2/Q^2$. A similar analysis is not yet available in general. What is 
done in practice, is to use recoil prescriptions to take into account some power corrections 
\cite{Catani:2015vma}.

In our code we work with a Lorentz-boost prescription of the Born-level 
amplitude which keeps the electroweak part of the amplitudes exact and transfers the transverse 
momentum to the electroweak bosons. More specifically, following ref.~\cite{Becher:2019bnm}, we 
start by generating the Born-level 
phase space and then boost this system to have transverse components 
$(q_T\cos\phi,q_T\sin\phi)$, where we now additionally integrate over the values of $q_T\geq0$ and 
$\phi\in[0,2\pi]$ using Monte 
Carlo methods. We use the boosted momenta
to evaluate the Born matrix elements and hard function and to perform the kinematical cuts. In the 
future, it would be interesting to investigate for which observables this provides the exact 
$\mathcal{O}(q_T/Q)$ power-suppressed terms. For the fiducial cross sections we compute in the 
following, we 
observe numerically that the power corrections are second order, as for the inclusive cross 
section, except for processes with photons.

\paragraph{Enhanced power corrections from photon isolation}
\label{par:photiso}
To separate direct photon production from photons arising in hadron decays, experiments impose that 
photons should be isolated from hadronic radiation. More precisely, only low-$q_T$ hadronic 
radiation is allowed inside a cone around the photon. In the limit $q_T \to 0$, and at leading 
power, photons are automatically isolated since all radiation has low $q_T$. This implies that the 
the leading-power factorization theorem \eqref{eq:fact} applies also to processes with photons 
in the final state. 

The photon isolation induces subleading power corrections, that are included via the matching to fixed-order predictions. However, the nature and size of these power corrections is different from what we encountered earlier since they are not imposed on the electroweak final state, but directly affect the hadronic matrix elements. 

For our studies of processes with photons, we adopt the smooth-cone isolation introduced by Frixione \cite{Frixione:1998jh}, which fully suppresses the collinear singularity from the $q\to q \gamma$ splitting in an  infrared-safe way, eliminating the need for fragmentation functions. It restricts the transverse energy inside a 
cone of size $R$ to 
\begin{equation}\label{eq:frixione}
E_T^\text{had} \equiv \sum_{j:\, d(j,\gamma)\leq r} E_T^j \leq \etiso\, \chi(r) \quad \forall r<R\, , 
\end{equation}
where $d(i,j)=\sqrt{(\phi_i-\phi_j)^2 + (\eta_i-\eta_j)^2}$ is the separation in azimuthal angle $\phi$
and rapidity $\eta$ between parton $i$ and photon $j$. 
The angular function is
\begin{equation}
 \chi(r) =  \left ( 
\frac{1-\cos r}{1 - \cos R} \right)^n  \approx \left (\frac{r^2}{R^2} \right)^n\,,
\end{equation}
where the approximation is valid for $R \ll 1$.
The isolation energy 
$\etiso$ can either be a fixed value or a fraction $\epsilon$ of the total photon 
transverse energy $\etiso=\epsilon \,E_T^\gamma = \epsilon\, q_T^\gamma$.

\begin{figure}
    \centering
    \includegraphics[width=0.495\textwidth]{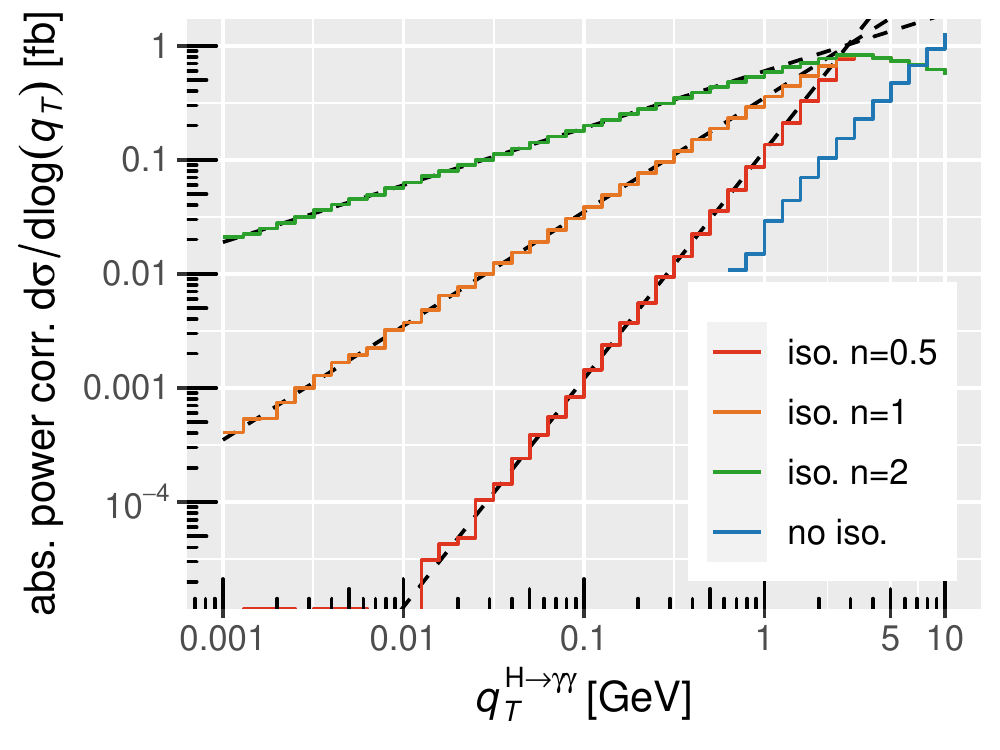}  
    \includegraphics[width=0.495\textwidth]{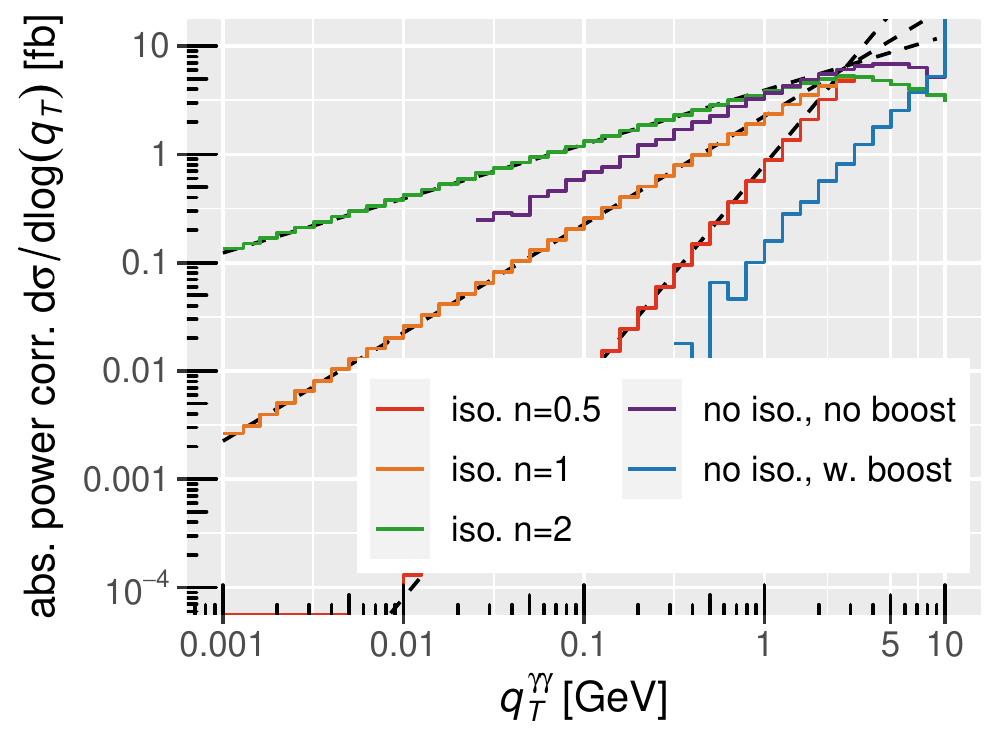}
    \caption{Power-suppressed matching corrections for $pp\to H\to \gamma\gamma$ (left) and $pp\to 
    \gamma\gamma$ (right). For diphoton production only the $u\bar{u}$ channel is shown,  with cuts $q_T^\gamma>\SI{25}{\GeV}$ on both photons. We plot results for the values  $n=0.5,1,2$ of the  isolation parameter $n$. Power corrections not from photon isolation are shown as purple and blue lines. The dashed lines show the scaling $(q_T/Q)^{1/n}$ derived in ref.~\cite{Ebert:2019zkb}.} \label{fig:powcorr_higgs}
\end{figure}

The effects of photon isolation on power-suppressed terms in $q_T$ factorization have been studied 
in ref.~\cite{Ebert:2019zkb}. These authors considered inclusive $H\to\gamma\gamma$ production
with photon isolation cuts and inclusive diphoton production restricted to the 
$q\bar{q}$ channel with photon isolation cuts and photon $q_T$ cuts. They showed that the smooth-cone isolation requirement induces subleading terms scaling as 
$(q_T/Q)^{1/n}$, where $n>0$ is the parameter in the isolation prescription above. We study this 
dependence in \cref{fig:powcorr_higgs} for $H\to\gamma\gamma$ and diphoton production. Our 
numerical results nicely confirm this scaling of the power corrections.  For comparison we also 
show the matching corrections not associated with photon isolation as blue and purple lines in 
\cref{fig:powcorr_higgs}. For $pp\to \gamma\gamma$ (right panel) we impose 
$q_T^\gamma>\SI{25}{\GeV}$. For the purple lines, Born-level kinematics are used for the photons, 
while for the blue one the recoil is taken into account using the boost prescription detailed 
above. We observe that these power corrections scale as the first power without the boost and as 
the second power with the boost. The recoil does not play a role for $pp\to H \to \gamma\gamma$ 
(left panel) since no fiducial cuts are employed.

While we reproduce the result of ref.~\cite{Ebert:2019zkb} for the $q\bar{q}$ partonic channel, we 
observe a different behavior if all partonic channels are included, due to fragmentation 
contributions. To explain the difference we consider the emission of a single soft particle with 
momentum $k$ into the isolation cone. For one emission $q_T = k_T$ so that the value of the 
transverse momentum of the particle is fixed. The momentum dependence of the squared, spin averaged 
amplitudes for soft gluon and soft quark emission are
\begin{equation}\label{eq:amps}
|\mathcal{M}_g|^2 \propto \frac{p_1 \cdot p_2}{p_1 \cdot k \,p_2 \cdot k } = \frac{1}{k_T^2} 
\,\qquad \text{and} \qquad  |\mathcal{M}_q|^2  \propto \frac{1}{2 p_\gamma \cdot k} \,,
\end{equation}
respectively. The momenta $p_1$ and $p_2$ are the momenta of the incoming partons and $p_\gamma$ is 
the photon momentum which defines the isolation cone. The result \eqref{eq:amps} shows that gluon 
emission is a leading-power effect while soft quark emissions are suppressed by one power of $k_T$. 
Writing the phase-space integral in terms of the transverse momentum, rapidity and azimuthal angle,
\begin{equation}
\frac{\mathrm{d}^3k}{E_k} = \mathrm{d}k_T k_T \,\mathrm{d}y\, \mathrm{d}\phi\,,
\end{equation}
we see that the soft gluon emission suffers from a soft divergence, while the quark emission has a 
collinear pole, which is regularized by the smooth-cone isolation requirement \eqref{eq:frixione}. For 
fixed transverse momentum $k_T$ and $R \ll 1$, the isolation requirement implies that the angular 
distance $r^2 = d(k,\gamma)^2 = \Delta y^2 +\Delta \phi^2$ must fulfill
\begin{equation}
r^2 \geq R_{\rm min}^2 = R^2 \left(\frac{k_T}{\etiso}\right)^{\frac{1}{n}}\,.
\end{equation}
The emitted particle can thus no longer be exactly collinear to the photon.

With these considerations we can now easily evaluate the power corrections associated with gluon and quark emission. Gluon emission is a leading-power effect and the power corrections are obtained by evaluating the difference between the isolated case and the  inclusive production 
\begin{equation}\label{eq:gluonPower}
\Delta \frac{\mathrm{d}\sigma}{\mathrm{d}q_T^2} \propto \int \!\! \mathrm{d}y\, \mathrm{d}\phi 
\left[ \theta(r-R_{\rm 
min}) - 1\right] |\mathcal{M}_g|^2  =- \frac{1}{q_T^2} \int_0^{R_{\rm min}}\mathrm{d}r\,r 
=-\frac{R_{\rm min}^2}{q_T^2}  =- \frac{R^2}{q_T^2} \left(\frac{q_T}{\etiso}\right)^{\frac{1}{n}}\,.
\end{equation}
This reproduces the result of ref.~\cite{Ebert:2019zkb}. Next, let us turn to fragmentation. In this case, the entire effect is a power correction, so we evaluate
\begin{equation}\label{eq:quarkPower}
\Delta \frac{d\sigma}{\mathrm{d}q_T^2} \propto \int_{R_{\rm min}}^R \!\! \mathrm{d}r\, r 
|\mathcal{M}_q|^2  
\approx \frac{1}{q_T p_T^\gamma} \int_{R_{\rm min}}^R\!\! \frac{\mathrm{d}r}{r}  = \frac{1}{q_T 
p_T^\gamma} 
\ln\frac{R}{R_{\rm min}}\,,
\end{equation}
where we approximated $2 p_\gamma \cdot k = k_T\, p_T^\gamma\, r^2+O(r^4)$. Here the dependence on 
the isolation requirement is logarithmic and the power correction is always first order. 
Furthermore, the effect in the gluon channel is suppressed by the cone radius $R^2$, while there is 
no such suppression in the fragmentation case. First-order power corrections will thus always be 
present and for small cone radius they will numerically dominate over the gluonic power 
corrections, even if these are larger than first order for $n>1$. We will present numerical results 
for the matching corrections for the sum of all partonic channels and including the fiducial cuts 
on the photons in \cref{sec:results}. The results in this section will confirm the presence of 
first-order power corrections. 

Linear or stronger power corrections lead to matching corrections which tend to a constant or even 
grow in $\mathrm{d}\sigma/\mathrm{d}q_T$ for $q_T\to 0$ and overwhelm the resummed leading-power 
result. 
This implies 
that it is not possible to obtain reliable predictions for very small $q_T$ in such cases, at least 
not without resumming also the power corrections. We will face this problem in \cref{sec:results} 
when studying processes with photons in the final state.

Having discussed the effect of photon isolation on power-suppressed corrections at small $q_T$, we should mention that photon isolation also leads to logarithmically enhanced contributions at large transverse momentum, since there is then a region of phase space, where the radiation is restricted by the isolation requirement. This is a typical situation in which non-global logarithms arise \cite{Dasgupta:2001sh} and their numerical effect in photon-production cross sections was studied in ref.~\cite{Balsiger:2018ezi} at leading-logarithmic accuracy. While the argument of the logarithms is large for the experimentally imposed photon isolation energies, the effect on the cross section is moderate, since it is suppressed by $R^2$. Similar conclusions were reached in ref.~\cite{Wiesemann:2020gbm}, which studied their size for the $Z\gamma$ transverse-momentum spectrum.

\paragraph{Implementation}
We have implemented the presented framework in a modular Fortran 2008 code, where hard function
evolution, beam functions and Fourier integrals are calculated separately and assembled
to the designated order for resummed result and its fixed-order expansion. All components
are combined with an easy to modify transition function in the \MCFM{} plotting routines. 
The phase-space parametrization routine for each process 
allows for an efficient integration down to very small $q_T$.
Since the essential resummation 
pieces are only loosely coupled to \MCFM{}, they could easily be reused or integrated into other codes, for example as a direct stand-alone extension of the interface to event files 
\cite{Becher:2019bnm} to \NNNLL{} and gluon-gluon initiated processes.

Resummation parameters that can and should be changed in the input file during normal use are
the integration range for the resummation and its expansion, and the hard cutoff below which the
matching corrections are turned off. Further details on how to use the code will be made available
in the manual together with the code. 

The \NNLO{} processes available to be matched with \NNNLL{} resummation are $H,Z,W^\pm$ \cite{Boughezal:2016wmq}, $W^\pm H$, $ZH$ 
\cite{Campbell:2016jau}, 
$\gamma\gamma$ \cite{Campbell:2016yrh} and $Z\gamma$ \cite{Campbell:2017aul}. Since \MCFM{}
implements several more processes at \NLO{}, these could easily be matched with \NNLL{} resummation 
and we would be 
happy to add these by request. All processes include all leptonic decay channels and Higgs 
production includes all major decay channels. Furthermore, for $Z$ production
electroweak corrections have been implemented \cite{Campbell:2016dks}.

\paragraph{Estimation of perturbative truncation uncertainty}

We estimate the perturbative truncation uncertainty by varying the renormalization, factorization
and resummation scales in our calculation with the multipliers 
\begin{equation}\label{eq:seven}
(k_F ; k_R) \in \{ 
(2,2), 
(0.5,0.5), (2,1), (1,1), (0.5,1), (1,2), (1,0.5) \}\,.
\end{equation}
For the fixed-order computation and the matching correction we use $\mu_F = k_F\, \hat{Q} $ and $\mu_R = 
k_R \hat{Q}$. We choose the default hard scale as $\hat{Q} = Q$ for the benchmark comparison to \CuTe{}, while we also use $ \hat{Q} =\sqrt{Q^2+q_T^2}$ in other cases, in line with the choice typically made in fixed-order computations at larger $q_T$. The seven-point prescription for scale variation resulting from \cref{eq:seven} is common practice in the fixed-order community. To set the 
resummation scale, we first 
calculate $q^*$ for each integration phase-space point (\enquote{event})  and then set 
\begin{equation}\label{eq:mures}
	\mu = \mathrm{max}\left( k_{F} \cdot q_T + q^*\exp(-q_T/q^*)\,,\, \SI{2}{\GeV} 
	\right)\,.
\end{equation}
This choice ensures that the scale is always in a perturbative and numerically stable regime, and 
for very small $q_T$ approaches $q^*$, while otherwise smoothly transitioning to $q_T$. For the 
hard scale, we use $\mu_h = k_R \hat{Q} $. With this prescription, we avoid the introduction of four 
different multipliers at the price of correlating some variations in the matching correction and 
the resummed result.

At small $q_T$ the logarithms resummed up to \NNNLL{} dominate,
and with the choice in \cref{eq:mures} the residual scale dependence can become small at very small $q_T$.
In this region the problem arises that varying the resummation scale leads to very low values of 
$\mu$ for which the Fourier-integral becomes numerically unstable. To avoid this, we have set a 
minimum value of $\mu = \SI{2}{\GeV}$ in \cref{eq:mures}, which restricts the scale variation but 
ensures that the scale $\mu$ always remains in the perturbative regime. 

A drawback of this approach is that at very small $q_T$ of a few GeV the downwards
variation for the resummation scale vanishes. To address this, one could symmetrize the 
uncertainties, if large asymmetries at small $q_T$ are observed.  We find this not to be an issue 
in practice 
and the variation of the hard renormalization scale generates the bulk of the scale uncertainty.
Furthermore, the overall uncertainty budget at such low values of $q_T$ should include non-perturbative effects that are not quantified here. Beyond that, various approaches have been used in the literature that argue for modifying the scale 
 variation procedure in combination with resummation \cite{Banfi:2015pju,Bizon:2017rah}. Also in 
 our case further variations could be considered. In addition to introducing a scale to estimate 
 uncertainties from different exponentiations of the rapidity logarithms, we could, for example, 
 introduce an additional evolution step to separate the scale in the perturbative kernels 
 $\bar{I}_{i\leftarrow j}$ in the beam functions in \cref{eq:beammod} from the \PDF{} scale and 
 then also vary this scale. Of course, ultimately one should simply compute the higher-order 
 corrections to know their size.

\paragraph{Beam-function grids}
While our setup can compute the beam functions on the fly by evaluating the convolution 
in \cref{eq:beammod} with the \PDF{}s for the relevant values of $\xi$ and $\mu$, it is 
computationally expensive to do so. It is much more effective to pre-compute 
\LHAPDF{} grids~\cite{Buckley:2014ana} for the beam functions. After doing so, the calculation of
the resummed component is no longer more time consuming than the other components. For each
individual \PDF{} grid five beam function grids are generated corresponding to the beam function coefficients
of different orders of $\alpha_s$ and $L_\perp$. The grid pre-computation is fully parallelized through {\abbrev OpenMP} and {\abbrev MPI} or Fortran Coarrays, and if \PDF{} uncertainties are enabled the eigenvector or replica \PDF{} set members can also be pre-computed accordingly. Through the infrastructure of \MCFM{}-9, matched results with multiple \PDF{} sets, including their respective uncertainties, can in this way be computed simultaneously.

\paragraph{Checks}
\label{sec:checks}

We have extensively compared all of our our resummation ingredients at \NNNLL{} against a private
 prototype implementation in Mathematica that resulted in the code \CuTe{}
\cite{Becher:2012yn} as well as against the \NNLL{} implementation in ref.~\cite{Becher:2019bnm} 
and find full agreement. 

Since our implementation is based on \MCFM{}-9, which employs jettiness subtractions
for the \NNLO{} calculations \cite{Gaunt:2015pea,Boughezal:2015dva}, all processes have been 
extensively checked and \IR{} cancellations have already been demonstrated to be numerically stable 
down to the per-mille level.

For all our presented results we checked that the fixed-order expansion of the resummed
result and the fixed-order predictions agree for $q_T\to 0$. We performed this
check down to values of \SI{0.01}{\GeV} with sub-permille precision in the cancellation,
depending on the process and cuts; see the individual process studies presented in the next section. 
With that we implicitly also tested numerically that these leftover power corrections to our 
$q_T$ resummation scale as predicted: For fiducial processes without photons we find that the power corrections without boosted Born kinematics are $\mathcal{O}(q_T/Q)$, while they are quadratic with a boost. For processes
with photons we can furthermore check the fixed-order result and our framework by testing that 
power corrections
due to smooth-cone photon isolation scale as $(q_T/Q)^{1/n}$, where $n>0$ is given as a parameter 
in the isolation prescription \cite{Ebert:2019zkb}. This asymptotic behavior sets in 
sufficiently below the photon isolation cone energy $\etiso$, which is typically just a few GeV.

We also compared our fully inclusive results against the \CuTe{} code: By default \CuTe{} makes
a series of choices that lead to power-suppressed differences. For example, it takes
into account a finite-$q_T$ modification of the phase-space. When setting the phase-space integration to use Born-level kinematics we find full agreement for fixed-order results in $W,Z$ and Higgs production as well as for the fixed-order
expansion of the resummed result at \NNNLL{}. For the resummed part, our results agree with \CuTe{} at
\NNNLL{} within the choices available in \CuTe{} for the expansion of the improved power counting scheme, see
the following section.

\section{Results}
\label{sec:results}

In this section we present resummed and matched results for a wide range of electroweak final 
states.  As a first step we perform benchmark computations for fully inclusive $Z$-boson and $H$-boson 
production and compare against the code \CuTe{}.\footnote{\CuTe{} is available at 
\url{https://cute.hepforge.org/}.} The code \CuTe{} is restricted to $Z,W$ and $H$ production and does 
not allow for fiducial cuts, but is based on the same formalism and ingredients as our 
implementation. While the ingredients were were individually cross checked against \CuTe{}, the 
numerical results for the cross sections differ through terms beyond the accuracy of the 
calculation. These include power-suppressed effects associated with a different treatment of phase 
space in \CuTe{}, as well as higher-order perturbative effects from different ways of organising 
the expansion. Given the different scheme choices, it is interesting to quantify the resulting 
differences that, in principle, should be covered by scale uncertainties.

After this benchmarking exercise, we impose experimental fiducial cuts and 
directly compare with
measurements from \ATLAS{} and \CMS{}. For $Z$ production we compare with studies at 
\SI{8}{\TeV} \cite{Aad:2015auj} and  \SI{13}{\TeV} 
\cite{Sirunyan:2019bzr}. For $W^\pm$-boson production, we compare with a transverse-momentum measurement at \SI{8}{\TeV} \cite{Khachatryan:2016nbe}.
 A high-precision theoretical description of the Higgs production process requires a careful 
 inclusion of top-quark mass effects which go beyond the scope of this study. For 
 the moment, we therefore present results in the strict heavy-top limit with fiducial cuts that are imposed in an experimental $H\to\gamma\gamma$ analysis. Currently the 
 experimental uncertainties in the Higgs transverse-momentum spectrum are still quite large, but it would be interesting to perform a detailed theoretical analysis in the future. Finally, we turn to diboson processes. For diphoton production we 
 show novel results at \NNNLL{} accuracy going beyond previous results at \NNLL{}. For 
 $Z\gamma$ production we also present novel results at \NNNLL{} that improve upon previous results 
 limited to fixed order.

In all cases we show fully matched \NTHREENNLO{} results, but usually refrain from showing results 
at a lower order or their scale uncertainties. For large $q_T$ the lower-order results are only 
Born-level accurate
and perturbative uncertainties are not properly estimated solely through the running of
$\alpha_s(\mu)$ and the \PDF{}s, without further intrinsic scale dependence from renormalized loop integrals.
Typically the first results that can
give reliable uncertainties at large $q_T$ are given by our \NTHREENNLO{} predictions that include the fixed-order results at large $q_T$ at a subleading order in $\alpha_s$.

Our results in the following are presented for random selections of some \NNLO{} \PDF{} sets with 
a fixed value of $\alpha_s(m_Z)=0.118$: {\abbrev ABMP16} 
\cite{ABMP16}, {\abbrev CT14} \cite{CT14}, {\abbrev CT18} \cite{CT18}, {\abbrev MMHT2014} \cite{MMHT2014}, {\abbrev MSTW2008}  
(this has $\alpha_s(m_Z)$=0.117) \cite{MSTW2008}, {\abbrev NNPDF30} \cite{NNPDF30} and {\abbrev NNPDF31} \cite{NNPDF31} interfaced to \LHAPDF{} \cite{Buckley:2014ana}. We also compute and compare the uncertainties associated with the different \PDF{} sets.

\subsection{Benchmark calculations and comparison with \CuTe{}}

As mentioned above, the implementations of the resummation formula in  \CuTe{} and \CuTeMCFM{} differ:
The default approach taken in \CuTe{} is to combine hard function and its evolution factor
into a common exponent and expand this exponent to a designated logarithmic accuracy in $\alpha_s$.
This approach thus exponentiates the higher-order corrections to the hard function.
\CuTe{} also implements certain higher-order beam function contributions which are relevant to obtain $\epsilon^5$ accuracy in the improved counting at very low $q_T$, while we only achieve $\epsilon^3$ accuracy, see the discussion in sec.\ \ref{sec:calculation}. A second difference arises because \CuTe{} modifies the phase-space integral to include power-suppressed effects. For the parton momentum fractions $\xi_{1,2} = \sqrt{\tau}\, e^{\pm Y}$ entering the beam functions \CuTe{} uses $\tau = (Q^2+q_T^2)/s$, while we work with the Born-level result $\tau = Q^2/s$. 

To compare with \CuTe{} we have ensured that all physical input parameters agree
and then checked that the fixed-order predictions and expansion of the resummed cross section agree. We work with the \NNPDFTO{}  \PDF{} set at 
$\sqrt{s}=\SI{13}{\TeV}$. To compare the resummed results, we work at $\epsilon^4$ in the improved power counting in \CuTe{}. This order resembles most closely our new implementation, since we include some terms beyond $\epsilon^3$. In \CuTeMCFM{} we integrate over $q_T$ and present the results bin-wise, while CuTe 
is limited to evaluating individual $q_T$ values. \CuTe{} can also parametrize non-perturbative effects and has different transition functions to choose from, but here we are only interested in the subleading differences of the resummed results for benchmarking purposes. 

We present benchmark results for $Z$ production as a quark-antiquark initiated process and for $H$ production for a gluon-gluon initiated
process. The other processes available in \CuTeMCFM{} are all based on the same resummation ingredients and only differ in the hard function and Born amplitudes. 

\subsubsection{Inclusive Higgs production}

\begin{figure}[t!]
	\centering
	\includegraphics{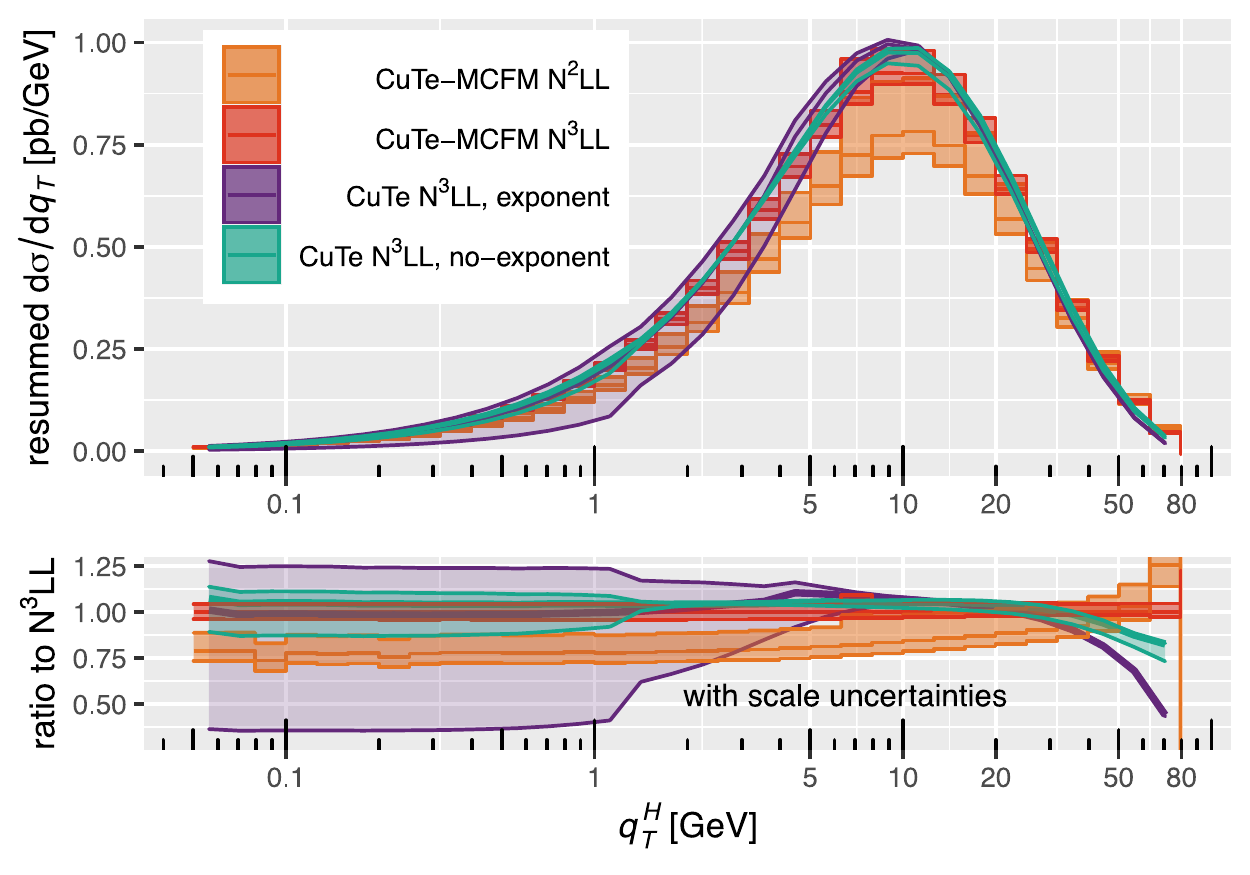}
	\caption{Resummed results without matching for inclusive Higgs production at \SI{13}{\TeV} 
		obtained using \CuTeMCFM{} and \CuTe{} at different logarithmic orders.
		For \CuTe{} we show results in two schemes: expanded in the exponent or on the level of the 
		cross section. The shaded bands display scale uncertainties.
		The bottom panel shows the ratio to the \NNNLL{} result in \CuTeMCFM{}.}
	\label{fig:H_resummation_benchmark}
\end{figure}

In \cref{fig:H_resummation_benchmark} we compare the resummed result for inclusive Higgs production without fixed-order matching obtained with \CuTe{}  to our new implementation \CuTeMCFM{}.  This comparison gives an indication of the uncertainties from subleading terms due to the different scheme choices. The first panel
shows the absolute distribution, while the second panel shows the ratio to our \NNNLL{} resummed 
result where scale uncertainties are also included.

Overall the predictions of \CuTe{} and \CuTeMCFM{} are within mutual scale uncertainties up to 
\SI{30}{\GeV}.
 Central values are also are well compatible and captured within one to two times the 
scale-uncertainty band of our own  \NNNLL{} prediction. 
The 
discrepancy beyond \SI{30}{\GeV} between \CuTe{} and \CuTeMCFM{} is solely due to the choice of 
$\tau=(Q^2+q_T^2)/s$ for the phase-space integral in \CuTe{}. Below we analyze this difference in detail for $Z$ production. Scale variation does not provide an estimate of the size of these 
power-suppressed differences, but performing the matching to fixed order would largely eliminate 
them. 

At large $q_T$ one observes almost zero scale uncertainties for \CuTe{} if the expansion is performed strictly in the exponent. At the same time, one sees a significant increase in
the scale uncertainties at tiny $q_T$, where also the improved expansion order plays a big role. It 
is perhaps a bit disconcerting that formally equivalent prescriptions give such different scale 
variation bands. It seems that there is an accidental cancellation of scale uncertainties at play, as evidenced by the fact that these uncertainties increase significantly when we impose fiducial cuts, see \cref{fig:higgs} below. We also observe that the N$^3$LL results are outside of the N$^2$LL result. This is a reflection of the well known fact that the Higgs cross section suffers from large perturbative 
corrections. If we instead considered the normalized distribution, the bands would 
overlap. The small scale uncertainties of \CuTeMCFM{} at tiny $q_T$ are a consequence of the choice 
in \cref{eq:mures} and not indicative of the true uncertainty, which would also need to include an 
estimate of non-perturbative effects.

\begin{figure}[!t]
	\centering
	\includegraphics{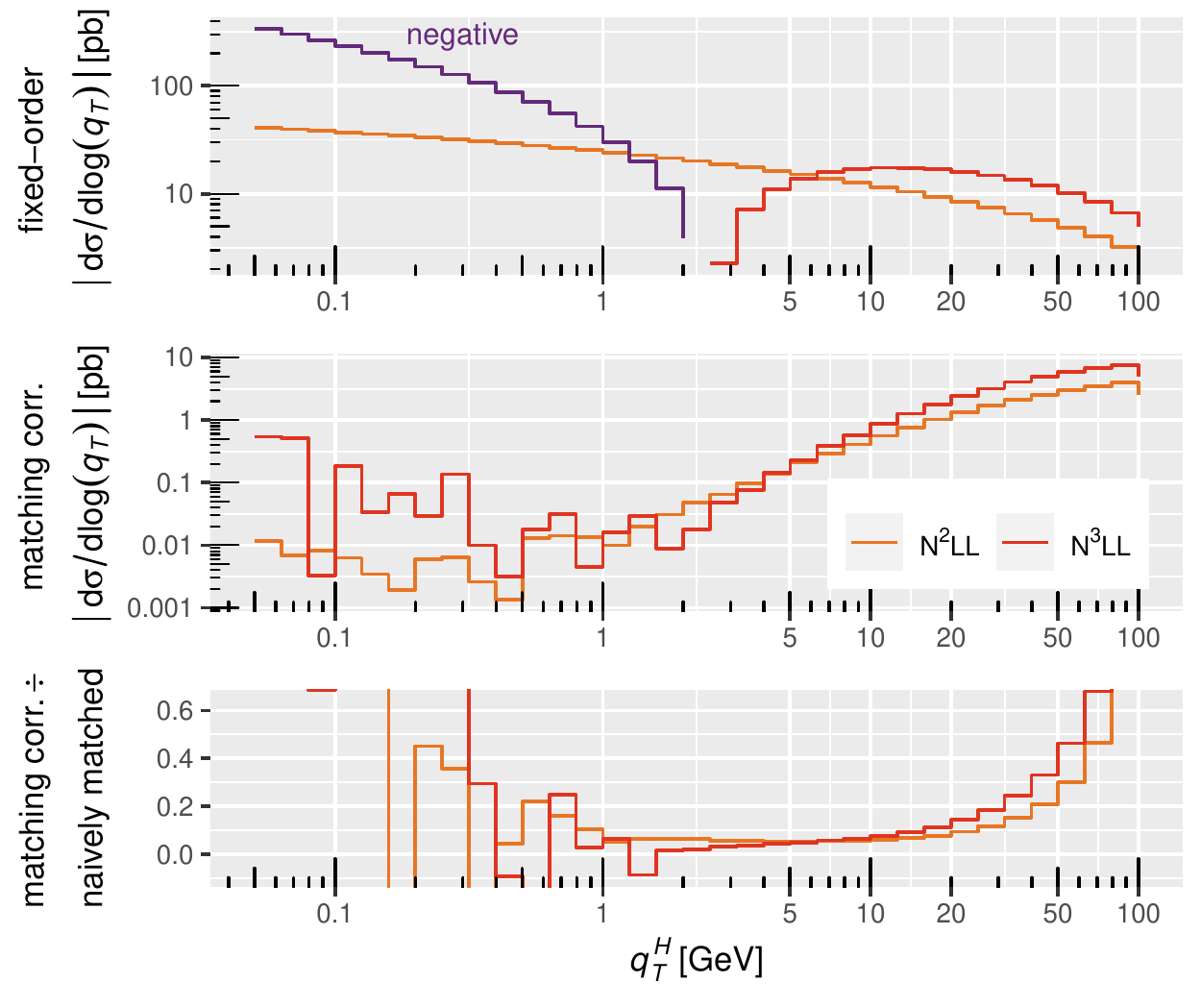}
	\caption{Top panel: Fixed-order prediction at \LO{} and \NLO{} for inclusive 
		Higgs production. Middle panel: Absolute value of the matching corrections to
		show the cancellation towards $q_T\to0$ and the resulting loss of 
		numerical accuracy at very small $q_T\lesssim \SI{1}{\GeV}$. Bottom panel: Matching corrections relative to the naively matched result.}
	\label{fig:H_resummation_benchmark_match}
\end{figure}

Having discussed the resummation, we now illustrate the numerical difficulties in computing the matching corrections in \cref{fig:H_resummation_benchmark_match}. The top panel shows the fixed-order predictions at $\alpha_s$ 
and $\alpha_s^2$ and their behavior  towards $q_T\to 0$. Note that the $\alpha_s^2$ prediction has a zero around \SI{2}{\GeV}. The matching corrections are shown in second panel.
Both cross sections are displayed as $\mathrm{d}\sigma/\mathrm{d}\log(q_T)= q_T\,
\mathrm{d}\sigma/d q_T = 2 q_T^2 \,\mathrm{d}\sigma/dq_T^2$.
 Since the 
matching corrections are suppressed by $\mathcal{O}(q_T^2)$ they should decrease quadratically as 
$q_T$ is lowered and we indeed observe this behavior for moderately small $q_T$. 
However, the fixed-order result for $q_T^2 \,\mathrm{d}\sigma/dq_T^2$ and the fixed-order expansion of the resummed 
result both go to a constant in the same limit so that we encounter large numerical cancellations when computing the matching in the region of very small $q_T$.

Indeed the quadratic behavior of the matching corrections is spoiled by numerical problems for 
$q_T\lesssim$\SI{1}{\GeV}. In this region one is 
limited by the Monte-Carlo integration, where, typically, relative uncertainties below $10^{-3}$ to 
$10^{-4}$ are computationally very expensive. Around \SI{1}{\GeV} for the \NNLL{} 
result, the cancellations in the computation of the matching correction already require a relative uncertainty of 
$10^{-4}$. The bottom panel in \cref{fig:H_resummation_benchmark_match} shows that for $q_T \lesssim 
\SI{1}{\GeV}$ the numerical noise in the matching corrections becomes large relative to the
Sudakov suppressed resummed result. 

Overall, the  above considerations imply that, for practical numerical reasons alone, one has to 
turn off the matching corrections below a certain value of $q_T$ to not spoil the results at small 
$q_T$ with an incomplete cancellation. For observables with quadratic power corrections, imposing 
this cutoff is completely unproblematic, but we will revisit the issue when discussing processes 
with photons, where the power suppression is weaker.
In any case, the computation of the power-suppressed matching terms using fixed-order perturbation 
theory is no longer viable in this region since the power corrections will involve large 
logarithms. On top of this, for such low values of $q_T$ also non-perturbative effects will play a 
role. For the remainder of this paper we switch off the matching corrections below \SI{1}{\GeV} 
unless otherwise noted. 

\begin{figure}[!t]
	\centering
	\includegraphics[width=0.8\textwidth]{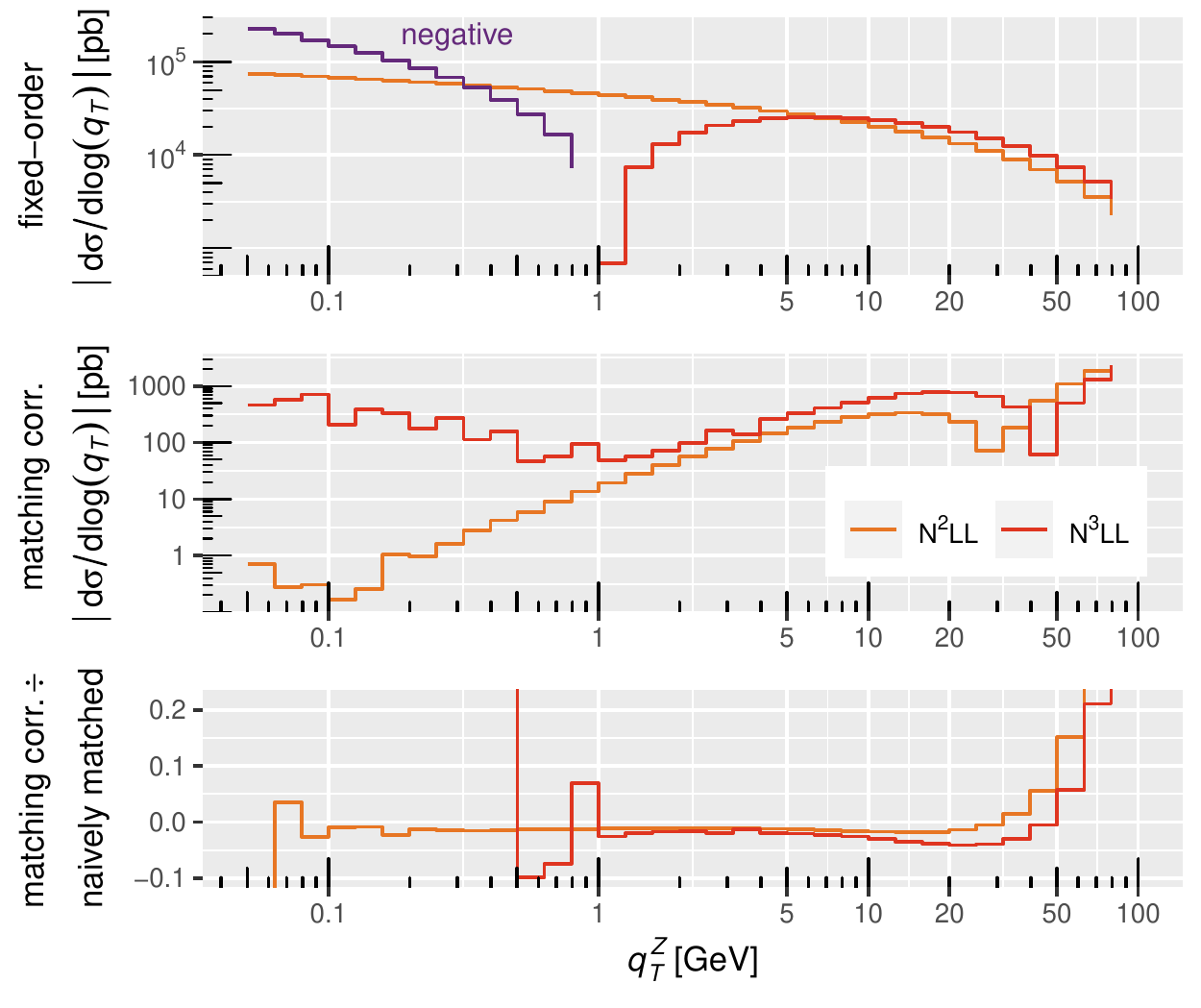}
	\caption{Top panel: Fixed-order prediction at \LO{} and \NLO{} for inclusive 
		$Z$ production. Middle panel: Absolute matching corrections to
		show the cancellation towards $q_T\to0$ and the effect of a limited 
		numerical accuracy at small $q_T$. Bottom panel: Matching corrections relative to the naively matched result.}
	\label{fig:Z_matching_benchmark}
\end{figure}

\subsubsection{Inclusive $Z$ production}

To benchmark a quark-antiquark initiated process we compare our fully inclusive predictions for $Z$ production with 
\CuTe{}. The results for the fixed-order expansion and the matching are
presented in \cref{fig:Z_matching_benchmark} and are qualitatively similar to the ones for Higgs production in \cref{fig:H_resummation_benchmark_match}.
However, the matching corrections are significantly smaller and almost negligible below 
$\SI{10}{\GeV}$. Up to \SI{50}{\GeV} they only reach few percent, but rapidly increase 
beyond that. 

\begin{figure}
	\centering
	\includegraphics{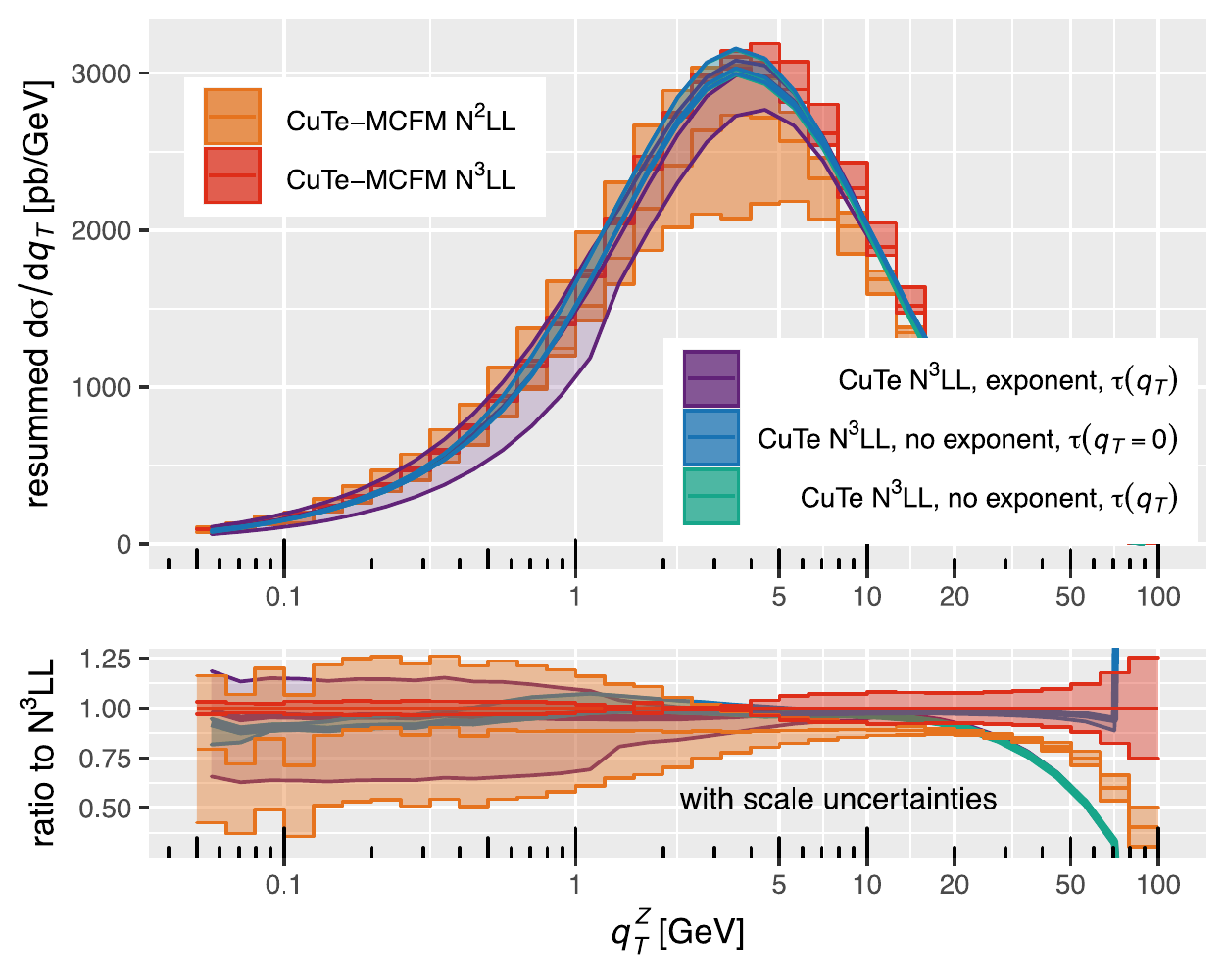}
	\caption{Resummed results without matching for inclusive $Z$ production at \SI{13}{\TeV} 
		obtained using \CuTeMCFM{} and \CuTe{} at different logarithmic orders.
		For \CuTe{} we show results in two schemes: expanded in the exponent or on the level of the 
		cross section.
		We furthermore present results with two different treatments of power-suppressed terms in the phase space related to the choice of $\tau(q_T)$, see text. The shaded band displays scale uncertainties.
		The bottom panel shows the ratio to the \NNNLL{} result in \CuTeMCFM{}.}
	\label{fig:Z_resummation_benchmark}
\end{figure}

Next, let us look at the resummed results shown in \Cref{fig:Z_resummation_benchmark}. For smaller 
$q_T$, we observe good agreement between  \CuTe{} and  \CuTeMCFM{}, but above \SI{20}{\GeV} 
there is again no overlap within scale uncertainties with the results from \CuTe{}. We have argued 
above that this is due to the inclusion of power-suppressed 
terms in the partonic momentum fractions in \CuTe{}, from setting $\tau \equiv \tau(q_T) = Q^2+q_T^2$. We verify 
this by including an additional curve where we have modified \CuTe{} to switch off the 
suppressed terms $\tau(q_T=0)$ and find good agreement also at large $q_T$. While both schemes are 
valid, we observe that the one used in \CuTe{}  leads to larger matching corrections.

\subsection{Fiducial $Z$ production}

We now turn to fiducial results, starting with $Z$ production, an experimental and theoretical standard candle. We compare with $Z\to l^+ l^-$ measurements presented in the \SI{8}{\TeV} 
\ATLAS{} study in ref.~\cite{Aad:2015auj} and the \SI{13}{\TeV} \CMS{} study in ref.~\cite{Sirunyan:2019bzr}.

\subsubsection{\ATLAS{} measurements at \SI{8}{\TeV}}
We first compare with the \ATLAS{} \SI{8}{\TeV} measurement \cite{Aad:2015auj}, which imposes the cuts listed 
in \cref{tab:zatlas8tev}.\footnote{As a 
side note, we strongly discourage the use of 
symmetric
	$q_T$ cuts, since this causes instabilities in higher-order calculations, and a slight
	asymmetry does not decrease the cross section much, see ref.~\cite{Campbell:2019dru}.}
All measurements are 
presented as normalized to the integrated fiducial cross section. Our predictions are 
calculated with a dynamic hard scale $\mu_h=\sqrt{Q^2+q_T^2}$ and the \NNPDFTO{} \PDF{} set.

\begin{figure}[t!]
	\centering
	\includegraphics{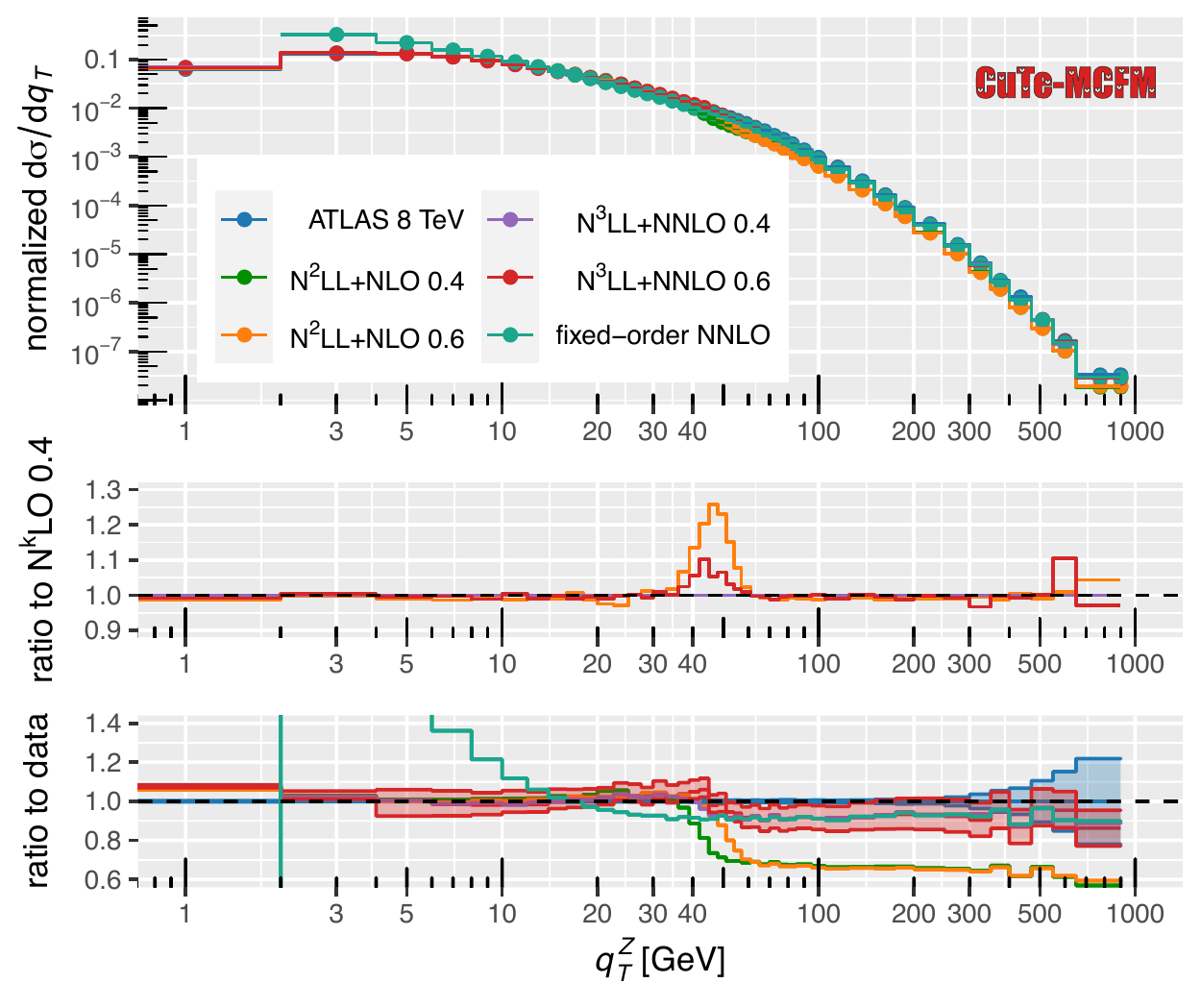}
	\caption{Predicted and measured normalized transverse-momentum distribution of the $Z$ boson 
		with fiducial cuts as in the \ATLAS{} study at \SI{8}{\TeV} in ref.~\cite{Aad:2015auj}. The middle panel shows the effect of varying the transition 
		function,
		while the bottom panel shows the ratio to data with estimated scale uncertainties. }
	\label{fig:z_atlas_8tev_ptll}
\end{figure}

In \cref{fig:z_atlas_8tev_ptll} we show
our matched prediction in comparison with the measurement.
We re-normalize all data to the $q_T$-integrated cross section with $q_T>\SI{2}{\GeV}$,
since the first bin is likely to receive non-perturbative contributions that we do not model. 
Including the first bin for the normalization would therefore skew the results.

\begin{table}
	\centering
	\caption{Fiducial cuts for $Z\to l^+ l^-$ at $\sqrt{s}=\SI{8}{\TeV}$, see ref.~\cite{Aad:2015auj}.}
	\vspace*{0.5em}
	\bgroup
	\setlength\tabcolsep{1em}
	\def\arraystretch{1.5}%
	\begin{tabular}{l|c}
		Lepton cuts & $q_T^{l} > \SI{20}{\GeV}, \abs{\eta^l} < 2.4$\\
		Separation cuts  & $\SI{66}{GeV} < m^{l^+l^-} < \SI{116}{GeV}$, $\abs{y^{l^+l^-}} < 
		2.4$
	\end{tabular}
	\egroup
	\label{tab:zatlas8tev}
\end{table}

The first panel shows the normalized distribution of the data, the \NNLL{}+\NLO{} and
\NTHREENNLO{} matched distributions with transition functions $\xmin=0.4,0.6$, and the
fixed-order prediction. The fixed-order prediction is normalized by the $q_T$-integrated matched 
result for $q_T>\SI{2}{\GeV}$ with $\xmin=0.6$.

The middle panel shows the difference between using transition function parameters $\xmin=0.4$ 
and $\xmin=0.6$ for the matched results at order $\alpha_s$ and $\alpha_s^2$.
At \NNLL{}+\NLO{} the matching effects are at the order of 10-25\% in the
region of \SIrange{40}{60}{\GeV}. A transition function that switches less rapidly than our choice
would wash out the effects to a broader range, so this has to be considered when estimating the
size of the matching effects. At \NTHREENNLO{} the matching effects are much 
smaller, as one might expect, and below 10\%.

The bottom panel shows the ratio to the experimental data and includes a
scale-uncertainty band for the \NTHREENNLO{} prediction. Due to the normalization, the experimental 
uncertainties
are at the sub-percent level for $q_T<\SI{150}{\GeV}$ and coincide with the dashed line on the 
displayed scale.
Overall our highest-order prediction at \NTHREENNLO{} describes the data very well up to
large $q_T$ within five to ten percent uncertainties.
At the largest shown $q_T$, relative \QCD{} $\alpha_s^2$ effects would increase the cross
section, but would have to be considered in addition to negative electroweak effects \cite{Becher:2013zua,Becher:2015yea,Dittmaier:2014qza}.

The fixed-order result has scale uncertainties of about $\pm10\%$, which we do not display, to keep the plot easily readable. Since the fixed-order result agrees well with the resummed 
results within mutual uncertainties down to \SI{10}{\GeV}, the transition to fixed-order could be 
induced  earlier than in range of \SIrange{40}{60}{\GeV} that we have used. Nevertheless, the resummation pushes the central prediction much closer to the data and results in agreement at the single-percent level. The presence of a large enough window for the matching is comforting and important to convince ourselves that we can combine the fixed-order and resummation results consistently and accurately.

\begin{figure}[!h]
	\centering
	\includegraphics{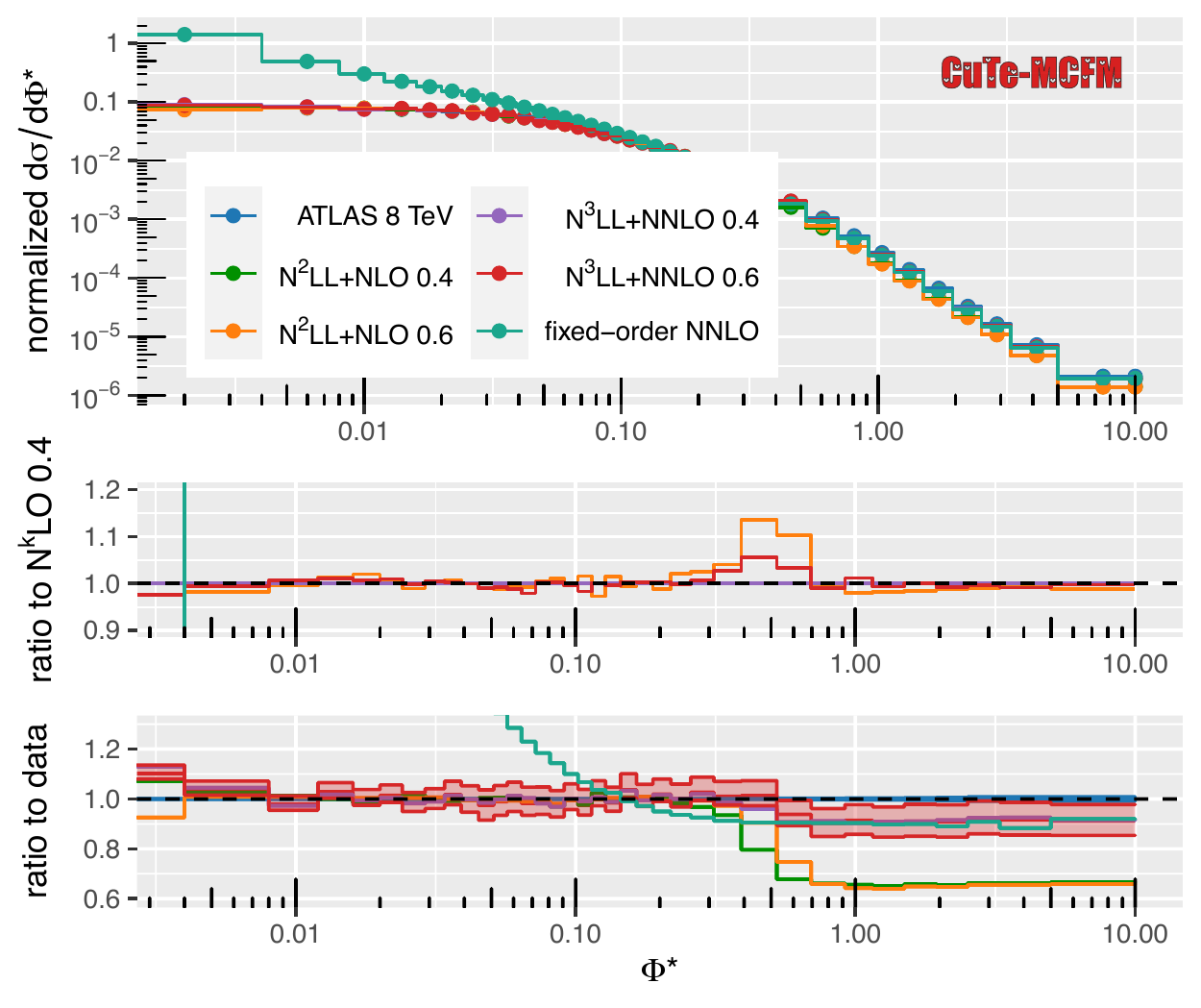}
	\caption{Predicted and measured $\phistar$ distribution of the $Z$ boson with fiducial cuts as 
		in the \ATLAS{} study \cite{Aad:2015auj} at \SI{8}{\TeV}. The middle panel shows the effect of varying the 
		transition 
		function,
		while the bottom panel shows the ratio to data with estimated scale uncertainties.}
	\label{fig:z_atlas_8tev_phistar}
\end{figure}

The resummation formula in \cref{eq:fact} is fully differential in the electroweak momenta and can be 
used to also resum logarithms in other observables related to $q_T$. An example is the observable 
\begin{equation}\label{phiDef}
\phistar = \tan\left( \frac{\pi - \delphi}{2} \right) \sin(\theta^*)\,,
\end{equation}
with $\cos(\theta^*) = \tanh\left( \frac{\Delta\eta}{2} \right)$, where $\Delta\eta$ is the 
pseudorapidity difference of the two charged leptons and $\delphi$ the azimuthal angle between 
them. This quantity was introduced in refs.~\cite{Banfi:2010cf, Banfi:2012du} and has the advantage over $q_T$ that it can be extracted purely based on angular measurements on the leptons.

Since $\phistar \propto q_T$ at small values, we also achieve full \NTHREENNLO{} accuracy for the 
$\phistar$ distribution as displayed in comparison with the measurement in \cref{fig:z_atlas_8tev_phistar}. We again 
exclude the region corresponding to small 
$q_T$ and normalize to the integrated result for $\phistar>0.004$.
The conclusions reached for the $q_T$ distribution discussed earlier apply also here, both qualitatively and quantitatively. The effects from the matching are overall smaller than $5\%$ for 
the
\NTHREENNLO{} prediction, as can be seen from the second panel. The third panel shows the ratio to the
experimental data and demonstrates a fantastic agreement with our prediction within scale uncertainties. Between $\phistar=0.1$ and $0.5$ fixed-order prediction and resummed prediction have a large window of agreement that indicates a well-behaved perturbative expansion.

\begin{table}[h]
	\centering
	\caption{Fiducial for $Z\to l^+ l^-$ at $\sqrt{s}=\SI{13}{\TeV}$, see ref.~\cite{Sirunyan:2019bzr}.}
	\vspace*{0.5em}
	\bgroup
	\setlength\tabcolsep{1em}
	\def\arraystretch{1.5}%
	\begin{tabular}{l|c}
		Lepton cuts & $q_T^{l} > \SI{25}{\GeV}, \abs{\eta^l} < 2.4$\\
		Separation cuts  & $\SI{76.2}{GeV} < m^{l^+l^-} < \SI{106.2}{GeV}$, $\abs{y^{l^+l^-}} < 
		2.4$
	\end{tabular}
	\egroup
	\label{tab:zcms13tev}
\end{table}

\subsubsection{\CMS{} measurements at \SI{13}{\TeV}}
\begin{figure}[!t]
	\centering
	\includegraphics{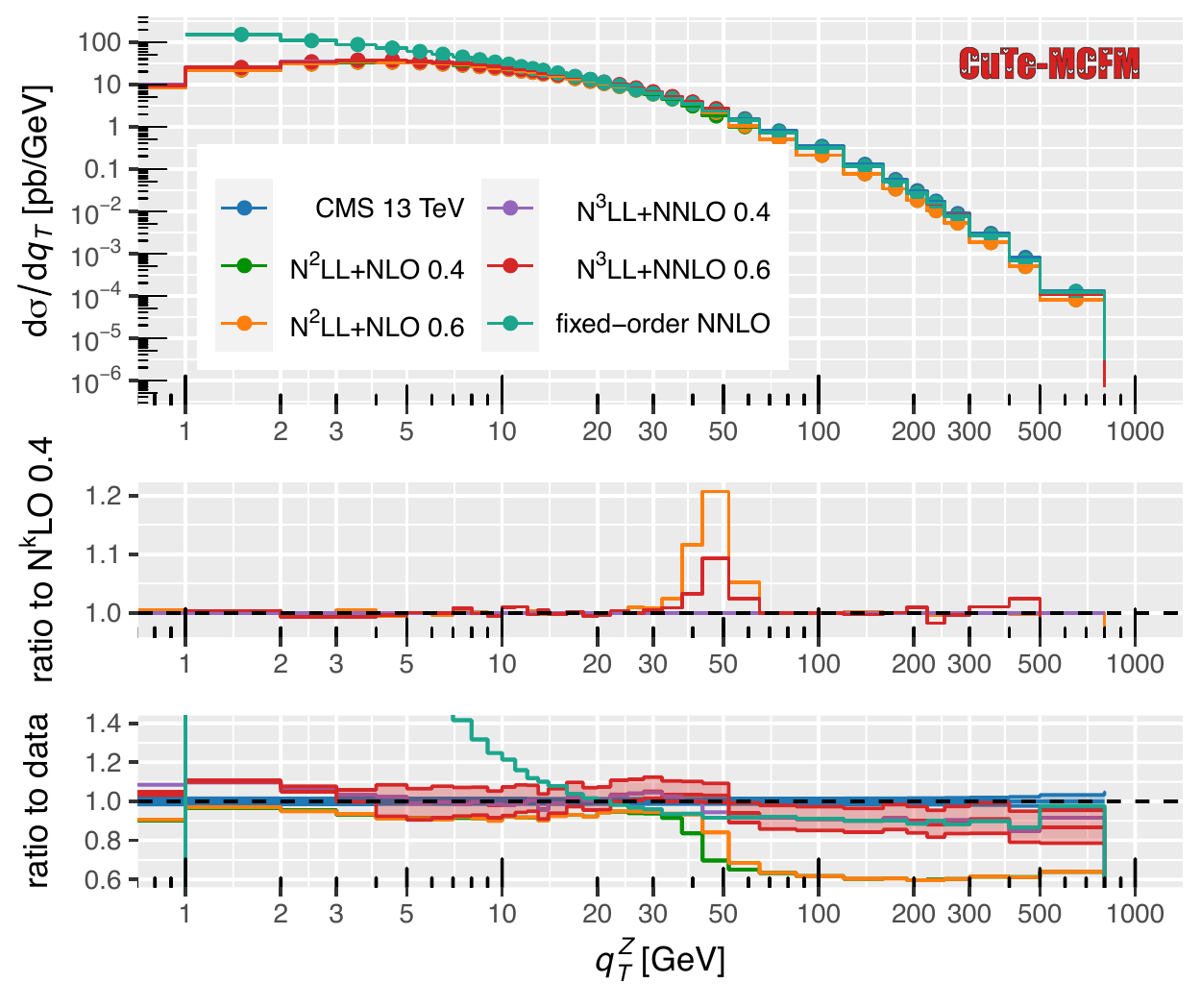}
	\caption{Predicted and measured transverse-momentum distribution of the $Z$ boson with fiducial cuts as in the \CMS{} study \cite{Sirunyan:2019bzr} at \SI{13}{\TeV}. The middle panel shows the effect of varying the 
		transition 	function, while the bottom panel shows the ratio to data with estimated scale uncertainties. }
	\label{fig:z13cms}
\end{figure}

As a second example, we directly compare with \SI{13}{\TeV} cross-section data from \CMS{} 
\cite{Sirunyan:2019bzr}
in 
\cref{fig:z13cms}, without normalizing the results. The applied cuts are presented in 
\cref{tab:zcms13tev}. We again choose
the dynamic hard scale as $\mu_h=\sqrt{Q^2+q_T^2}$ and use the \NNPDFTO{} \PDF{} set.

Overall the conclusions are similar to our findings for the normalized 
predictions shown at \SI{8}{\TeV} before. Up to \SI{40}{\GeV} the resummed result (matched with small matching 
corrections)
agrees at the percent level with data. Only in the first bins the small scale uncertainties 
and a deviation of up to 10\% hint towards non-perturbative effects.

\begin{figure}[!h]
	\centering
	\includegraphics{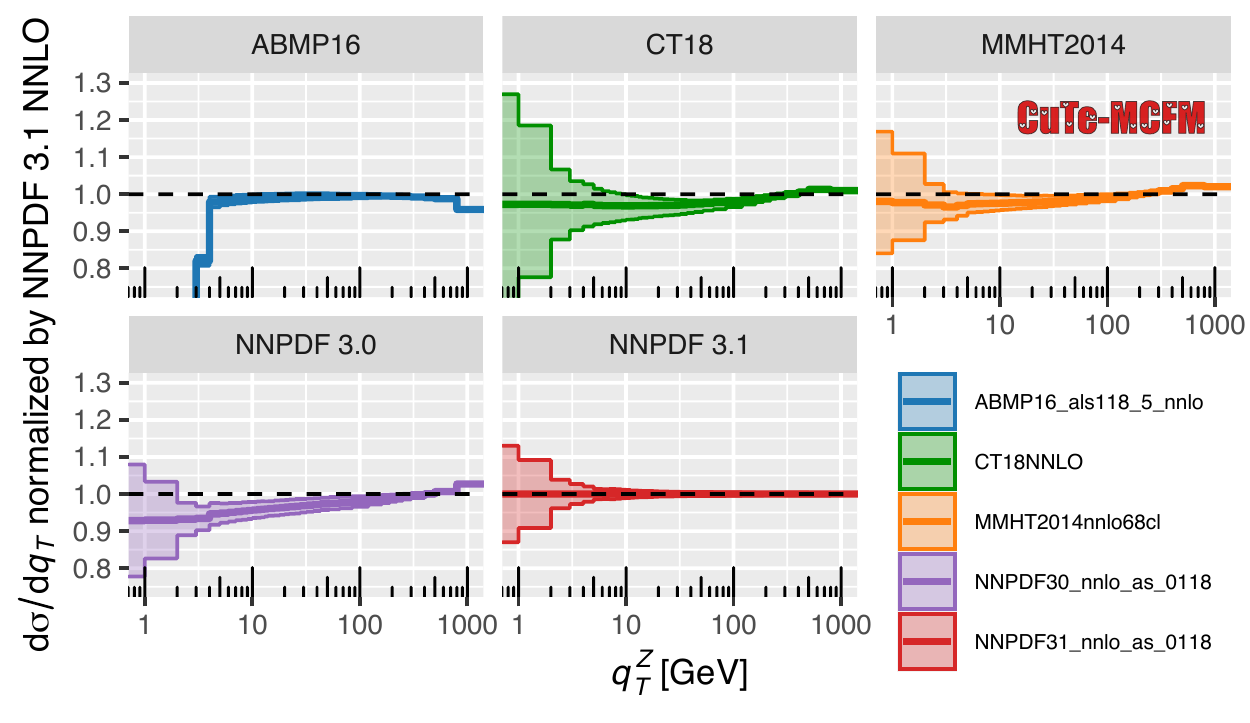}
	\caption{$Z$-boson transverse-momentum distribution including
		\PDF{} uncertainties for various \PDF{} sets normalized to the prediction with the 
		NNPDF3.1 \NNLO{} central value. 
		See \cref{fig:z13cms}.}
	\label{fig:z13cmsPDF}
\end{figure}

While non-perturbative transverse-momentum effects would be captured by fitting transverse-momentum dependent \PDF{}s, also the
standard \PDF{}s encode non-perturbative physics. To study the associated uncertainties, we computed the \PDF{} uncertainties for multiple \PDF{} sets and show the result in \cref{fig:z13cmsPDF}. On a technical level, this demonstrates the efficient
and accurate evaluation of \PDF{} uncertainties in \MCFM{}-9 and consequently also in our setup 
\CuTeMCFM{}.

The minimum scale value of $Q^\text{min}=\SI{4.47}{\GeV}$ for the ABMP16 \PDF{} set causes the predictions to break down when our scale is set to a value lower than this. To fix this issue, one
could, in principle, perform a {\abbrev DGLAP} evolution below this scale, or enforce a minimum scale of $Q^\text{min}$ in our resummation code. Instead we deliberately show the result with the default settings of \LHAPDF{} and our default minimum safety scale of $\SI{2}{\GeV}$. 

The other \PDF{}s broadly predict uncertainties above $10\%$ below $q_T=\SI{2}{\GeV}$, and CT18 
even predicts uncertainties of more than $20\%$. The $10\%$ difference between our prediction and 
data in the first bin of \cref{fig:z13cms} is therefore well within even just \PDF{} uncertainties.

\subsection{Fiducial $W$ production as measured by \CMS{} at \SI{8}{\TeV}}

While the transverse-momentum distribution of the charged lepton in $W$ production
enters many precision analyses, the fully reconstructed $W$ boson transverse-momentum distribution
has also been presented by \CMS{} at \SI{8}{\TeV} \cite{Khachatryan:2016nbe} and at \SI{7}{\TeV} by 
\ATLAS{} \cite{Aad:2011fp}.

Here we compare with the normalized \SI{8}{\TeV} \CMS{} data, where both
$W^-$ and $W^+$ channels are added.
The only applied cuts are a minimum $q_T$ of $\SI{25}{\GeV}$ and a maximum
absolute pseudorapidity of $2.5$ on the electron/positron. We furthermore choose a central hard 
scale 
of $\mu_h=\sqrt{Q^2 + 
q_T^2}$ and the \PDF{} set \CTEIGHT{}. 
As for $Z$ production, the resummed logarithms describe the full result at an impressive
level with matching corrections that stay just at the few percent-level for
$q_T^W \lesssim \SI{40}{\GeV}$, reaching about $30$ percent for \SI{60}{GeV} (not shown).

\begin{figure}[!t]
	\centering
	\includegraphics{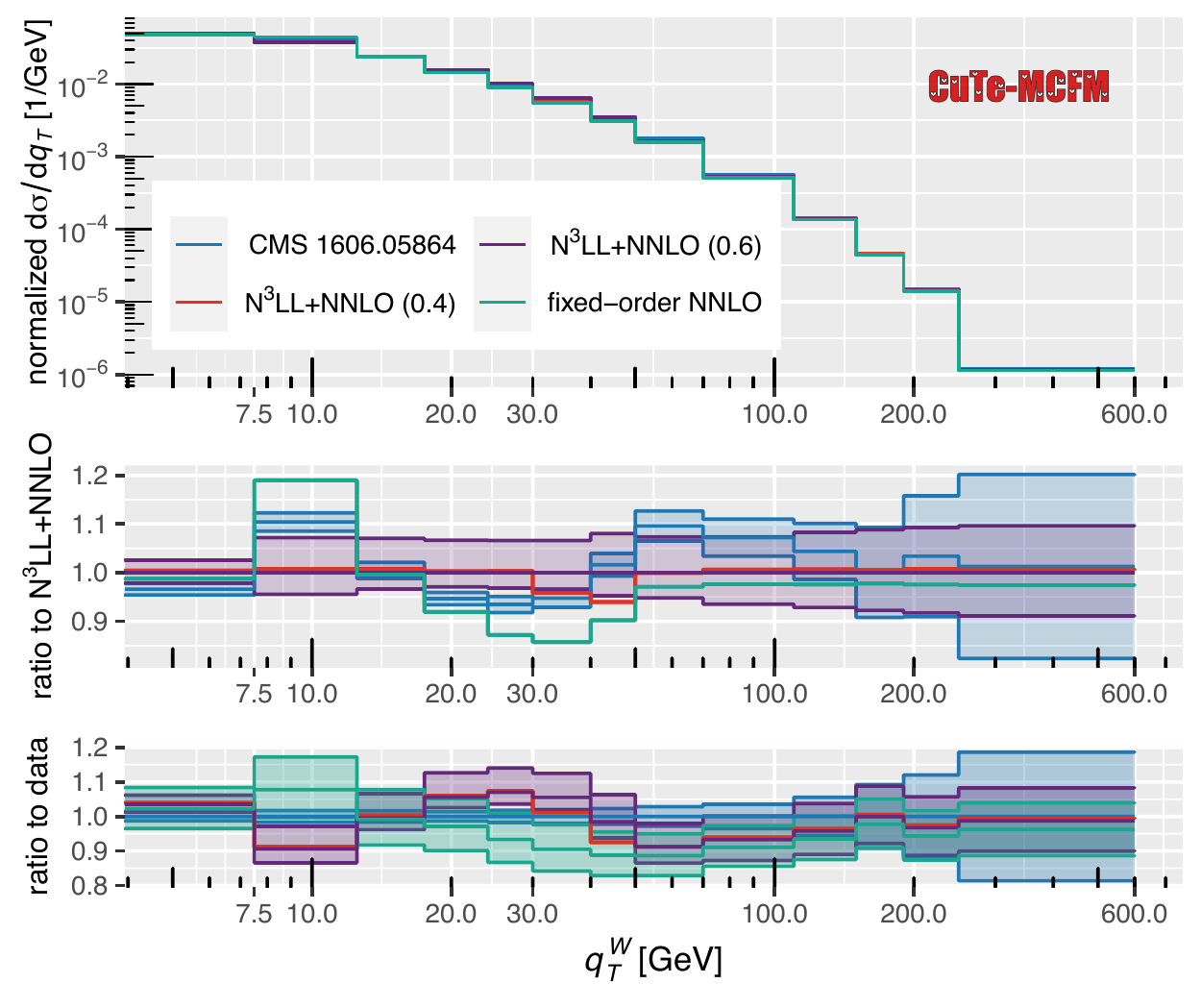}
	\caption{Comparison to normalized $W$ transverse-momentum data from \CMS{} at \SI{8}{\TeV} with predictions at \NTHREENNLO{} including uncertainties associated with scale variation.\label{fig:1606_05864}}
\end{figure}

We show our matched results in \cref{fig:1606_05864}, where the first panel shows the normalized distribution. The second panel shows the ratio to our matched result with $\xmin=0.6$ to guide the eye on the difference between the two transition functions with $\xmin=0.4,0.6$.
The resulting difference between our two choices of $\xmin$ can be seen around 
\SIrange{30}{50}{\GeV}
where the two distributions differ by about five percent. This estimates the size and position of 
the matching uncertainty.
The second panel also shows the experimental uncertainties and scale uncertainties for the 
matched result. The scale-variation band is shown after normalizing by the central-value 
cross-section.

The third panel shows the ratio to the \CMS{} data and now also includes a scale-variation band for the fixed-order prediction. While we would have expected to find
very good agreement in the region where the matching corrections are small ($\lesssim\SI{40}{\GeV}$),
the overall agreement to the data is not better than the fixed-order prediction, but overall we find agreement within scale uncertainties.

The most striking difference of $10\%$ between central prediction and data is in the second bin from \SIrange{7.5}{12.5}{\GeV}. This deviation can also be observed in the \CMS{} publication
\cite{Khachatryan:2016nbe} when the data is compared with with \NNLL{} resummed predictions. This 
difference and the overall shape of the data for $q_T\lesssim \SI{40}{\GeV}$ is perhaps indicative 
a systematic issue with the experimental analysis at low $q_T$, but it is not possible to make a 
definite statement since the results are compatible within mutual uncertainties.

\subsection{Fiducial $H\to\gamma\gamma$ benchmark} 

The Higgs transverse-momentum distribution has been measured by \CMS{} and \ATLAS{}
at \SI{8}{\TeV} and \SI{13}{\TeV} in various production and decay channels. But even
after a combination the overall uncertainties are at the order of 40\% or worse 
\cite{Sirunyan:2018sgc}.

For a precise study and prediction, one should at the least take into account
top-quark-mass effects and consider the resummation of $\pi^2$ terms \cite{Ahrens:2008qu,Ahrens:2008nc}. Further contributions like bottom-quark-mass effects have also been studied at low $q_T$ \cite{Caola:2018zye} and become relevant at the percent level for the resummation. 
Within the \MCFM{} framework top-quark-mass effects have been included 
throughout \NLO{}
accuracy for $q_T \gg m_t$ and $q_T\ll m_t$ \cite{Neumann:2018bsx,Neumann:2016dny,Budge:2020oyl} 
and \NNLO{} 
corrections have been presented in the \EFT{} for large $q_T$
\cite{Campbell:2019gmd,Boughezal:2015aha}.
Including these mass effects goes beyond the scope of our paper and we only show results in the heavy-top-quark limit.
For now we present results without comparison to data, but include a set of cuts as used in experiments, see \cref{tab:higgscuts}.

\begin{table}[!t]
	\centering
	\caption{Fiducial cuts for $H\to\gamma\gamma$ at $\sqrt{s}=\SI{13}{\TeV}$.}
	\vspace*{0.5em}
	\bgroup
	\setlength\tabcolsep{1em}
	\def\arraystretch{1.5}%
	\begin{tabular}{l|c}
		Photon cuts & $q_T^{\gamma} > \SI{40}{\GeV}, \SI{30}{\GeV}$, $\abs{\eta_\gamma} < 2.5$\\
		Smooth-cone photon isolation & $\etiso = \SI{10}{\GeV}, R=0.3, n=1$
	\end{tabular}
	\egroup
	\label{tab:higgscuts}
\end{table}

\begin{figure}
	\centering
	\includegraphics{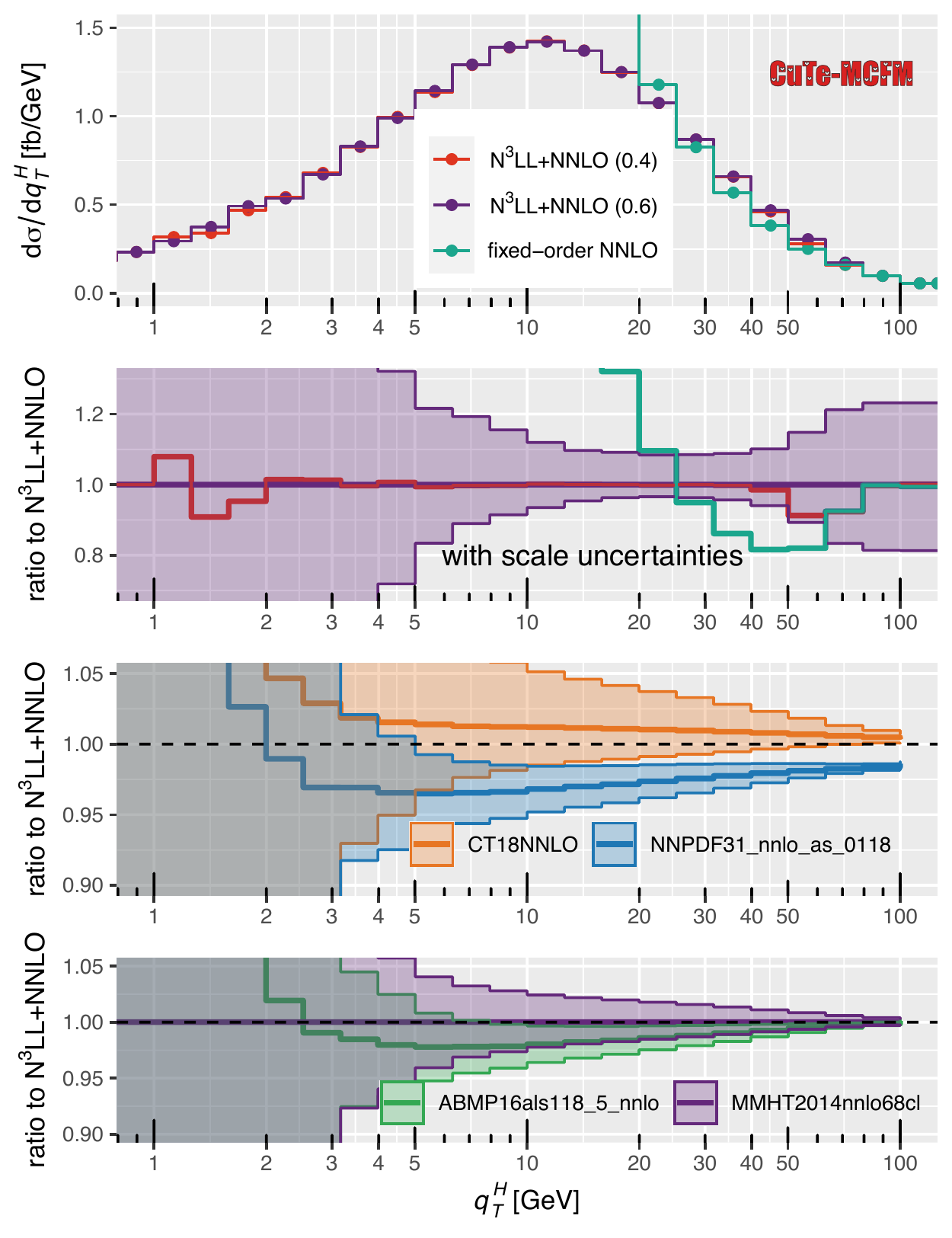}
	\caption{Higgs transverse-momentum distribution with fiducial cuts as in \cref{tab:higgscuts}. 
	The second panel shows the effect of a different transition function and scale uncertainties, 
	while the bottom panels show \PDF{} uncertainties for different sets.}
	\label{fig:higgs}
\end{figure}

In \cref{fig:higgs} we show matched results for the Higgs transverse-momentum distribution
with fiducial cuts as in \cref{tab:higgscuts} using the \MMHTFOUR{} \PDF{} set and a central hard 
scale of $\mu_h=\sqrt{m_H^2 + q_T^2}$.
The second panel shows the effect of the matching
to fixed order by switching between the transition function parameters $\xmin=0.4$ and $0.6$.
Matching effects are about \SI{10}{\%} in the region of \SIrange{50}{80}{\GeV} and the resummation
stabilizes the fixed-order predictions below such values. At small values of $q_T\lesssim\SI{2}{\GeV}$
the cancellations within the matching corrections are numerically difficult and reflect in
the larger fluctuations.

The bottom two panels show the effect of \PDF{} uncertainties relative to our central prediction.
At $q_T\lesssim \SI{5}{\GeV}$ uncertainties of more than \SI{10}{\%} have to be added to the 
already sizeable
scale uncertainties. While these uncertainties add up to just give an order of magnitude
prediction, the uncertainties from $\alpha_s$ itself, to which gluon fusion Higgs production is highly sensitive,
are not even included yet. The road towards precision Higgs transverse-momentum measurements and 
predictions is therefore a long one, but using the normalized distribution would mitigate some of these additional uncertainties.

\subsection{Fiducial $\gamma\gamma$ production}
\label{sec:diphoton}

In this section we present results for fiducial diphoton production. The fixed-order \NNLO{}
result in \MCFM{} is based on ref.~\cite{Campbell:2016yrh}. \NNLO{} results have also been presented
in ref.~\cite{Catani:2011qz}, which were subsequently interfaced with \NNLL{} $q_T$ resummation \cite{Cieri:2015rqa}.
Resummation at \NNLL{} has also been presented in ref.~\cite{Balazs:2007hr} matched to \NLO{}.

Perturbative \NNLO{} corrections in diphoton production are large and increase \NLO{} results by 50-75\%, depending on 
cuts.
The gluon-gluon channel, which first appears at \NNLO{} through a
quark-box diagram, also constitutes a noticeable part of these corrections. Therefore, only at 
\NNNLO{} one has
control at the \NLO{} level over all partonic channels. Two-loop \NLO{} corrections for the 
gluon-gluon channel have been
calculated in refs.~\cite{Bern:2001df,Bern:2002jx}. These have later been implemented in \MCFM{} 
together with
the \NNLO{} corrections to the $q \bar{q}$ channel \cite{Campbell:2016yrh} and constitute a part of 
the {\abbrev N$^3$LO}
corrections. We also discuss these and show the effect of including them in the following.

\begin{figure}
	\centering
	\includegraphics{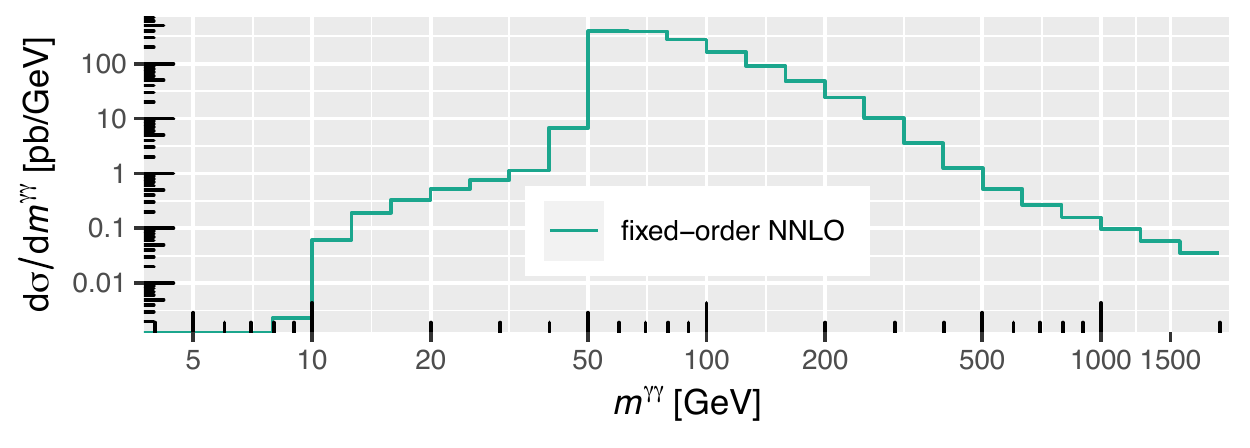}
	\caption{Diphoton invariant mass distribution at \NNLO{} with  $q_T^{\gamma,1}>\SI{25}{\GeV}$, 
	$q_T^{\gamma,2} >\SI{22}{\GeV}$ and further cuts specified in \cref{tab:diphoton7tev}.}
	\label{fig:atlas7-diphoton-m34}
\end{figure}

A common requirement for diphoton production is that both photons have specific minimum 
transverse momenta, $q_{T,\text{min}}^{\gamma,1}$ (harder) and 
$q_{T,\text{min}}^{\gamma,2}$ (softer), where 
$q_{T,\text{min}}^{\gamma,1}>q_{T,\text{min}}^{\gamma,2}$.
For transverse momenta larger than 
$q_{T,\text{min}}^{\gamma,1}+q_{T,\text{min}}^{\gamma,2}$
both photons can be aligned in the same direction and recoil against hadronic radiation. This threshold
can be seen for example in the diphoton invariant mass distribution shown in 
\cref{fig:atlas7-diphoton-m34}, which strongly peaks above  
$\sim q_{T,\text{min}}^{\gamma,1}+q_{T,\text{min}}^{\gamma,2}=\SI{47}{\GeV}$. 
(As a side remark, we note that the cusp in this distribution could be removed by an appropriate soft gluon resummation \cite{Catani:1997xc,Catani:2018krb}.) Transverse-momentum resummation is no longer valid for $q_T$ values above this threshold
and becomes numerically unstable, so that one wants to fully switch
to the fixed-order prediction above this threshold. To not introduce a discontinuity,
the transition function has to be chosen to give negligible contributions from $q_T > 
q_{T,\text{min}}^{\gamma,1}+q_{T,\text{min}}^{\gamma,2}$. Of course, from the resummation point of view, it would be best to impose a lower cut on the invariant mass of the two photons, which would avoid these problems.

\subsubsection{\ATLAS{} measurements at \SI{7}{\TeV}}
\label{sec:atlas7-diphoton}
We first compare with the \SI{7}{\TeV} \ATLAS{} diphoton measurement \cite{Aad:2012tba}.
The fiducial phase space is defined by the cuts in \cref{tab:diphoton7tev}.
Our results here are presented using the \PDF{} set \MSTW{} \cite{MSTW2008} and a
central hard scale of $\mu_h=m_{\gamma\gamma}$, following the choice in the previous study at \NNLL{} \cite{Cieri:2015rqa}
to allow for a direct comparison. 

\begin{table}[]
	\centering
	\caption{Fiducial in diphoton production at $\sqrt{s}=\SI{7}{\TeV}$, see ref.~\cite{Aad:2012tba}.}
	\vspace*{0.5em}
	\bgroup
	\setlength\tabcolsep{1em}
	\def\arraystretch{1.5}%
	\begin{tabular}{l|c}
		\multirow{2}{*}{Photon cuts } &  $q_T^{\gamma,1}>\SI{25}{\GeV}$, $q_T^{\gamma,2} 	
		>\SI{22}{\GeV}$ \\ &
		$|\eta_\gamma|<2.37$,  $1.56 < |\eta_\gamma| < 1.37$\\
		Photon separation & $\Delta R(\gamma,\gamma)>0.4$, \\
		Smooth-cone photon isolation & $\etiso=\SI{4}{\GeV}$, $n=1$, $R=0.4$
	\end{tabular}
	\egroup
	\label{tab:diphoton7tev}
\end{table}

Our detailed discussion in \cref{par:photiso} (\cpageref{par:photiso}), showed that for smooth-cone isolation (and $n=1$) linear power corrections are present. To account 
for this, we could modify our transition function to be a function of  $q_T/Q$ without spoiling 
power 
corrections. Instead, we keep it as a function of $q_T^2/Q^2$, but choose a sufficiently smaller parameter $\xmin$ determining the transition region. We 
find that 
$\xmin=0.1,0.2$ are sufficiently small such that we can can fully switch to the fixed-order result above \SI{47}{\GeV},
but can also study the effect of the transition to the fixed-order result.

\begin{figure}
    \centering
    \includegraphics[width=0.495\textwidth]{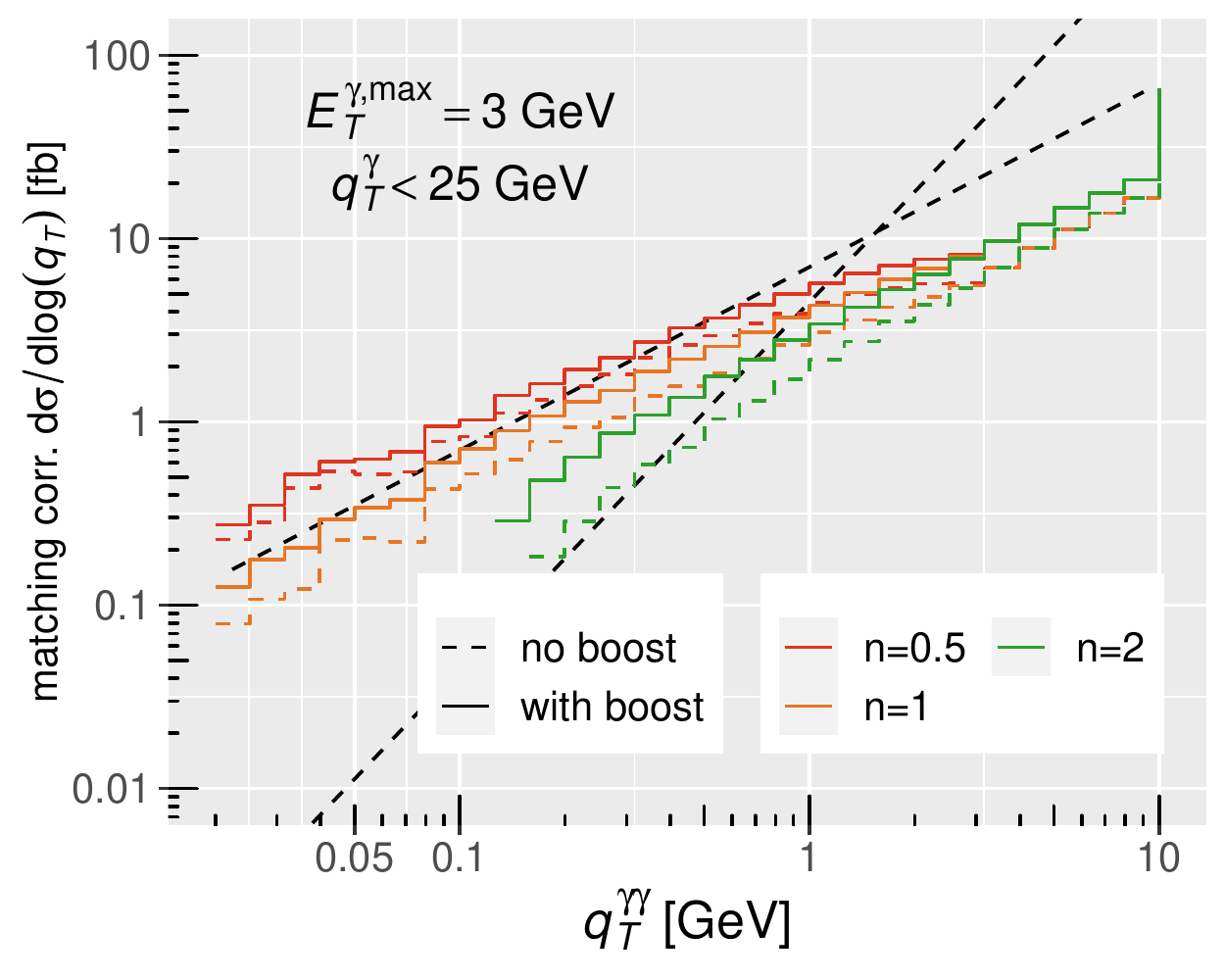}  \includegraphics[width=0.495\textwidth]{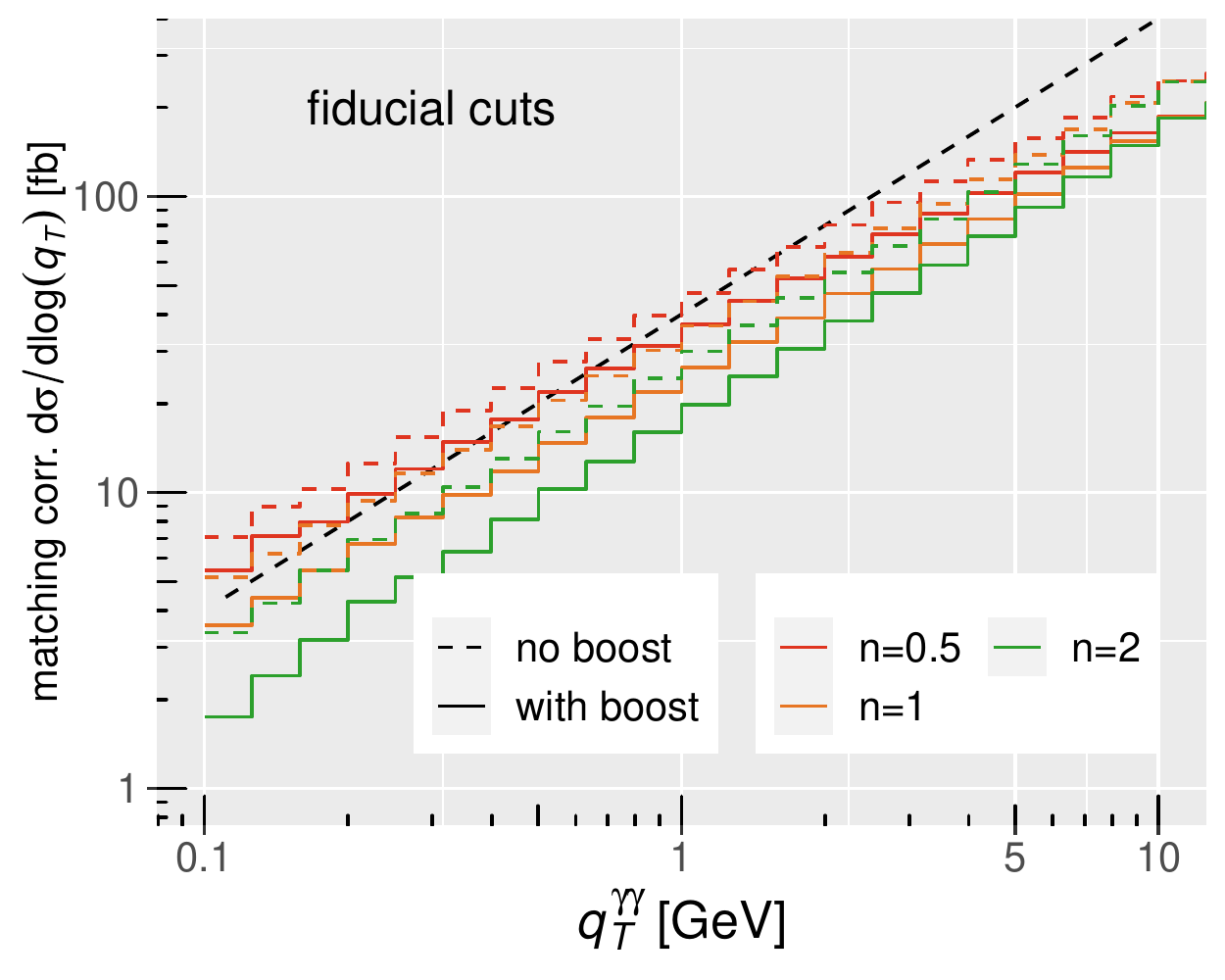}
    \caption{Matching corrections at \NNLL{} for diphoton production with photon $q_T$ and photon isolation 
    cuts only (left) 
    and all fiducial cuts in \cref{tab:diphoton7tev} (right). The different colors correspond to different values of $n$. The 
    solid lines are the power corrections after accounting for recoil, the dashed lines are without recoil. The black dashed lines indicate first and second order scaling in $q_T$}
    \label{fig:powcorr_diphoton}
\end{figure}

In \cref{fig:powcorr_higgs} in \cref{sec:calculation}, we have discussed the asymptotic scaling of the power corrections with the isolation parameter $n$ and have shown numerical results for the  partonic $q\bar{q}$ channel. We now show the sum of all partonic channels, with and without fiducial cuts, in 
\cref{fig:powcorr_diphoton}. The behavior is qualitatively different than in 
the $q\bar{q}$ channel shown earlier in \cref{fig:powcorr_higgs}:  At least in the $q_T$ range we consider, the power corrections are approximately linear, and relatively insensitive to the choice of $n$. Furthermore, even for $n=2$, where power corrections scale like $\sqrt{q_T}$ in the $q\bar{q}$ channel, they are somewhat more suppressed when considering all partonic channels.

The reason for the different behavior is that the power corrections associated with a gluon radiated inside the cone, as present in the $q\bar{q}$ channel, are suppressed by $R^2$, in contrast to the ones associated with the fragmentation correction, see \cref{eq:gluonPower,eq:quarkPower}. These two contributions also enter with different
signs, so that for non-asymptotic values of $q_T$ cancellation effects occur.
Eventually, for $n=2$ and sufficiently small $q_T$, the $R^2$ suppression is overcome and the asymptotic behavior should set in and one would expect to observe $\sqrt{q_T}$ scaling again.

\begin{figure}
	\centering
	\includegraphics{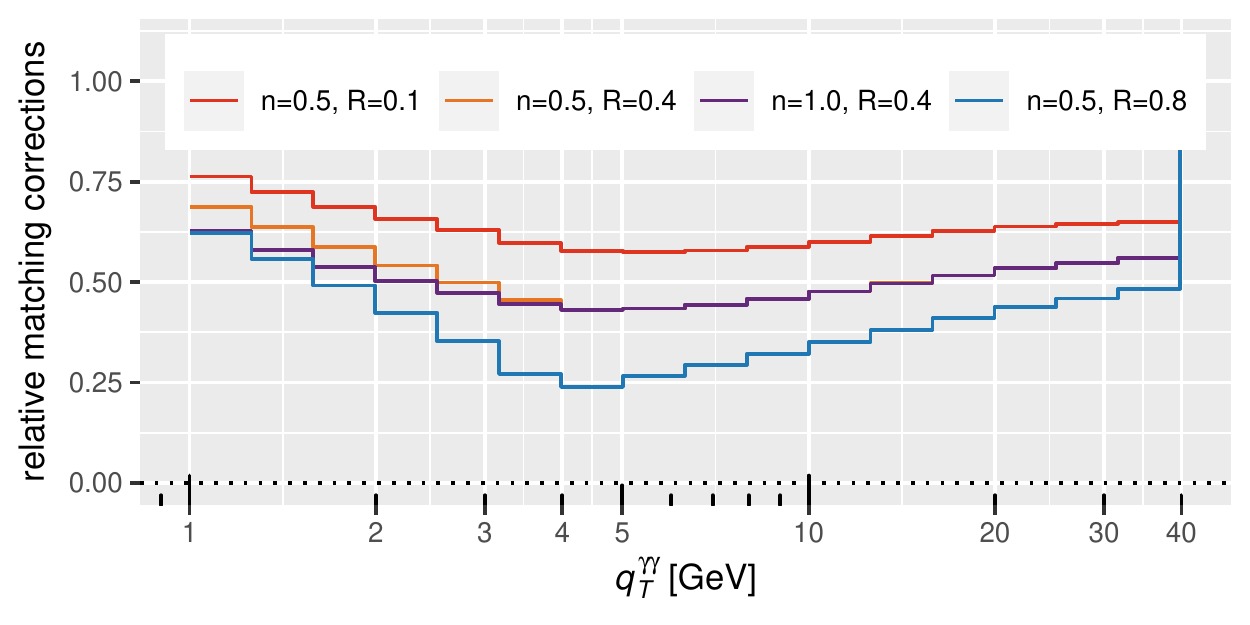}
	\caption{Diphoton maching corrections at \NNLL{} relative to naively matched results with cuts as in \cref{tab:diphoton7tev}, but for different choices of the photon isolation parameters $n$ and $R$. }
	\label{fig:atlas7-diphoton-matchcorr}
\end{figure}

As discussed in \cref{sec:calculation}, the presence of linear power 
corrections implies that the matching corrections no longer go to zero in $d\sigma/dq_T$. We show these corrections at \NNLL{} in \cref{fig:atlas7-diphoton-matchcorr} relative to the naively matched result. Here, we are interested in larger $q_T$ values of practical relevance and not in the asymptotic behavior. We include different
choices of the photon isolation parameters $n$ and $R$. The purple curve with $n=1$ and $R=0.4$
corresponds to the default fiducial cuts in \cref{tab:diphoton7tev}.

Already at small $q_T$ the purple matching corrections start at about $60\%$ and never dip below
$40\%$. Since the matching corrections go to a constant for $q_T\to0$, a characteristic of the linear power corrections, and the resummed result approaches zero, the relative matching
corrections eventually approach 100\% for $q_T\to0$.
Note that this does not signify an issue with our implementation of the $q_T$~resummation. We have checked that the fixed-order expansion of the resummed result 
cancels with the fixed-order result down to the sub per-mille level for very small $q_T$.
The correct cancellation can indeed be observed as the (linearly) vanishing matching corrections in \cref{fig:powcorr_diphoton}. 

With \cref{fig:atlas7-diphoton-matchcorr} we can now also discuss the effect of cancellations between photon-isolation power corrections associated with fragmentation and associated with gluon emission for moderate values of $q_T$. These contributions enter with opposite signs, which
has a peculiar effect on the $R$-dependence: Naively one might
expect smaller power corrections with a smaller $R$, since the gluon radiation power corrections
scale like $R^2$. But these negative power corrections have to be added to the positive
and larger power corrections from the fragmentation contribution. One therefore observes a cancellation and overall smaller matching corrections for larger~$R$.

\begin{figure}[t!]
	\centering
	\includegraphics{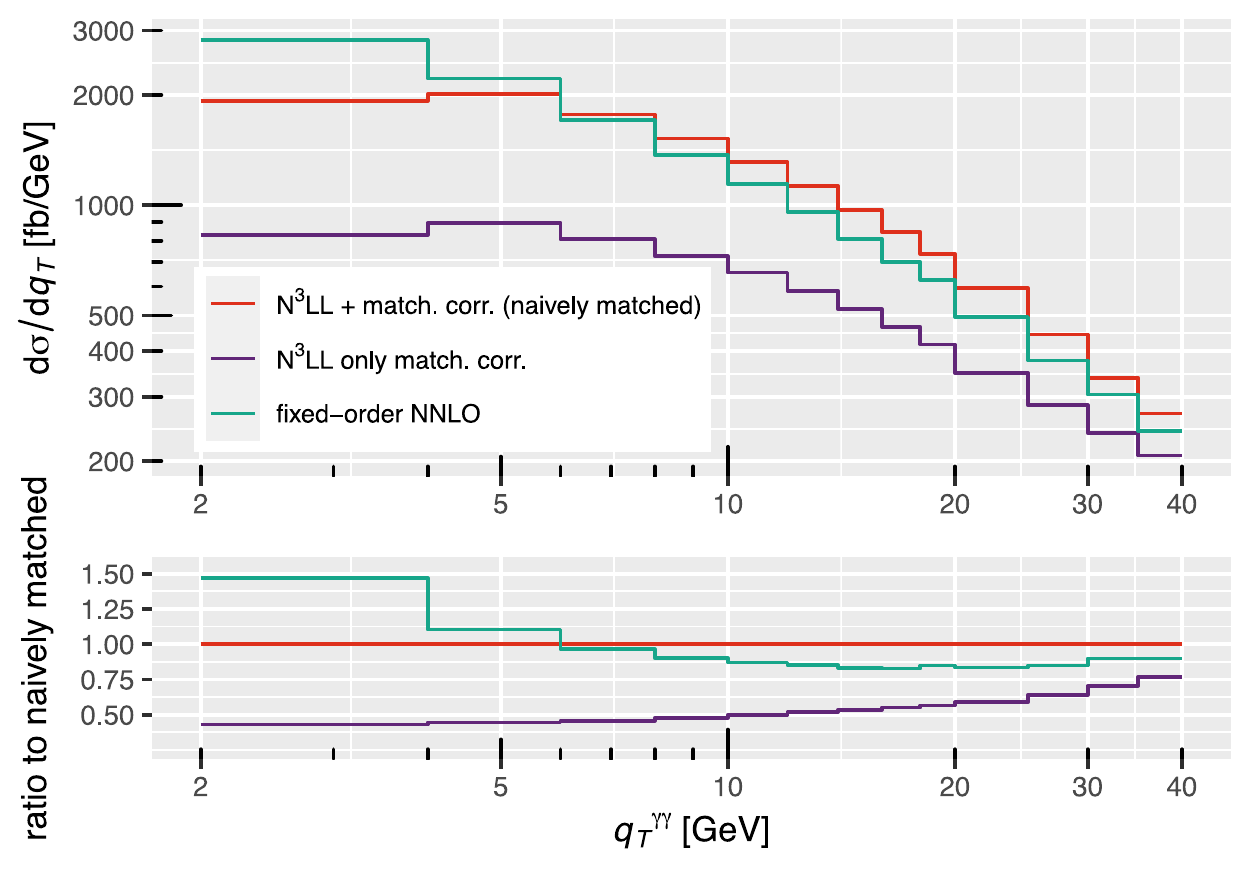}
	\caption{Comparison of the naively matched result at \NTHREENNLO{}, the matching corrections and the fixed-order result for 
		diphoton production with the cuts used by \ATLAS{} \cite{Aad:2012tba}, see \cref{tab:diphoton7tev}. The second panel
		displays the ratio to the naively matched result to demonstrate the large matching corrections even at small $q_T$. }
	\label{fig:atlas7-diphoton-pt34-one}
\end{figure}

At \NNNLL{} the observed large matching corrections do not change qualitatively, as shown in \cref{fig:atlas7-diphoton-pt34-one}. The matching corrections now start just below 50\% around $q_T=\SI{2}{\GeV}$ and reach 75\% just before the resummation validity threshold of $\SI{47}{\GeV}$.

Since we do not include matching corrections below \SI{1}{\GeV} for numerical stability, we neglect sizeable effects below this value. Taken at face value, the matching would amount to a$~\SI{50}{\%}$ effect. A resummation of power-suppressed terms would likely suppress the matching corrections, but their true size is difficult to estimate. The situation is different from processes where the matching corrections are quadratic and such a safety cutoff of \SI{1}{\GeV} leads to small effects. While 
such a hard cutoff is relatively unproblematic for the $q_T$ distribution itself, it is more difficult for other observables that benefit from resummation, like $\phistar$ or the azimuthal angle difference between the photons $\delphi$. The cutoff may affect a broader spectrum for such observables, and not just one bin. Fortunately, at least for $\phistar$ only low values are affected since $\phi^* \leq q_T/Q$. To do better than this somewhat arbitrary cutoff prescription, we would need to determine and include the Sudakov suppression factor for the power-suppressed terms.

Overall, since the first few GeV in $q_T$ are likely to receive non-perturbative effects (see
for example the parametrization of non-perturbative effects in ref.~\cite{Cieri:2015rqa}), we can disregard the first experimental bin (from \SIrange{0}{2}{\GeV}) for a meaningful comparison in this study. For observables like $\phistar$ or $\delphi$ similar arguments regarding non-perturbative corrections hold so that predictions for the regions corresponding to values $q_T\sim 0$ need to be studied very carefully.

\begin{figure}
    \centering
    \includegraphics{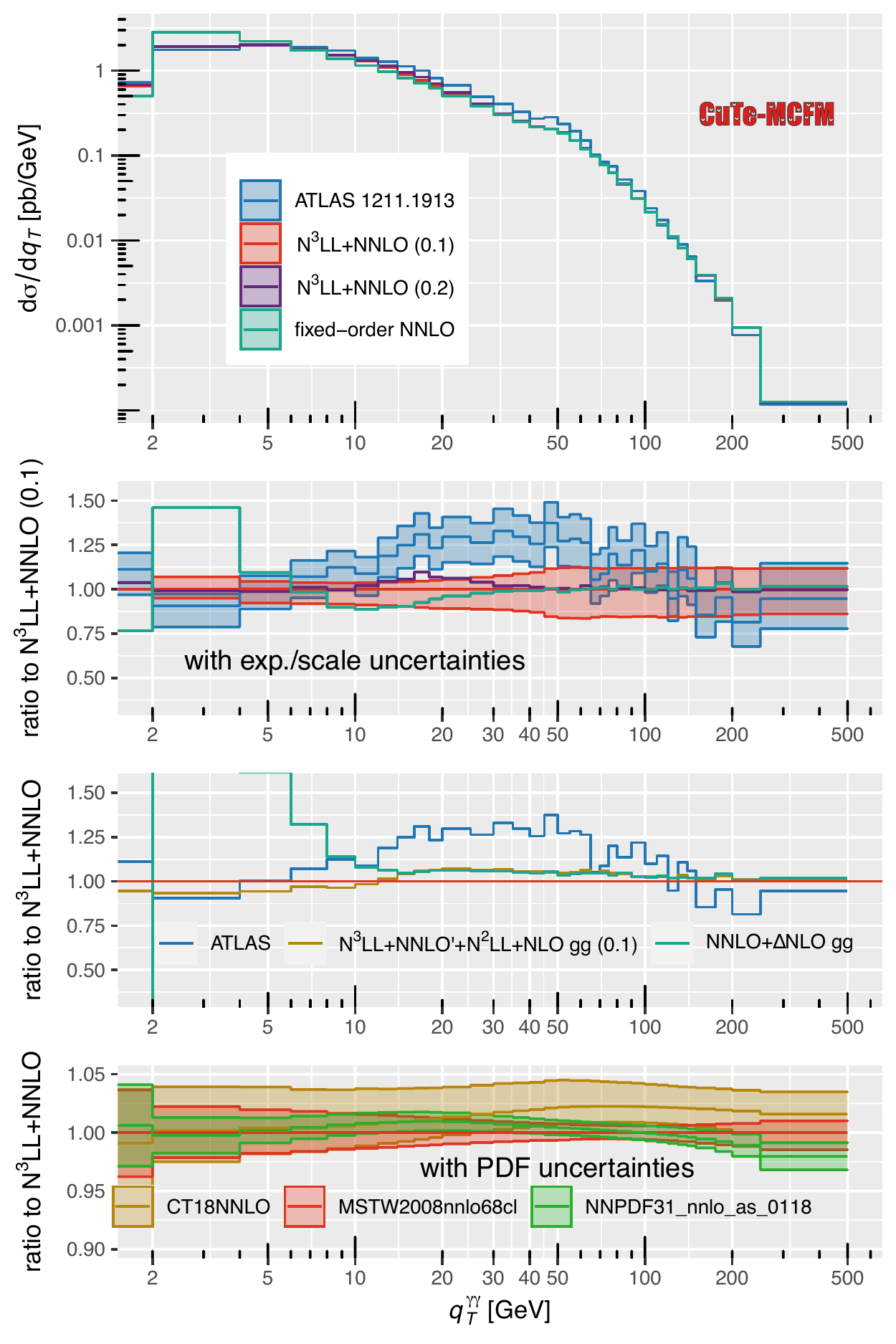}
    \caption{Comparison of \SI{7}{\TeV} \ATLAS{} diphoton results with predictions in various approximations including uncertainties associated with scale variation, \PDF{}s and \NLO{} contributions to the gluon-gluon channel; see text for details. The labels $0.1$ and $0.2$ in the plots refer to the value of $\xmin$.}
    \label{fig:atlas7-diphoton-pt34-two}
\end{figure}

Having analyzed the matching corrections in detail, we now choose the transition function with $\xmin=0.1,0.2$, as indicated earlier, and present our matched \NTHREENNLO{} results in
\cref{fig:atlas7-diphoton-pt34-two}. The upper
panel in this figure shows the absolute \NTHREENNLO{} matched distribution with the two choices of $\xmin$,
the \NNLO{} fixed-order result, as well as the measurement.

The second panel shows corresponding ratios to the matched \NTHREENNLO{} result with $\xmin=0.1$. The agreement
with data in the region of up to \SI{30}{\GeV} is clearly improved, with resummation effects of 
up to 13\% around \SI{15}{\GeV}. Beyond $\sim\SI{45}{\GeV}$ practically only the
fixed-order result contributes. The filled regions denote the experimental uncertainties for the \ATLAS{} data and 
uncertainties from scale variation for the matched result, respectively. For brevity we do not include
uncertainties for the fixed-order prediction, which agree above
$q_T^{\gamma\gamma}\gtrsim\SI{50}{\GeV}$ with the matched result and are similarly sized below.

Including the gluon-gluon channel at \NLO{} has been found to be an important contribution at fixed-order
perturbation theory \cite{Campbell:2016yrh}. Therefore, in the third panel we additionally display the
ratios where the gg-channel is included at \NLO{} and \NNLL{}+\NLO{}. The fixed-order result is obtained by
adding the \NLO{} gg-channel $\text{gg}\to\gamma\gamma \text{g}$ to the \NNLO{} fixed-order result without the gg-channel.
The resummed result is obtained by matching at \NNNLL{} with the \NNLO{} result without gg-channel,
and adding the \NNLL{}+\NLO{} resummed gg-channel. 

While the fixed-order result indeed receives sizeable corrections from the gg$\to\gamma\gamma\text{g}$ channel in
the region up to $\sim\SI{50}{\GeV}$, the corrections from the matched result change little compared to the 
overall uncertainties and agreement with data. In fact, the improved fixed-order result (cyan) and improved matched result (yellow) agree above \SI{15}{\GeV}. This indicates a significant stabilization
of the perturbative series through the \NLO{} corrections in the $gg$-channel.

Finally, the fourth panel displays the comparatively small \PDF{} uncertainties at the level of a few
percent (with a fixed value of $\alpha_s$).

\clearpage
\subsubsection{\ATLAS{} measurements at \SI{8}{\TeV}}

Next, we compare against the most recent diphoton \ATLAS{} measurement at \SI{8}{\TeV} \cite{Aaboud:2017vol}
which also considers $\phistar$ for this process, as defined in \cref{phiDef}, but using the 
photon instead of the lepton directions. For this study we impose the cuts listed in \cref{tab:diphoton8tev}.
We now choose a central hard scale of
 $\mu_h=\sqrt{m_{\gamma\gamma}^2 + (q_T^{\gamma\gamma})^2}$ and \NNPDFTZ{} as
 our default \PDF{} set. Given the minimum transverse momenta of the photons we ensured that 
 our  transition function
 fully switches to the fixed-order result beyond \SI{70}{\GeV}.
 
 \begin{table}
 	\centering
 	\caption{Fiducial cuts in diphoton production at $\sqrt{s}=\SI{8}{\TeV}$, see ref.~\cite{Aaboud:2017vol}.}
 	\vspace*{0.5em}
 	\bgroup
 	\setlength\tabcolsep{1em}
 	\def\arraystretch{1.5}%
 	\begin{tabular}{l|c}
 		\multirow{2}{*}{Photon cuts } &  $q_T^{\gamma,1}>\SI{40}{\GeV}$, $q_T^{\gamma,2} 	
 		>\SI{30}{\GeV}$ \\ &
 		$|\eta_\gamma|<2.37$,  $1.56 < |\eta_\gamma| < 1.37$\\
 		Photon separation & $\Delta R(\gamma,\gamma)>0.4$, \\
 		Smooth-cone photon isolation & $\etiso=\SI{11}{\GeV}$, $n=1$, $R=0.4$
 	\end{tabular}
 	\egroup
 	\label{tab:diphoton8tev}
 \end{table}

\begin{figure}
	\centering
	\includegraphics{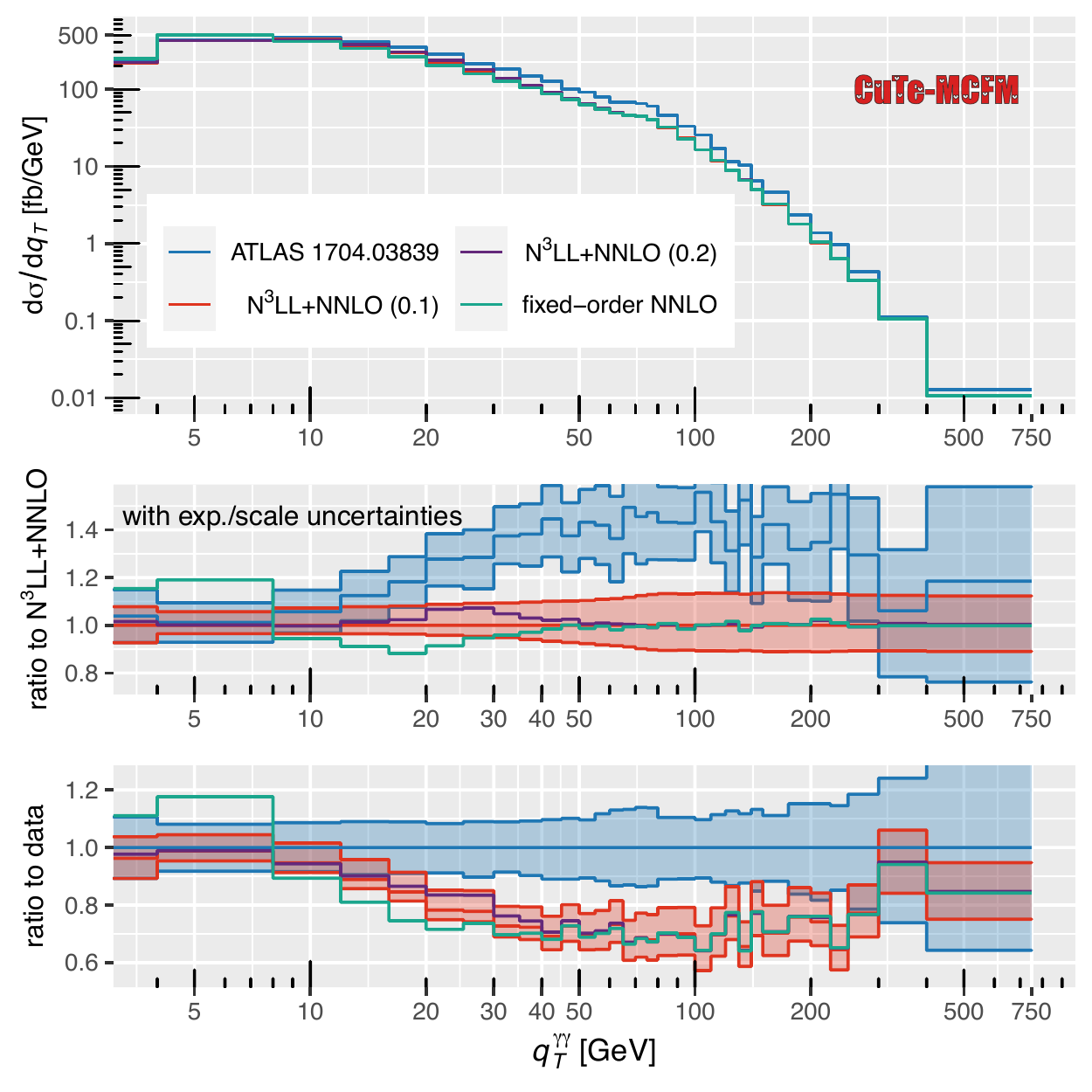}
	\caption{Comparison of the \ATLAS{} measurement of the diphoton transverse-momentum distribution at \SI{8}{\TeV} with predictions at \NTHREENNLO{} including uncertainties associated with scale variation. The labels $0.1$ and $0.2$ in the plots refer to the value of $\xmin$.}
	\label{fig:1704_03839_pt}
\end{figure}

\begin{figure}
    \centering
    \includegraphics{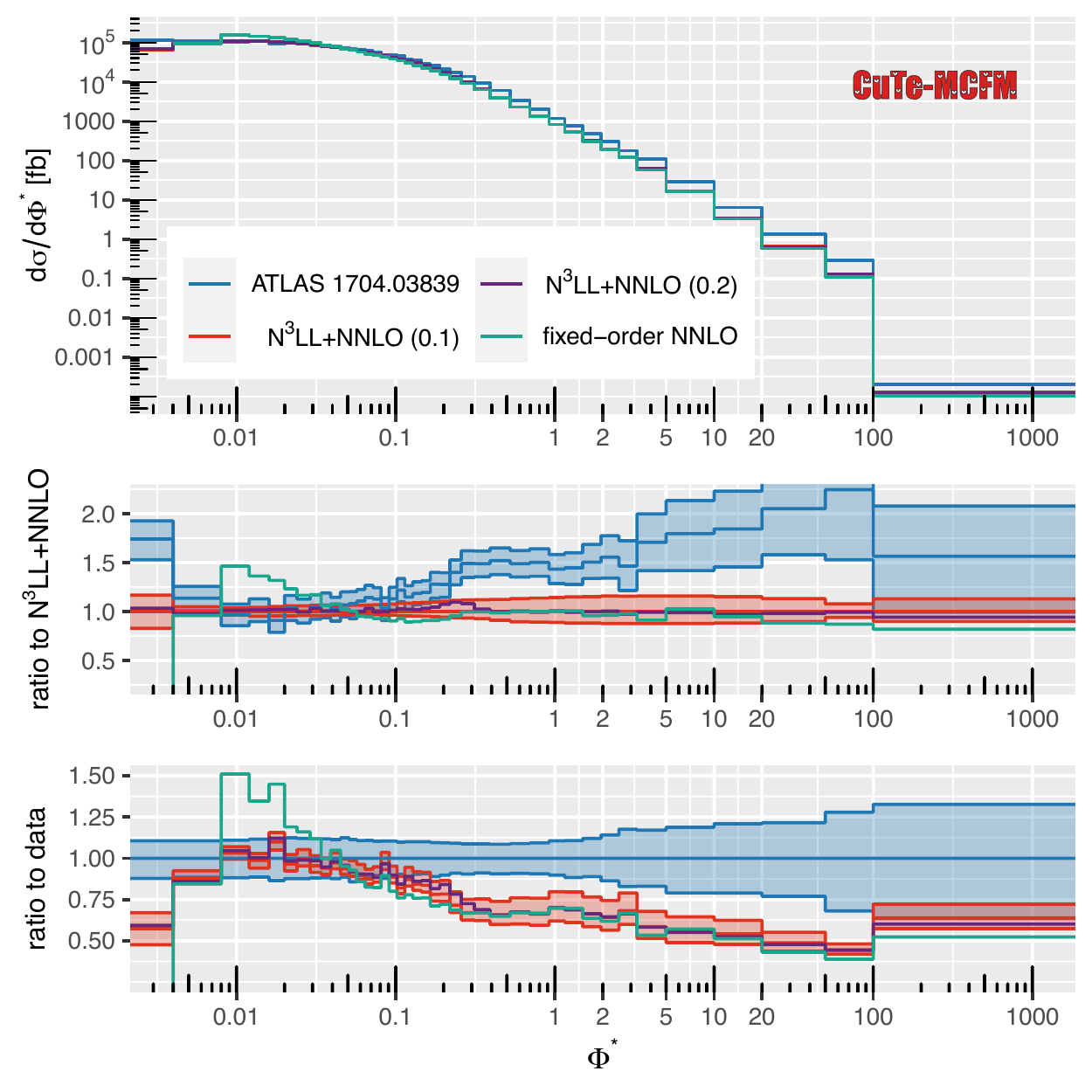}
    \caption{Comparison of the \ATLAS{} measurement of the diphoton $\phistar$ at \SI{8}{\TeV} with predictions at 
    \NTHREENNLO{} including 
    uncertainties associated with scale variation.}
    \label{fig:1704_03839_phistar}
\end{figure}

\begin{figure}
    \centering
    \includegraphics{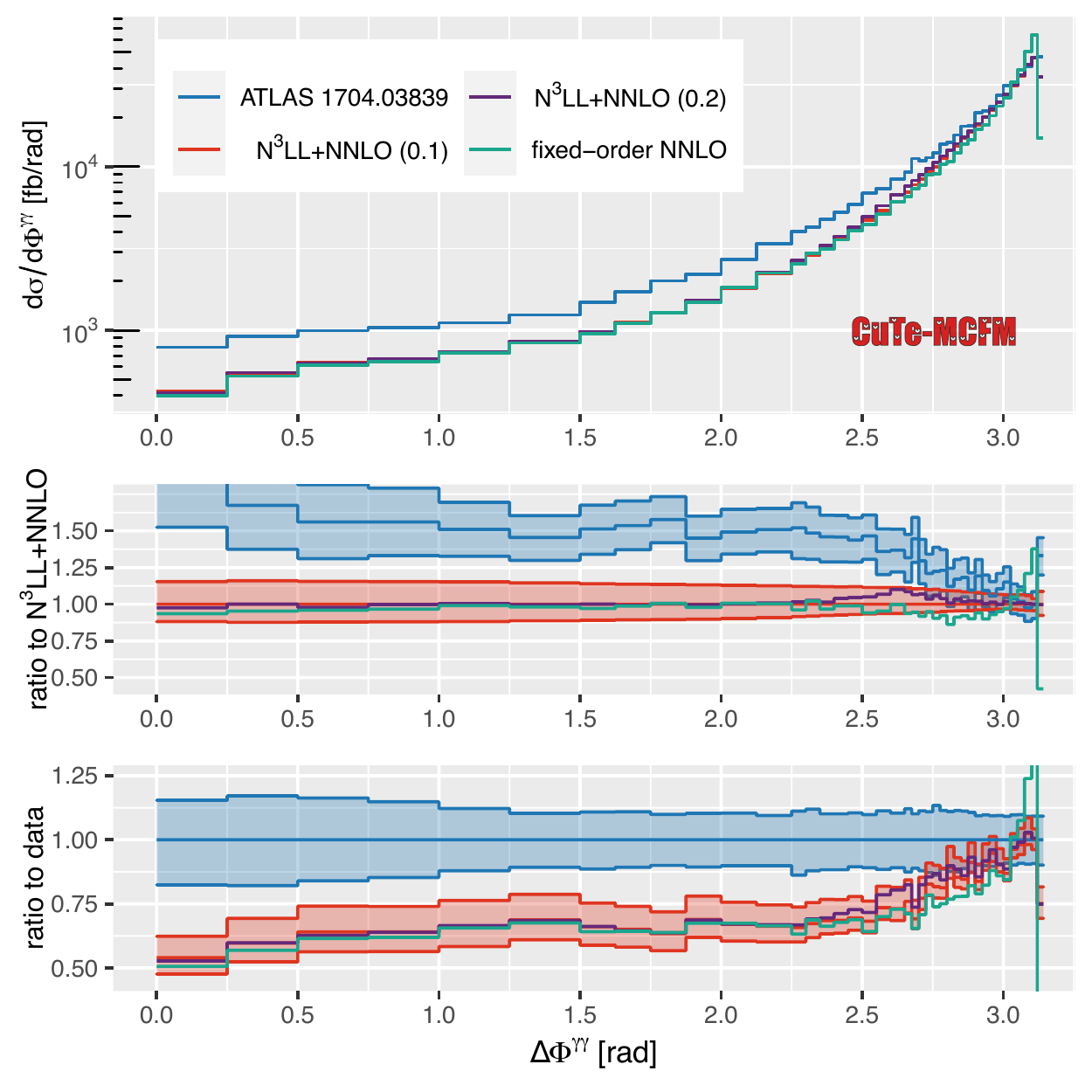}
    \caption{Comparison of the \ATLAS{} measurement of the diphoton $\delphi$ at \SI{8}{\TeV} with predictions at  \NTHREENNLO{} including uncertainties associated with scale variation. The labels $0.1$ and $0.2$ in the plots refer to the value of $\xmin$.}
    \label{fig:1704_03839_delphi}
\end{figure}

 Our results for the observables $q_T$,
$\phistar$ and $\delphi^{\gamma\gamma}$, the azimuthal-angle difference between the two photons, 
are shown in
in \cref{fig:1704_03839_pt,fig:1704_03839_phistar,fig:1704_03839_delphi}, respectively.
For each figure the first panel shows the absolute distribution, the second panel the
results in ratio to the matched result with $\xmin=0.1$, and the third panel the results in ratio
to the experimental data. The uncertainties associated with the matching to fixed order
can be read off from the second panel and are about $5-10\%$ in the region of \SIrange{15}{40}{\GeV}
in the $q_T$ distribution and an equivalent amount in the $\phistar$ distribution around $0.1-0.4$.
For $\delphi^{\gamma\gamma}$ they correspond to a region of $\sim2.4 - 2.8$.
 
The resummation of course stabilizes predictions for all observables in the region below 
\SI{10}{\GeV}.
Beyond that the resummed result improves the agreement with data up to $15\%$. For the $q_T$ 
distribution we find agreement of predictions with 
data within  uncertainties below \SI{20}{\GeV}. Unfortunately at large $q_T$ the fixed-order \NLO{} predictions are still insufficient to fully describe the data.
 We do not show distributions at the lower precision \NNLL{}+\NLO{}
since these \emph{significantly} underestimate the size of the cross sections. We therefore expect 
that
$\alpha_s^3$ corrections are necessary to achieve agreement with data in the region of large $q_T$. 
Similar conclusions hold for the $\phistar$ and $\delphi^{\gamma\gamma}$ distributions.

While our \SI{1}{\GeV} cutoff for the matching corrections (about 40\% relative to the matched 
result)
has a relatively small impact in the \SIrange{0}{4}{\GeV} bin, is is clearly visible in the first
and last bins of $\phistar$ and $\delphi^{\gamma\gamma}$, respectively. We decided to keep these
bins in our plot to demonstrate this effect which is unavoidable due to the large matching corrections, unless these can be calculated reliably also at small $q_T$. 

Lastly, we show \PDF{} uncertainties for the $q_T$, $\phistar$ and $\delphi$ distributions in 
\cref{fig:1704_03839_pdfs} in the appendix on \cpageref{fig:1704_03839_pdfs}. Uncertainties 
are generally at the few percent level for each \PDF{} set,
but when taking into account the span of multiple \PDF{} sets like {\abbrev CT18} and {\abbrev NNPDF3.1}
they can reach up to 10\%. The {\abbrev ABMP16} set is undefined below scales of \SI{4.47}{\GeV}
and the \LHAPDF{} grid-based prediction therefore breaks down. In principle, as mentioned before, one could abandon
the grid-based approach in this region and {\abbrev DGLAP} evolve further downwards or just
set \SI{4.47}{\GeV} as a minimum scale value.


\clearpage
\subsection{Fiducial $Z\gamma$ production}
\label{sec:zgamma}
We now present results for fiducial $Z\gamma$ production in the decay channel $Z\to e^+e^-$
in comparison with the  \SI{13}{\TeV} \ATLAS{} measurement \cite{Aad:2019gpq}.
The fixed-order \NNLO{} result in \MCFM{} is based on ref.~\cite{Campbell:2017aul}, but \NNLO{} results
have also been been computed in refs.~\cite{Grazzini:2013bna,Grazzini:2015nwa}. Results for 
$q_T$ resummation of $Z\gamma$ production at the same accuracy considered here have very recently been 
presented \cite{Wiesemann:2020gbm}. They are based on a different resummation framework 
\cite{Bizon:2017rah} implemented in the {\abbrev RadISH} code.

We fully reproduce the \NLO{} and \NNLO{} fixed-order fiducial
cross sections calculated in the \ATLAS{} study \cite{Aad:2019gpq} after applying the parton-to-particle factor 
$C_\text{theory} = 0.934\pm0.005$.\footnote{While ref.~\cite{Aad:2019gpq} gives a factor of 
$0.934$ when electroweak 
	corrections are included in the partonic prediction and $0.915$ when they are not, we find that 
	the former factor 
	reproduces the fixed-order \NLO{} and \NNLO{} results in ref.~\cite{Aad:2019gpq} table 6. It 
	seems possible that these 
	factors have been mixed-up.}
We furthermore fully reproduce the presented fixed-order differential distributions at \NNLO{} when 
including
the differential parton-to-particle factors. Since we consider the use of parton-to-particle factors problematic, we do not directly include them in our results. 

\begin{table}[!h]
	\centering
	\caption{Experimental cuts in $Z\gamma$ production with $Z\to e^+e^-$ decay at a center of mass 
		energy $\sqrt{s}=\SI{13}{\TeV}$.}
	\vspace*{0.5em}
	\bgroup
	\setlength\tabcolsep{1em}
	\def\arraystretch{1.5}%
	\begin{tabular}{l|c}
		Leptons & $q_T^{l} > \SI{30}{\GeV}, \SI{25}{\GeV}$, $\abs{\eta^l} < 2.47$ \\
		Photon & $q_T^\gamma > \SI{30}{\GeV}$ and $\abs{\eta^\gamma}<2.37$ \\
		Smooth-cone isolation & $\epsilon_\gamma=0.1, R=0.1, n=2$ \\
		Separation & $m_{l^+l^-}>\SI{40}{\GeV}$, $m_{l^+l^-}+m_{l^+l^-\gamma}>\SI{182}{\GeV}$, $\Delta R(\gamma,l) > 0.4$
	\end{tabular}
	\egroup
	\label{tab:cuts-13TeV-Zepem}
\end{table}

For $Z\gamma$ production two different contributions arise: an $s$-channel mode,
where the photon is radiated from the charged leptons in the $Z$ decay, and a $t$-channel mode, where the
photon is radiated from the initial state.
The photon isolation enters differently in these channels. In the $s$-channel, the only isolation-cone power corrections are associated with gluon emission into the cone and suppressed by $R^2$,
while the fragmentation part is absent. On the other hand, the $t$-channel has fragmentation
contributions which are not suppressed by $R^2$. While the linear power corrections of the
fragmentation contribution are asymptotically smaller than corrections from gluon emission
(for isolation parameter $n>1$), they could still predominate for any reasonably small value
of $q_T\to0$ when $R\ll 1$, see discussion on \cpageref{eq:frixione} and following.
Since neither of these power corrections are included in our resummation, one expects larger matching corrections when the cuts allow for significant $t$-channel contributions.

\paragraph{Impact of $s$-channel vs. $t$-channel contributions} The fiducial cuts chosen in the \ATLAS{} study, see \cref{tab:cuts-13TeV-Zepem}, almost entirely
suppress the $s$-channel contribution to enhance the $Z$ peak in $m_{l^+l^-}$ of the signal. This
is primarily achieved by applying a selection cut $m_{l^+l^-}+m_{l^+l^-\gamma}>\SI{182}{\GeV}$,
which can be nicely seen in figure~2 of ref.~\cite{Aad:2019gpq}.
For those cuts, the matching corrections are large, as 
elaborated later on. To numerically test the impact of photon-isolation power corrections from $s$-channel and $t$-channel contributions we consider benchmark cuts as defined in \cref{tab:Zepem-benchmarkcuts}. For these cuts we enhance the $s$-channel contribution by reversing the separation cut $m_{l^+l^-}+m_{l^+l^-\gamma}<\SI{182}{\GeV}$ and relaxing the photon and lepton $q_T$ 
cuts.

\begin{figure}[!t]
	\centering
		\includegraphics{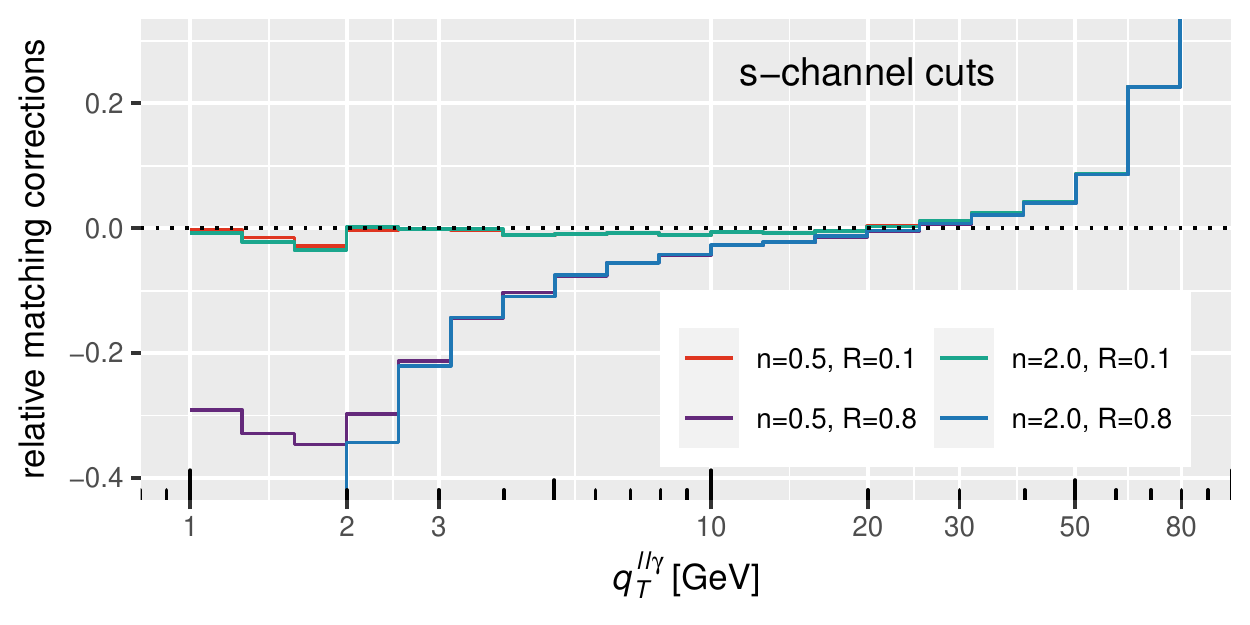}
		\includegraphics{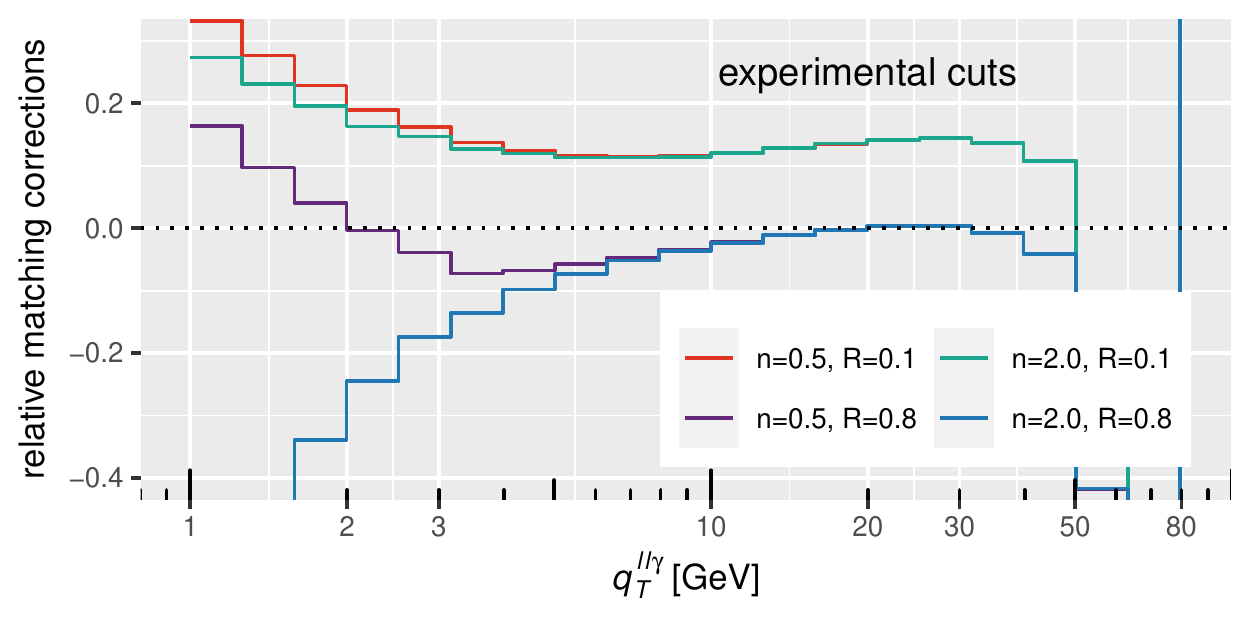}
	\caption{Matching corrections for $Z\gamma$ production at \NNLL{}, relative to the naively matched results for different choices of smooth-cone photon isolation parameters $n$ and $R_0$.  The upper panel shows the results with $s$-channel benchmark cuts, see \cref{tab:Zepem-benchmarkcuts}. The lower panel  is relevant for the experimental cuts that enhance the $t$-channel contribution, see \cref{tab:cuts-13TeV-Zepem}.}
	\label{fig:diphoton_matchcorr_bench}
\end{figure}

\begin{table}[!t]
	\centering
	\caption{Benchmark cuts enhancing the $s$-channel contribution in $Z\gamma$ production with $Z\to e^+e^-$ decay at a center of mass energy 
		$\sqrt{s}=\SI{13}{\TeV}$.}
	\vspace*{0.5em}
	\bgroup
	\setlength\tabcolsep{1em}
	\def\arraystretch{1.5}%
	\begin{tabular}{l|c}
		Leptons & $q_T^{l} > \SI{25}{\GeV}, \SI{20}{\GeV}$, $\abs{\eta^l} < 2.47$ \\
		Photon & $q_T^\gamma > \SI{20}{\GeV}$ and $\abs{\eta^\gamma}<2.37$ \\
		Smooth-cone isolation & $\epsilon_\gamma=0.1, R \text{ and } n \text{ set individually }$ \\
		Separation & $m_{l^+l^-}>\SI{40}{\GeV}$, $m_{l^+l^-}+m_{l^+l^-\gamma}<\SI{182}{\GeV}$, $\Delta R(\gamma,l) > 0.1$
	\end{tabular}
	\egroup
	\label{tab:Zepem-benchmarkcuts}
\end{table}

In \cref{fig:diphoton_matchcorr_bench} we consider the \NNLL{} matching corrections relative to the naively matched result with $s$-channel benchmark cuts and experimental cuts for different choices of photon
isolation parameters $R$ and $n$. For these benchmark cuts one observes exactly the behavior predicted for the gluonic corrections in \cref{eq:gluonPower}, namely negative effects scaling with $R^2$.
This is different from diphoton production, where no $s$-channel mode exists and one cannot easily separate the power corrections associated with soft gluon emission from the fragmentation contribution.  

The experimental cuts almost exclusively select the $t$-channel contribution. The nature of the power corrections changes and they become qualitatively similar to what we observed for diphoton production, except that they are smaller in size, because we only have a single photon in the final state. Even for $R=0.1$ the matching corrections are relatively large and positive around $10-20\%$ and fully dominate over the Sudakov-suppressed resummed result towards $q_T\to0$, since they scale linearly in $q_T$ and therefore approach a finite constant in the $q_T$ distribution. The matching corrections accidentally 
decrease for larger $R$ since the negative gluonic photon-isolation power corrections  increase like $R^2$ and cancel against the fragmentation contributions. In all cases, the $n$-dependence only becomes relevant below
at small $q_T<\etiso$. For \NNNLL{} the same conclusions hold qualitatively and quantitatively.

\subsubsection{\ATLAS{} measurements at \SI{13}{\TeV}}

Having demonstrated that matching corrections are at the percent to sub-percent level for the
$s$-channel benchmark cuts and even moderate $q_T\lesssim \SI{40}{\GeV}$, we now compare with the \SI{13}{\TeV} Z$\gamma$ measurement by \ATLAS{}
\cite{Aad:2019gpq} with fiducial cuts in \cref{tab:cuts-13TeV-Zepem}. We use a central hard scale
of $\mu_h = \sqrt{Q^2+q_T^2}$ and the \CTFOUR{} \PDF{} set.

\begin{figure}[!t]
	\centering
	\includegraphics{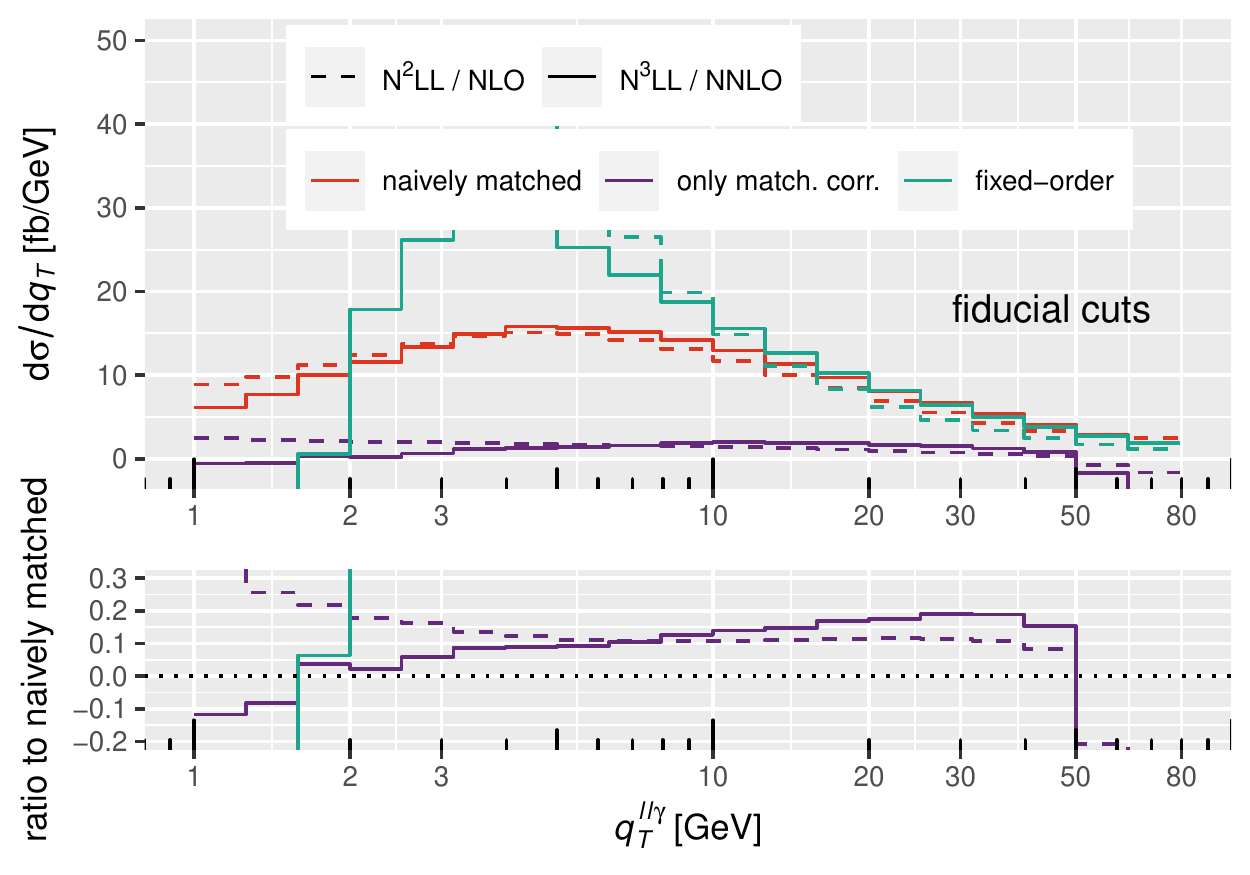}
	\caption{$Z(\to e^+e^-) \gamma$ transverse-momentum distribution with experimental fiducial cuts as in \cref{tab:cuts-13TeV-Zepem}. Shown are results with naive matching, the matching corrections by themselves, and the fixed-order predictions, each at order $\alpha_s$ and $\alpha_s^2$, respectively. The bottom panel shows the ratio
		of the matching corrections relative to the naively matched result for each order in $\alpha_s$.}
	\label{fig:zga_naivematch}
\end{figure}

To see the effect of the experimental fiducial cuts at \NNNLL{} on the size of the matching corrections, we show
the naively matched result and matching corrections in \cref{fig:zga_naivematch}.
With a strong suppression of the $s$-channel
contribution, the matching corrections at \NNNLL{} are at the order of $10-20\%$. Fortunately
the matching corrections are quite a bit smaller at \NNNLL{} than at \NNLL{} for $q_T$ values in the few-\si{\GeV} range.

To mitigate numerical issues due to a root in the matching corrections around \SI{2}{\GeV}
and required cancellations of more than 0.1 per-mille, we save computational resources
and cut off the matching corrections for the remaining $Z\gamma$ results below \SI{2}{\GeV} and, as a consequence, neglect matching corrections of $\lesssim5\%$ below this value. The numerical
issues can be seen in \cref{fig:zga_naivematch} for $q_T$ less than $\SI{2}{\GeV}$ as a discontinuity, or rather larger numerical noise.

To account for the safety cutoff on the matching corrections at \SI{2}{\GeV}, one should assign a larger uncertainty for the first few bins in the $q_T$ distribution. To estimate the effects for other variables, one can vary the cutoff value over the range of a few GeV. We have done so for the $\delphi$ distribution by increasing the cutoff from \SI{2}{\GeV} to  \SI{5}{\GeV}. The resulting changes are small compared to the other uncertainties which affect this distribution.

\begin{figure}[!t]
	\centering
	\includegraphics{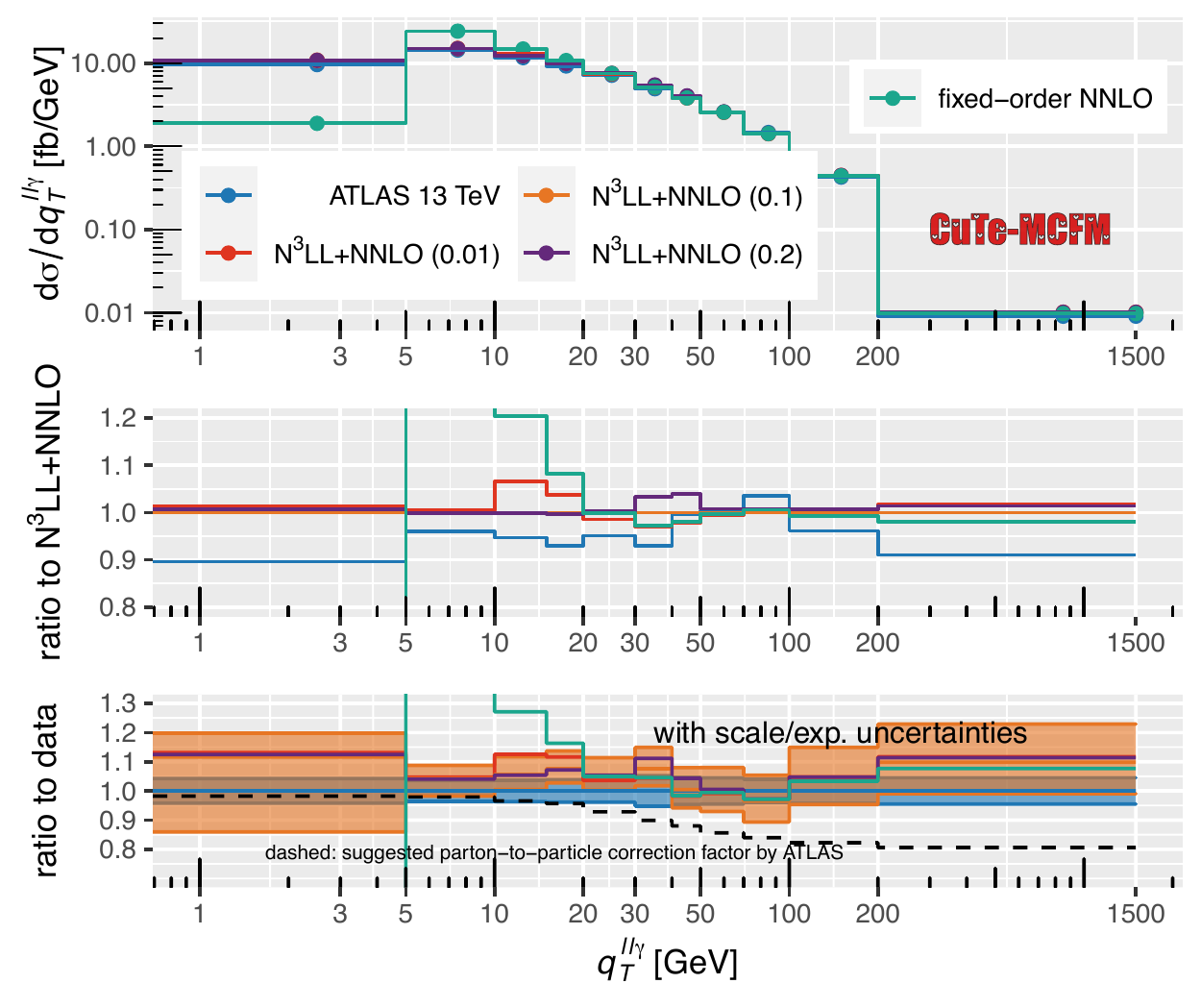}
	\caption{Comparison of \NTHREENNLO{} predictions with the \ATLAS{} measurement of 
		$q_T^{ll\gamma}$ at \SI{13}{\TeV}. 
		Note that this plot includes the first three bins individually that are only shown combined for the \NNLO{} 
		fixed-order comparison in ref.~\cite{Aad:2019gpq}. The labels $0.01$, $0.1$ and $0.2$ in the plots refer to the value of $\xmin$.}
	\label{fig:cms_zgamma_ptllgamma}
\end{figure}

\paragraph{$q_T$ distribution}
In \Cref{fig:cms_zgamma_ptllgamma} we show our predictions for the $q_T^{ll\gamma}$ 
distribution
in comparison with the experimental data. The top panel shows the absolute distributions for
data, resummed predictions matched to \NNLO{} fixed-order results using transition function arguments $\xmin=0.01,0.1$ and 
$\xmin=0.2$, and the fixed-order \NNLO{} prediction alone. 
The middle panel shows these distributions relative to the matched result with $\xmin=0.1$ to
demonstrate matching effects to fixed order from the transition function. The bottom panel shows
the results relative to the experimental data and includes uncertainties from scale-variation for the matched prediction ($\xmin=0.1$) and experimental uncertainties for the experimental data.

The reason for choosing relatively small $\xmin$ for the transition function is that we want to minimize matching effects beyond $\sim\SI{60}{\GeV}$, see \cref{fig:zga_naivematch}.
The choice of $\xmin=0.01$ performs the matching in the region of $\sim\SI{10}{\GeV}$, where
the validity of the fixed-order result is questionable and is entering the divergent regime. 
 With $\xmin=0.1$ the transition region moves to between
\SI{30}{\GeV} and \SI{50}{\GeV}, while $\xmin=0.2$ stretches further into the region with large matching corrections. 
Taking $\xmin=0.1$ as a central choice, we estimate that the overall matching effects are about five percent around \SIrange{30}{50}{\GeV}.

Despite the large matching corrections, the matched \NTHREENNLO{} predictions show  good
agreement with data within scale uncertainties. The results using the resummation framework {\abbrev RadISH} obtained in ref.~\cite{Wiesemann:2020gbm} have similarly sized scale uncertainties and show a similar agreement with data. But unlike here, where we advocate to switch off the matching corrections with $\xmin=0.1$ in the region of \SIrange{30}{50}{\GeV}, they state that resummation and matching are crucial also in the region $\SI{40}{\GeV} \lesssim q_T^{ll\gamma} \lesssim \SI{200}{\GeV}$. While this choice could potentially have a positive effect on the agreement with data,
the matching corrections and resummation spoil other kinematics, as we will see in the case of the the $\delphi$ observable. It is possible that the different matching procedure alleviates such issues, but the $\delphi$ distribution has not been considered in ref.~\cite{Wiesemann:2020gbm}. 

In the bottom panel we have additionally indicated
the suggested parton-to-particle factor in the \ATLAS{} study that, when applied, would decrease agreement.
Note that in addition to the large scale uncertainties\footnote{The uncertainty is
	asymmetric in the first bin because we do not vary the resummation scale into the 
	non-perturbative regime.} of $10-20\%$ also \PDF{} 
uncertainties and unknown non-perturbative effects contribute, as well as a small effect from 
neglected matching corrections. 

\begin{figure}[!h]
	\centering
	\includegraphics{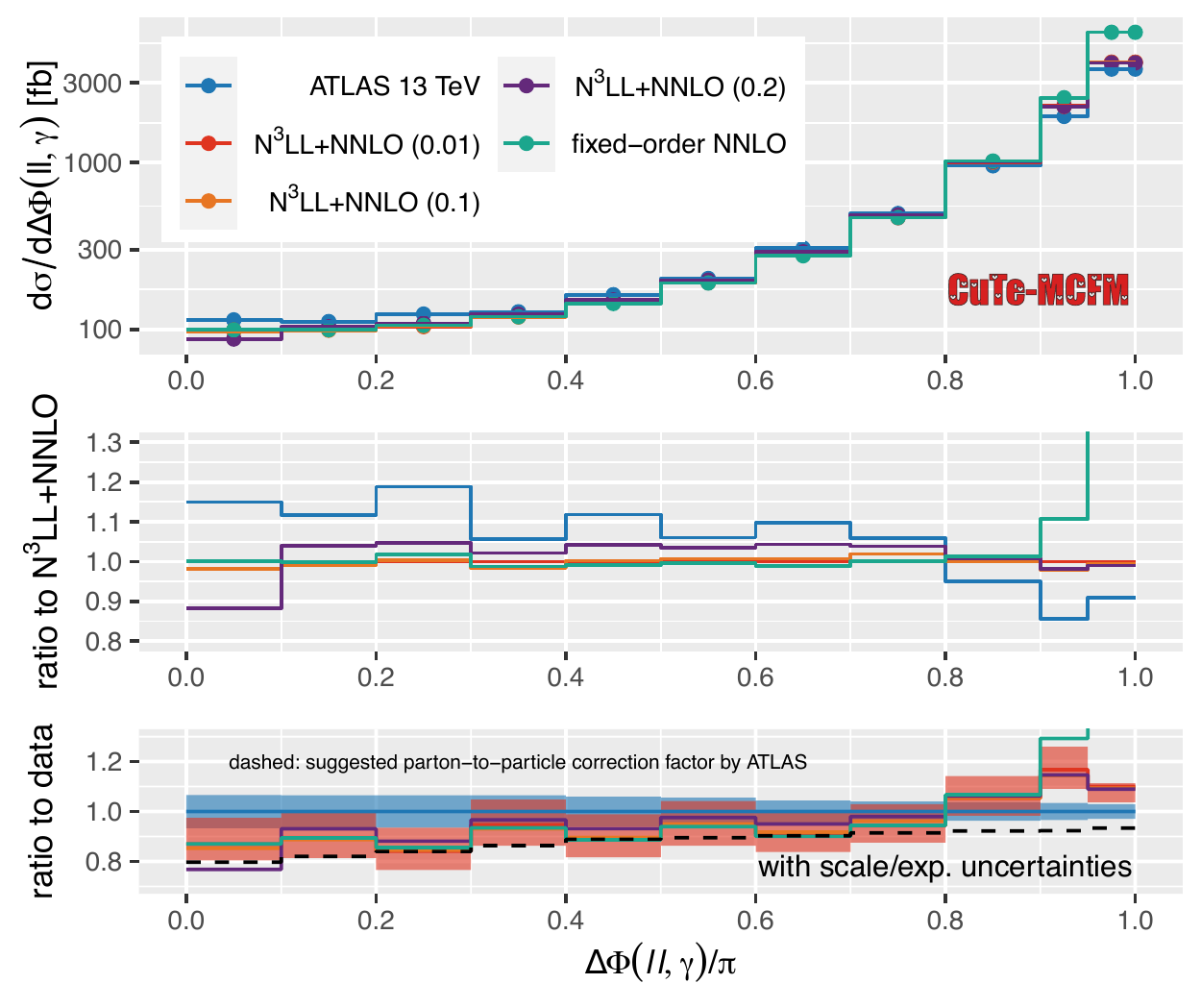}
	\caption{Comparison of \NTHREENNLO{} predictions with \ATLAS{} measurement of 
		$\delphi(ll,\gamma)$ at \SI{13}{\TeV}. Note that this plot includes the last two bins 
		individually that are
		only shown combined for the \NNLO{} fixed-order comparison in ref.~\cite{Aad:2019gpq}. The labels $0.01$, $0.1$ and $0.2$ in the plots refer to the value of $\xmin$.}
	\label{fig:cms_zgamma_delphi}
\end{figure}

\paragraph{$\delphi$ distribution}
In \cref{fig:cms_zgamma_delphi} we compare our predictions with \ATLAS{} data for the 
$\delphi(ll,\gamma)$ 
observable. The first panel shows the absolute distributions, the middle panel the ratio to our  matched prediction with $\xmin=0.01$, and the bottom panel the ratio to data with scale uncertainties for
our predictions and experimental uncertainties for the measurement, respectively.

With this observable it becomes clear that with $\xmin=0.2$ we are tapping into a region where
the resummation breaks down. This is most visible in the first bin, where the matched results
should agree with the fixed-order prediction. Instead one sees a 10\% difference. The first bin
corresponds to azimuthally aligned $Z$ and $\gamma$, which due to the fiducial cuts have to recoil
against at least $\sim\SI{60}{\GeV}$. This is exactly the region where the resummation breaks down
and which a transition function with $\xmin=0.2$ includes. While this is effect is most prominent
for the first bin, also the larger azimuthal angles are affected at the percent-level. The effect
worsens quickly with an even later transition.

The transitions with $\xmin=0.1$ and $\xmin=0.01$ ensure that the region where the resummation breaks down
is excluded with negligible remaining effects. Therefore fixed-order and matched results agree
up to $\delphi=0.9\cdot\pi$. The benefits of the $q_T$ resummation come into play in the last two 
bins,
which are stabilized compared to the differentially diverging fixed-order prediction. Our prediction
does not directly agree with the measurement at large $\delphi$, but \PDF{} uncertainties and
non-perturbative effects are not yet included. In this region, the differential parton-to-particle
 correction used by \ATLAS{} might be used as an estimate of 
 non-perturbative effects. Including it  would lead to better agreement with the data, at least in this region.
 
 We present \PDF{} uncertainties for the $q_T$ and $\delphi$ 
 distributions in \cref{fig:zgamma_pdferrs} in the appendix on \cpageref{fig:zgamma_pdferrs}.

\clearpage
\section{Conclusions}
\label{sec:conclusions}

The transverse momentum of electroweak bosons is one of the cornerstone observables at the
\LHC{} and future colliders.
It constitutes a precision probe of the Standard Model and is therefore
one of the key observables to find or constrain new physics.
The experimental precision reached today ranges from the per-mille level for $Z$ production,
to the percent level for many diboson processes.  While for Higgs production the uncertainties are currently still large, also these will diminish with more luminosity in the future. Such high-precision measurements are a huge challenge for theory that we help addressing with this study.

We presented a framework for $q_T$ resummation at \NTHREENNLO{} for color-singlet processes based 
on a factorization theorem in \SCET{}. Our implementation \CuTeMCFM{} provides precise predictions
 with uniform accuracy at small and large $q_T$ through a systematic power counting in $\alpha_s$ and large logarithms. Predictions can be calculated for the boson processes $W^\pm, Z, H$, as well as for the diboson processes $\gamma\gamma$, $Z\gamma$, $ZH$ 
and $W^\pm H$. These resummed and matched predictions are fully differential in the Born kinematics 
including decays and therefore also provide predictions for other observables benefiting
from resummation at small $q_T$. Uncertainties from the perturbative \QCD{} truncation,  resummation, and \PDF{}s can be evaluated efficiently using the possibility to pre-generate beam-function grids.

We first benchmarked our predictions for inclusive $Z$ and $H$ production with the code \CuTe{} and then directly compared with fiducial experimental
data for $Z$, $W$, $\gamma\gamma$ and $Z\gamma$ production. For $\gamma\gamma$ our results improve 
upon previous predictions at a lower logarithmic accuracy, and for $Z\gamma$ we presented novel 
results, previously only available at fixed-order accuracy. For $Z$ production, we observe excellent agreement at the few-percent level with the experimental measurements. The agreement is also quite good for $W$ and $Z\gamma$ production, while there are some tensions for diphoton production, which would likely ease after including  $\alpha_s^3$ fixed-order corrections at large $q_T$. For Higgs production, where experimental uncertainties in the $q_T$ distribution are still large, we have presented results in the $H\to\gamma\gamma$ decay channel with realistic fiducial cuts as a first application. Furthermore, also the
processes $W^\pm H$ and $ZH$ can be calculated at \NTHREENNLO{} with our code, which could become useful in the high-luminosity phase of the \LHC{}.

All of our results are shown with estimates of higher-order effects through scale variations of the hard scale, renormalization scale, factorization scale and resummation scale. We furthermore presented and 
discussed \PDF{} uncertainties. We transition to fixed-order predictions at large $q_T$
through a simple sigmoid-type function which can easily be varied. Through this variation we estimated the uncertainty on the matching to fixed-order predictions. 

We find that matching corrections are suppressed by $q_T^2$ for processes without photon isolation, if recoil effects are taken into account. For photon processes we showed that the necessary isolation requirements enhance the matching corrections and studied the form of the leading-order power corrections analytically and numerically. In the case of soft gluons radiated into the isolation cone, previous results are available that predict a power dependence on the smooth-cone isolation parameter $n$, which we confirm. For the power
corrections associated with quark fragmentation, we find that they are always first order in $q_T$ and are not suppressed by the size of the isolation cone. Therefore, the resulting presence of large power corrections can make it difficult to find a window in which the fixed-order and resummed predictions are both valid and can be matched reliably.

While our implementation is part of \MCFM{}, the resummation code is highly modular and could easily be decoupled and interfaced to other codes supplying the fixed-order process-dependent pieces, i.e. the hard function for the resummation itself, and the process with additional radiation recoiling at large $q_T$. Our code \CuTeMCFM{} will be made publicly available shortly.

Using our existing framework one could, with limited effort, match with $\alpha_s^3$ predictions at large $q_T$
\cite{Boughezal:2015dva,Campbell:2019gmd,Boughezal:2015ded,Campbell:2016lzl,Campbell:2017dqk}
to provide predictions at N$^3$LL$^\prime$+N$^3$LO. To do so, one will need to implement the recently computed three-loop beam functions \cite{Ebert:2020yqt}. However, apart from the case of Higgs and Drell-Yan production this would mean neglecting the $\alpha_s^3$ hard function.  We could furthermore easily include non-perturbative effects either through a form-factor modification in the resummation or through swapping out the beam functions for transverse-momentum dependent \PDF{}s. With this, even fits for these generalized \PDF{}s can be envisioned as long as precise control over matching corrections is maintained when they
are sizeable at small $q_T$. The inclusion of electroweak effects in the resummation
and in fixed-order results is another issue that should be tackled together with other higher-order 
effects. For Higgs production the inclusion of heavy-quark mass effects is another interesting avenue to pursue.

\paragraph{Acknowledgments.}

We would like to thank John Campbell for useful suggestions (electroweak corrections) and Markus Ebert for providing additional details on the results in ref.~\cite{Ebert:2019zkb}. The research of T.B.\ is supported by the Swiss National Science Foundation (SNF) under grant 200020\_182038. This work was supported by the U.S.\ Department of Energy under award
No.\ DE-SC0008347.  This document was prepared using the resources of
the Fermi National Accelerator Laboratory (Fermilab), a
U.S. Department of Energy, Office of Science, HEP User
Facility. Fermilab is managed by Fermi Research Alliance, LLC (FRA),
acting under Contract No.\ DE-AC02-07CH11359.
\appendix
\section{\PDF{} uncertainties }

In this appendix we present plots with \PDF{} uncertainties for diphoton and for $Z\gamma$ 
production. We evaluate these for the following \NNLO{} \PDF{} sets with 
fixed value of $\alpha_s(m_Z)=0.118$: {\abbrev ABMP16}
\cite{ABMP16}, {\abbrev CT14} \cite{CT14}, {\abbrev CT18} \cite{CT18}, {\abbrev MMHT2014} \cite{MMHT2014}, {\abbrev NNPDF30} \cite{NNPDF30} 
and {\abbrev NNPDF31} \cite{NNPDF31} interfaced to \LHAPDF{} \cite{Buckley:2014ana}. Overall the \PDF{} 
uncertainties are broadly at the few percent level, but can become larger when taking into account 
multiple sets. The individual central values are mostly compatible within mutual uncertainties.
The {\abbrev ABMP16} set is not defined below
scales of $\mu=\SI{4.47}{\GeV}$ and breaks down with the default grid-based interpolation in 
\LHAPDF{}. In principle one could switch to {\abbrev DGLAP} evolution to circumvent this or use a larger 
minimum scale of $\SI{4.47}{\GeV}$ in \CuTeMCFM{}. 

\begin{figure}
	\centering
	\includegraphics{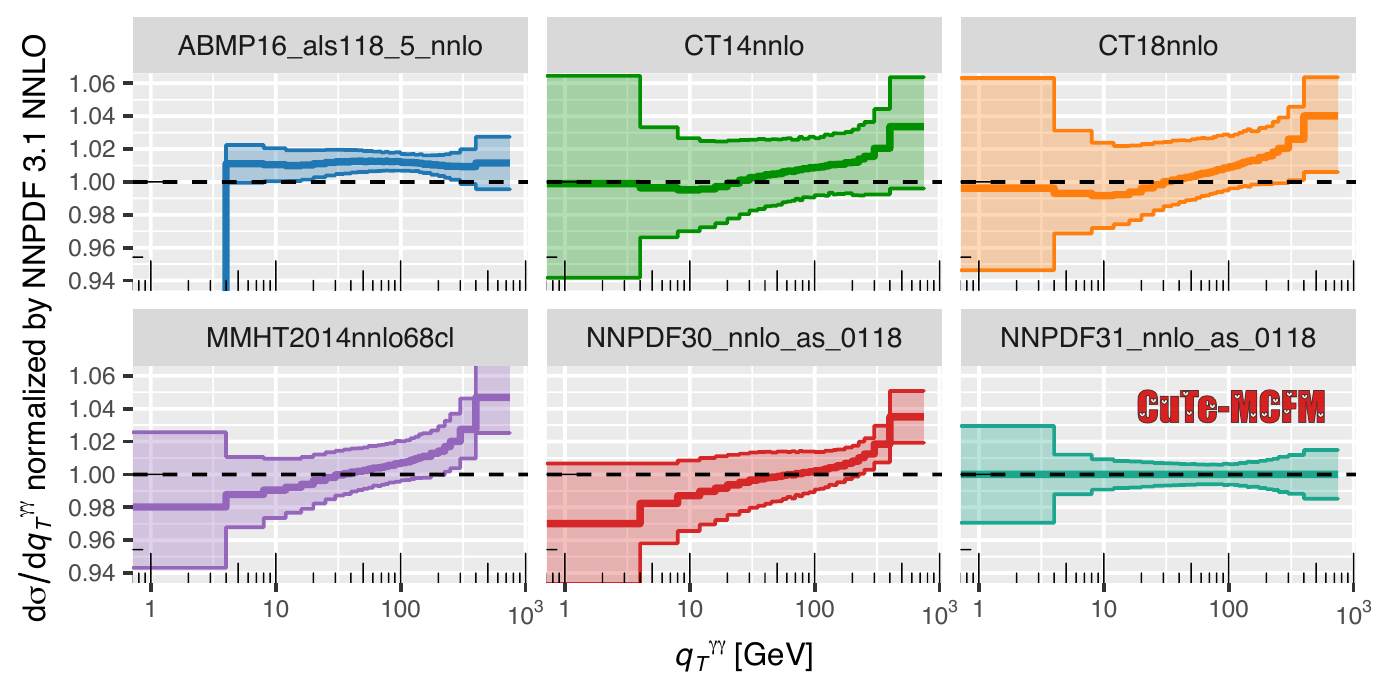}
	\includegraphics{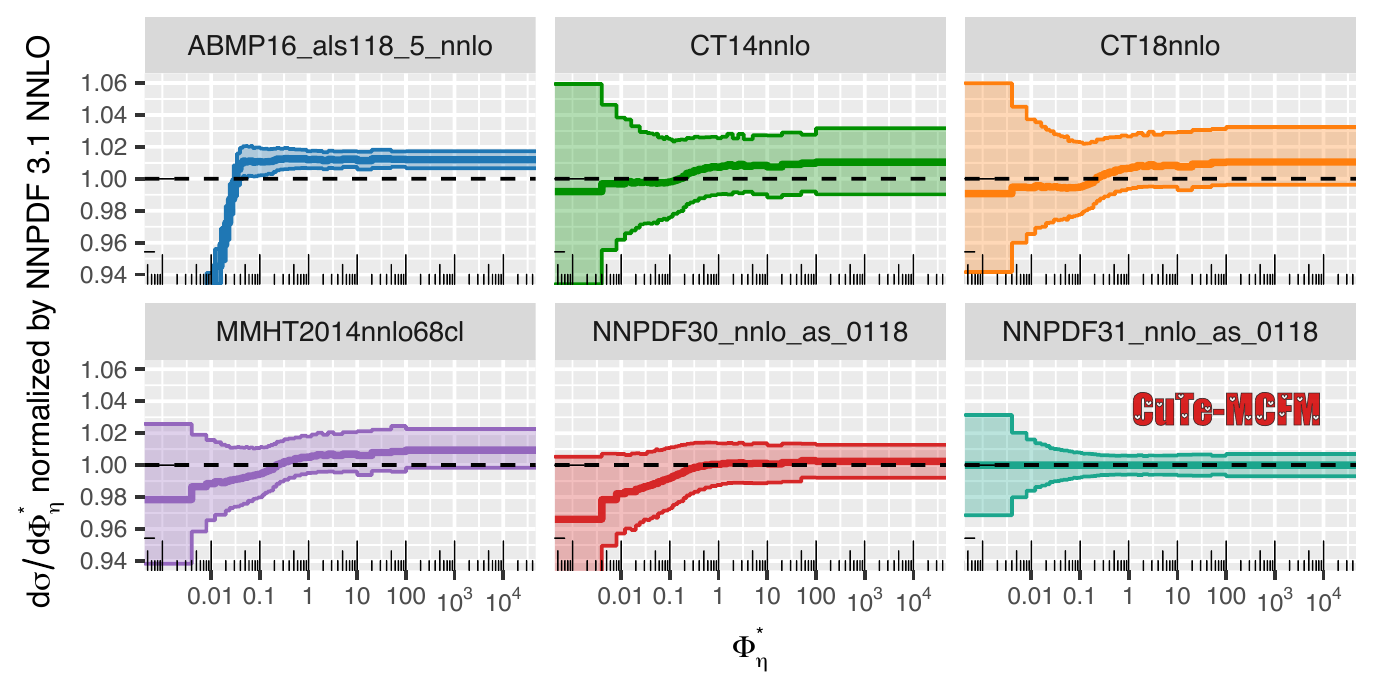}
	\includegraphics{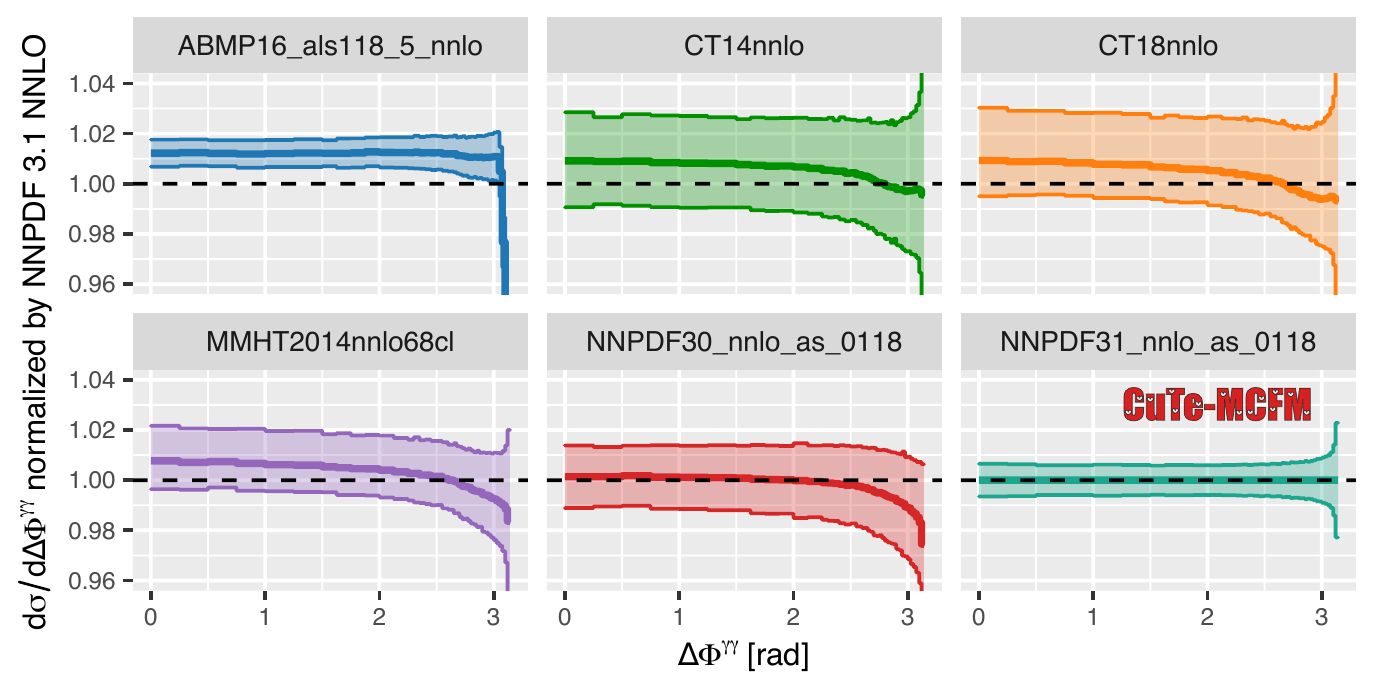}
	\caption{\PDF{} uncertainties for the $q_T$, $\phistar$ and $\delphi$ distributions
		in \cref{fig:1704_03839_pt,fig:1704_03839_phistar,fig:1704_03839_delphi}. The {\abbrev 
		ABMP16} set is not 
		defined below $\mu=\SI{4.47}{\GeV}$, which can be seen as breakdown in the prediction. }
	\label{fig:1704_03839_pdfs}
\end{figure}

\begin{figure}
	\centering
	\begin{subfigure}[b]{0.45\textwidth}
		\includegraphics[]{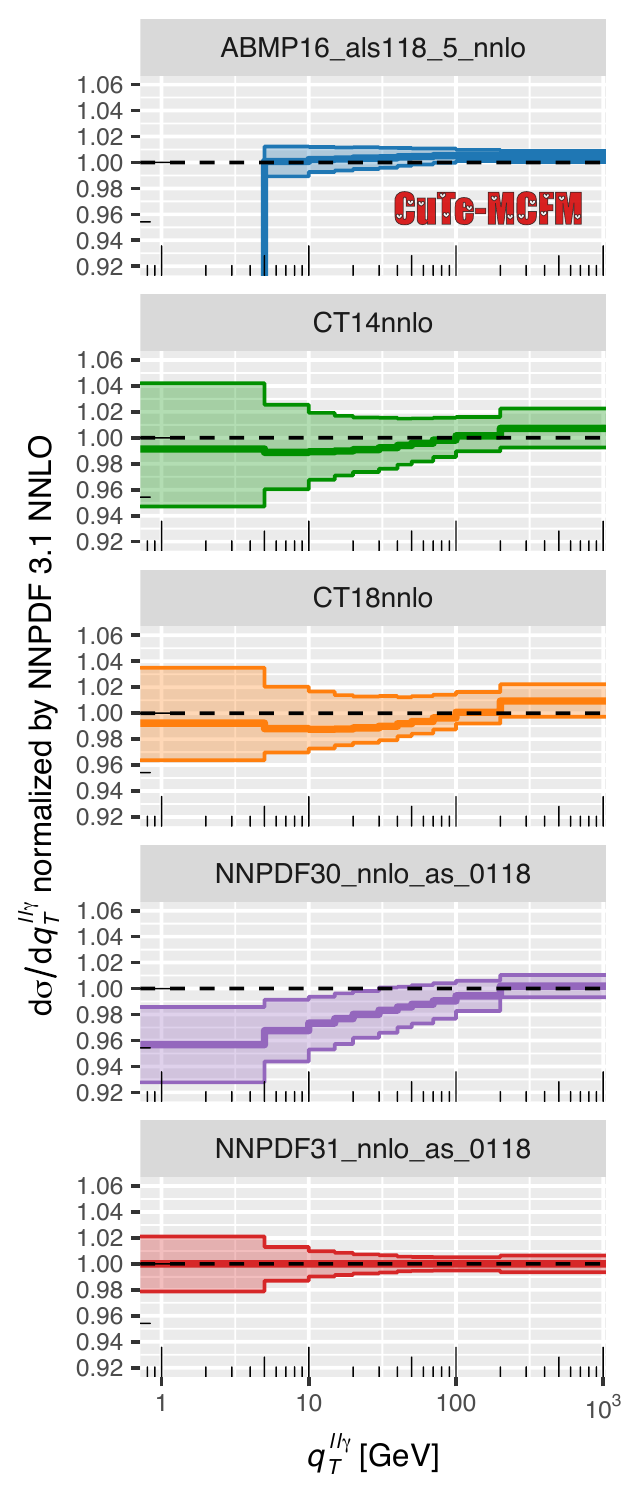}
		\caption{For $q_T$ distribution}
	\end{subfigure}
	\begin{subfigure}[b]{0.45\textwidth}
		\includegraphics[]{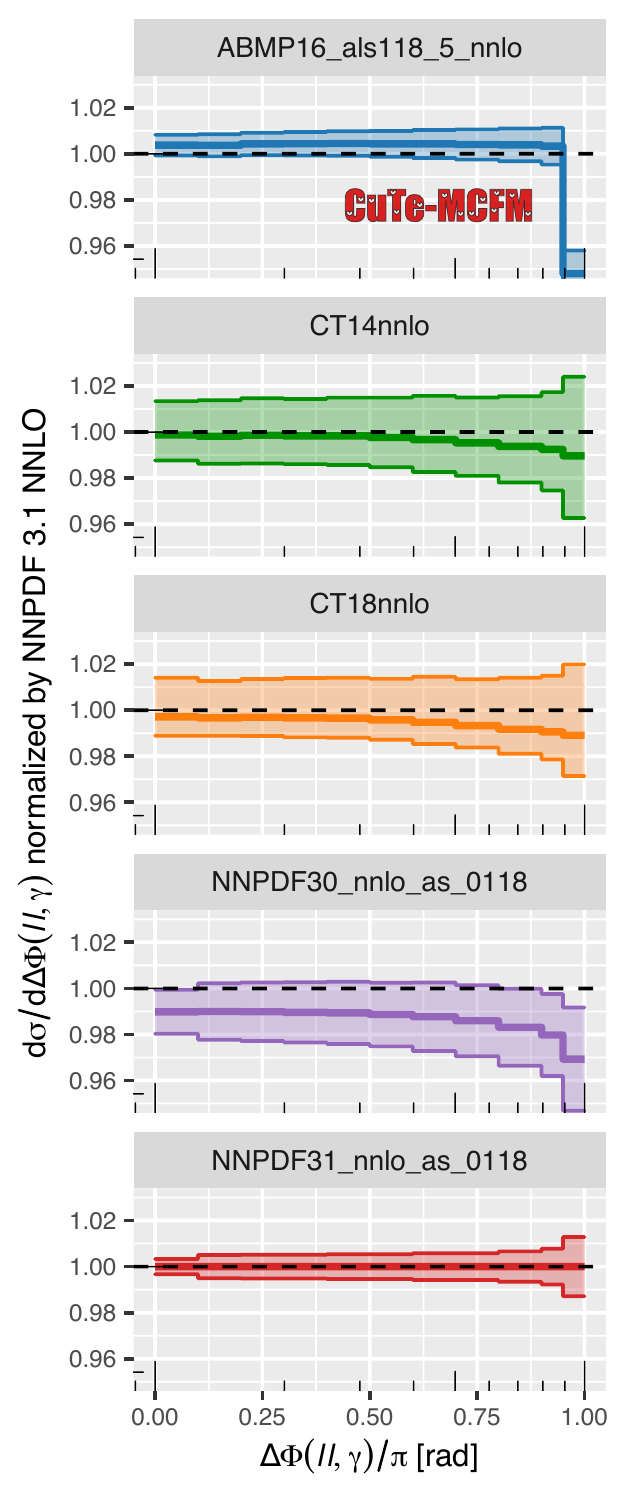}
		\caption{For $\delphi$ distribution}
	\end{subfigure}
	
	\caption{\PDF{} uncertainties for the $q_T$ (a) and $\delphi$ (b) 
		distributions in $Z\gamma$
		production corresponding to the distributions with experimental selection 
		cuts in \cref{fig:cms_zgamma_ptllgamma} 
		and \cref{fig:cms_zgamma_delphi}, respectively.  The {\abbrev ABMP16} grid 
		is not 
		defined below $\mu=\SI{4.47}{\GeV}$, which can be seen as breakdown in the 
		prediction. }
	\label{fig:zgamma_pdferrs}
\end{figure}

\clearpage

\bibliographystyle{JHEP}
\bibliography{refs}

\end{document}